\documentclass{aa}
\usepackage{txfonts}
\usepackage{graphicx}
\begin{document}

\title{The circumstellar envelopes of the Cepheids $\ell$\,Car and RS\,Pup\thanks{Based on observations made with ESO telescopes at Paranal Observatory, under ESO programs 073.D-0142(B), 077.D-0500(A), 078.D-0739(B), 078.D-0739(D) and 081.D-0165(A).} }
\subtitle{Comparative study in the infrared with \emph{Spitzer}, VLT/VISIR and VLTI/MIDI}
\titlerunning{Comparative study of the circumstellar envelopes of $\ell$\,Car and RS\,Pup}

\authorrunning{P. Kervella et al.}
\author{
P.~Kervella\inst{1}
\and
A.~M\'erand\inst{2}
\and
A.~Gallenne\inst{2,1}
}
\offprints{P. Kervella}
\mail{Pierre.Kervella@obspm.fr}
\institute{
LESIA, Observatoire de Paris, CNRS, UPMC, Universit\'e Paris Diderot, 5 Place Jules Janssen,
F-92195 Meudon Cedex, France
\and
European Southern Observatory, Alonso de C\'ordova 3107, Casilla 19001, Santiago 19, Chile
}
\date{Received ; Accepted}
\abstract
{Compact circumstellar envelopes (CSEs) have now been detected around several Cepheids by interferometry. These envelopes are particularly interesting for two reasons: their presence could impact the Cepheid distance scale, and they could imply stellar mass loss.}
{ Our goal is to establish the spatial and spectral properties of the CSEs of $\ell$\,Car and RS\,Pup. This is done through a parametrization of the envelopes in terms of fractional flux (with respect to the star) and angular size.}
{We retrieved archival \emph{Spitzer} images of the two stars ($\lambda = 3.5 - 70\,\mu$m), and obtained new diffraction-limited imaging with the VLT/VISIR camera in BURST mode ($\lambda = 8.6 - 11.9\,\mu$m), as well as interferometric observations with VLTI/MIDI ($\lambda = 8 - 13\,\mu$m). This combination of single-telescope and interferometric techniques allows us to probe the envelopes of the two Cepheids over a broad range of angular scales, from arcminutes to milliarcseconds.}
{The circumstellar envelope of RS\,Pup is resolved spatially at 24 and 70\,$\mu$m by \emph{Spitzer}, and around 10\,$\mu$m by MIDI and VISIR. The envelope of $\ell$\,Car appears much more compact, and is resolved only in the VISIR and MIDI observations. The infrared excesses we detect around RS\,Pup and $\ell$\,Car are both very significant, but differ considerably in spectral and spatial properties. We detect a warm component in the CSE of both stars at a spatial scale of a few 100 to a few 1\,000\,AU. In addition, RS\,Pup presents a very large (several 100\,000\,AU) and cold ($\approx 40$\,K) dusty envelope.}
{The observed properties of the CSEs lead us to propose that the cold dust content of the large reflection nebula surrounding RS\,Pup has an interstellar origin, while the warm CSEs of the two stars were created by ongoing stellar mass loss. We also speculate that the NGC 7023 reflection nebula surrounding the Herbig Be star HD\,200775 is an analogue of RS\,Pup at an age of 100\,000 years. The presence of CSEs around the two brightest long-period Cepheids indicates that warm CSEs is probably common around Cepheids. However, very large dusty envelopes such as that of RS\,Pup are probably rare, as according to our scenario, they require the presence of a high dust density in the interstellar medium at the time of the formation of the Cepheid progenitor.}
\keywords{Stars: circumstellar matter; Stars: variables: Cepheids; Stars: individual: $\ell$ Car, RS Pup; Techniques: high angular resolution; Techniques: interferometric; Infrared: stars}

\maketitle

\section{Introduction\label{intro}}

Exactly a century after the discovery of their Period-Luminosity (P--L) relation (Leavitt~\cite{leavitt08}), and a few years from the centennary of its first calibration (Leavitt \& Pickering~\cite{leavitt12}), Cepheids appear as increasingly complex objects. 
The first IRAS observations in the 1980s led to the discovery of infrared (IR) excess around many Cepheids (McAlary \& Welch~\cite{mcalary86}; Deasy \& Butler~\cite{deasy86}). Based on these observations and IUE ultraviolet spectra, Deasy~(\cite{deasy88}) identified mass loss in a number of Cepheids. A very significant mass loss rate is attributed by this author to RS~Pup ($10^{-6}\,M_{\odot}.{\rm yr}^{-1}$). B\"ohm-Vitense \& Love~(\cite{bohm94}) obtained an even higher value of $2.10^{-5} M_{\odot}.{\rm yr}^{-1}$ for $\ell$\,Car.
Recent interferometric observations revealed compact circumstellar envelopes around several nearby Cepheids (Kervella et al.~\cite{kervella06}; M\'erand et al.~\cite{merand06}; M\'erand et al.~\cite{merand07}). Before these observations, only one Cepheid was known with certainty to be surrounded by a circumstellar envelope (CSE), RS\,Pup. This large nebula ($\approx 2\arcmin$) scatters the light from the central star, and hosts one of the most striking examples of light echoes known to date (Kervella et al.~\cite{kervella08}).
Another Cepheid, SU\,Cas, is close to a nebula, but the physical association is uncertain. The presence of CSEs around Cepheids raises questions about the mass loss mechanisms of these stars. As shown by M\'erand et al.~(\cite{merand07}), long period Cepheids tend to show brighter CSEs in the near-IR $K$ band than short periods, indicating that the mass-loss mechanism could be linked to the pulsation of the star. Interestingly, pulsation driven mass loss models by Neilson \& Lester~(\cite{neilson08a}) predict lower mass loss rates for long periods compared to short periods. Using the same models, Neilson et al.~(\cite{neilson08c}) predicted a sufficiently large IR excess to impact the near and mid-IR P--L relations of LMC Cepheids.

In the present article, we report our observations of the two long-period Cepheids $\ell$\,Car (\object{HD 84810}, \object{SAO 250683}) and RS\,Pup  (\object{HD 68860}, \object{SAO 198944}) using single-telescope and interferometric techniques in order to search for and characterize the mid-IR excess of these stars.
We detail the data sets obtained with the \emph{Spitzer}, VLT/VISIR and VLTI/MIDI instruments in Sect.~\ref{observations}. Sections~\ref{spitzer-analysis} to \ref{midi-analysis} are dedicated to the analysis of the data, while we summarize in Sect.~\ref{vinci-2006} the VINCI data of $\ell$\,Car, previously published by Kervella et al.~(\cite{kervella06}).
We present in Sect.~\ref{properties-sect} the properties of the excess emission present in the spectral and spatial energy distribution of the two stars. The current standard explanation of the presence of Cepheid envelopes is that they were created by mass loss from the the stars themselves, through a mechanism probably associated with pulsation. We propose an alternative scenario for the formation of the CSE of RS\,Pup in Sect.~\ref{discussion}, as well as a critical evaluation of the existing Cepheid mass loss models.
The detailed physical modeling of the CSEs is beyond the scope of the present paper and is not addressed.

In the following, we will phase the observations using reference epochs $T_0$ (expressed in Modified Julian Date) and periods of $T_0 = 52289.916$ and $P = 35.5513$\,days for $\ell$\,Car (Szabados~\cite{szabados89}), $T_0 = 54090.836$ and $P = 41.4389$\,days for RS\,Pup (Kervella et al.~\cite{kervella08}).

\section{Observations and data reduction methods\label{observations}}

\subsection{Spitzer observations\label{spitzer-observ}}

We retrieved the observations of $\ell$\,Car and RS\,Pup obtained with the InfraRed Array Camera (IRAC; Fazio et al.~\cite{fazio04}) and the Multiband Imaging Photometer for \emph{Spitzer} (MIPS; Rieke et al. \cite{rieke04}; Werner et al.~\cite{werner04a}) from the archive\footnote{http://ssc.spitzer.caltech.edu/} at the \emph{Spitzer} Science Center (SSC).
The observations of $\ell$\,Car with IRAC took place on 13 August 2006 (MJD\,53\,960.984, $\phi=0.00$), and with MIPS on 19 July 2006 (MJD\,53\,935.328, $\phi = 0.28$), while RS\,Pup was observed with IRAC on 29 December 2006 (MJD\,54\,098.477, $\phi=0.18$), and with MIPS on 4 December 2006 (MJD\,54\,073.425, $\phi = 0.58$). The IRAC and MIPS wavelengths complement very well our MIDI and VISIR observations around 10\,$\mu$m towards shorter and longer wavelengths.

The IRAC images were taken in the four channels of the camera: 3.8, 4.5, 5.8 and 8.0\,$\mu$m. We used the MOPEX software suite v18.1.5 (Makozov \& Marleau~\cite{makozov05}) to assemble mosaics of the individual images, with a pixel scale of $1.22\arcsec$/pixel.
The MIPS images were obtained in photometry mode at 24\,$\mu$m (silicon-based array) and 70\,$\mu$m (germanium-based detector).
The pipeline recipes that were applied by the archive to convert the raw data into these BCD files are described by Gordon et al.~(\cite{gordon05}). At 70\,$\mu$m, we selected the non-filtered versions of the BCD images. We then used MOPEX to assemble mosaics of the individual images, with a pixel scale of $2.45\arcsec$/pixel (close to the original BCD images) at 24\,$\mu$m and $4\arcsec$/pixel at 70\,$\mu$m (about half of the original pixel scale).

\subsection{VISIR observations}

We used the VISIR instrument (Lagage et al.~\cite{lagage04}), installed at the Cassegrain focus of the Melipal telescope (UT3) of the VLT (Paranal, Chile). VISIR is a mid-IR imager that also provides a slit spectrometer. As it is a ground-based instrument, its sensitivity is severely limited by the high thermal background of the atmosphere, compared for instance to \emph{Spitzer}, but its resolving power is ten times higher, thanks to the 8\,m diameter of the primary mirror. VISIR is therefore very well suited for our programme to search for compact, spatially resolved mid-IR emission around $\ell$\,Car and RS\,Pup.

However, under standard conditions at Paranal (median seeing of $0.8\arcsec$ at 0.5\,$\mu$m), the 8\,m telescope is not diffraction limited at 10\,$\mu$m (seeing $\approx 0.4\arcsec$ vs. $0.3\arcsec$ diffraction). Instead of a pure Airy diffraction pattern, several moving speckles and tip-tilt usually degrade the quality of the image (see e.g. Tokovinin, Sarazin \& Smette~\cite{tokovinin07}). To overcome this limitation, a specific mode of the instrument, called the BURST mode, was introduced by Doucet et al.~(\cite{doucet06}, \cite{doucet07}). Its principle is to acquire very short exposures ($\Delta t \lesssim 50$\,ms), in order to keep the complete integration within a fraction of the coherence time ($\approx 300$\,ms at Paranal in the mid-IR). The detector is therefore read very quickly, and the resulting images freeze the turbulence. It is subsequently possible to select the best images that present a single speckle (``lucky imaging''), and are thus diffraction-limited. Our processing consists of three main steps: 1) integer-pixel recentering of the chop-nod background corrected frames, 2) oversampling of each frame by a factor of 4 using spline interpolation (resampled scale of 19\,mas/pixel), 3) fine recentering with a Gaussian fitting procedure, 4) averaging of the best frames with respect to the Strehl ratio. A more detailed description of the processing is presented in Kervella \& Domiciano de Souza~(\cite{kervella07}). 

The observations of $\ell$\,Car and RS\,Pup were performed during the nights of 23 and 24 May 2008. Table~\ref{visir_log} lists the sequence of our VISIR observations of the Cepheids and their associated calibrator stars. They were observed immediately before and after the Cepheids to monitor the evolution of the sky transparency (photometry) and thermal infrared seeing (PSF calibration). 
The observations were obtained in three filters\footnote{http://www.eso.org/sci/facilities/paranal/instruments/visir/inst/}: PAH1, PAH2 and SiC covering respectively the following wavelength domains: $8.59 \pm 0.42$\,$\mu$m, $11.25 \pm 0.59$\,$\mu$m and $11.85 \pm 2.34$\,$\mu$m.

For an unknown reason, possibly a temporary burst of very bad seeing or thin cirrus clouds, observation \#c suffers from a low photometric SNR, and was not included in the analysis. Due to the lower sensitivity of VISIR in the SiC filter, we could not recenter the individual VISIR images for observation \#p.

\begin{table}
\caption{Log of the VISIR observations of $\ell$\,Car and RS\,Pup. MJD$^*$ is the modified Julian date of the start the exposures on the target minus 54\,000, and $\phi$ the phase of the Cepheid. The Detector Integration Time (DIT) is given in milliseconds for one BURST image. $\theta$ is the seeing in the visible ($\lambda=0.5\,\mu$m) measured by the observatory DIMM sensor, in arcseconds, and AM is the airmass of the observation. The duration of each observation was 360\,s on target.} 
\label{visir_log}
\begin{tabular}{lcclcccc}
\hline \hline
\# & MJD$^*$ & $\phi$ & Star & Filter & DIT & $\theta$\ & AM \\
\hline
\noalign{\smallskip}
a & 609.953 & & HD\,76110 & PAH1 & 16 & 1.5 & 1.09 \\
b & 609.961 & & HD\,76110 & PAH2 & 8 & 1.4 & 1.11 \\
c$^{*}$ & 609.980 & 0.53 & RS\,Pup & PAH1 & 16 & 1.6 & 1.26 \\
d & 609.987 & 0.53 & RS\,Pup & PAH2 & 8 & 1.3 & 1.31 \\
e & 610.002 & 0.26 & $\ell$\,Car & PAH1 & 16 & 1.5 & 1.34 \\
f & 610.009 & 0.26 & $\ell$\,Car & PAH2 & 8 & 1.3 & 1.37 \\
g & 610.035 & & HD\,89682 & PAH1 & 16 & 1.6 & 1.26 \\
h & 610.042 & & HD\,89682 & PAH2 & 8 & 1.8 & 1.29 \\
i & 610.081 & & HD\,98118 & PAH1 & 16 & 1.7 & 1.31 \\
j & 610.089 & & HD\,98118 & PAH2 & 8 & 1.6 & 1.36 \\
\hline
\noalign{\smallskip}
k & 610.960 & & HD\,76110 & PAH1 & 16 & 0.7 & 1.11 \\
l & 610.967 & & HD\,76110 & SIC  & 20 & 0.9 & 1.13 \\
m & 610.979 & 0.29 & $\ell$\,Car & PAH1 & 16 & 1.0 & 1.31 \\
n & 610.986 & 0.29 & $\ell$\,Car & SIC  & 20 & 0.8 & 1.32 \\
o & 610.998 & 0.55 & RS\,Pup & PAH1 & 16 & 0.8 & 1.38 \\
p$^{*}$ & 611.005 & 0.55 & RS\,Pup & SIC  & 20 & 0.9 & 1.44 \\
q & 611.016 & & HD\,89682 & PAH1 & 16 & 0.6 & 1.23 \\
r & 611.024 & & HD\,89682 & SIC  & 20 & 0.9 & 1.24 \\
s & 611.073 & & HD\,124294 & PAH1 & 16 & 0.8 & 1.06 \\
t & 611.081 & & HD\,124294 & SIC  & 20 & 0.8 & 1.05 \\
\hline
\end{tabular}
\begin{list}{}{}
\item[$^{*}$] Observations \#c and \#p suffer from low photometric SNR.
\end{list}
\end{table}

\subsection{MIDI observations\label{midi_obs}}

MIDI (Leinert et al.~\cite{leinert03}; Ratzka et al.~\cite{ratzka07}) is the mid-infrared two-telescope beam combiner of the Very Large Telescope Interferometer (VLTI; Glindemann et al. \cite{glindemann04}). This instrument is a classical Michelson interferometer working in the astronomical $N$ band (7.6--13.3\,$\mu$m). It can combine the light from two VLT Unit Telescopes (8.2\,m, UTs) or two Auxiliary Telescopes (1.8 m, ATs). For the reported observations of $\ell$\,Car and RS\,Pup, we used a prism with a spectral resolution of $R=\lambda/\Delta \lambda \simeq 30$ to obtain spectrally dispersed fringes.
During the observations, the secondary mirrors of the two Unit Telescopes were chopping with a frequency of 2\,Hz to properly sample the sky background. 
MIDI has two photometric calibration modes: {\tt SCI\_PHOT}, in which the photometry is measured simultaneously with the interferences fringes, and {\tt HIGH\_SENS}, in which the flux is measured separately before the interferometric observations. The visibility is derived from the ratio of correlated flux (modulation in the interferometric fringes) with the total flux (given by the photometry).
Due to the relatively low thermal-IR brightness of the two Cepheids, the observations presented here were obtained in {\tt HIGH\_SENS} mode, using the low dispersion setup (spectral resolution of $\lambda/\Delta \lambda \approx 30$).

For the raw data processing, we used two different software packages: {\tt MIA} developed at the Max-Planck-Institut f\"ur Astronomie and {\tt EWS} developed at the Leiden Observatory ({\tt MIA+EWS}\footnote{http://www.strw.leidenuniv.nl/$\sim$nevec/MIDI/index.html}, version 1.5.2) in order to extract first the instrumental squared coherence factors and then the calibrated visibilities $V(\lambda)$ (Chesneau~\cite{chesneau07}). We found a good agreement between the results of the MIA and EWS packages within the error bars. In the following we will refer to the results obtained with the EWS package.
To remove the instrumental and atmospheric signatures, three calibrators were observed together with the Cepheids : \object{HD 70555} ($\theta_{\rm LD} = 2.54 \pm 0.04$\,mas) for RS\,Pup, \object{HD 67582} ($\theta_{\rm LD} = 2.39 \pm 0.06$\,mas) and \object{HD 94510} ($\theta_{\rm LD} = 2.23 \pm 0.01$\,mas) for $\ell$\,Car. They were chosen in the Cohen et al.~(\cite{cohen99}) catalog, except \object{HD 94510} which was selected from the {\tt CalVin}\footnote{http://www.eso.org/observing/etc/} database. The calibrators are nearly unresolved by MIDI on the selected baselines ($V \geqslant 95$\%), and the systematic uncertainty associated with their {\it a priori} angular diameter error bars is negligible compared to the typical precision of the MIDI visibilities.

The log of the MIDI observations of $\ell$\,Car and RS\,Pup is presented in Table~\ref{midi_log}. The new observations took place on 17 April 2006, and 7-9 November 2006, on the UT1-UT4 baseline. To complete this new data set, we retrieved from the ESO archive the observations of $\ell$\,Car obtained on the night of 8-9 April 2004 on the UT2-UT3 baseline, and we reprocessed them using the {\tt EWS} package. These observations were originally presented by Kervella et al.~(\cite{kervella06}), but they were processed at the time using the MIDI Data Reduction Software developed by the Paris Observatory. The present reprocessing is intended to ensure the consistency of the visibilities with the more recent observations. The 2004 observations were obtained with natural seeing, while the 2006 observations employed the MACAO adaptive optics system (Arsenault et al.~\cite{arsenault03}), that provides an improved stability of the photometry in MIDI.

The mean photospheric angular diameters of $\ell$\,Car and RS\,Pup are respectively $\theta_{\rm LD} = 2.992 \pm 0.012$\,mas (measured by Kervella et. al.~\cite{kervella04a}), and $\theta_{\rm LD} \approx 1.015$\,mas (predicted by Moskalik \& Gorynya~\cite{moskalik05}, \cite{moskalik06}) . Considering the projected baseline lengths and the wavelength of the MIDI observations, the photosphere of $\ell$\,Car is only slightly resolved at 8\,$\mu$m with $V \approx 93\%$ and RS\,Pup is mostly unresolved, with $V \approx 99\%$. A measured visibility significantly lower than these figures will therefore be characteristic of the presence of additional emission close to the star. The analysis of the MIDI visibilities is presented in Sect.~\ref{lcar-midi} and \ref{rspup-midi} respectively for $\ell$\,Car and RS\,Pup. The presence of a strong ozone atmospheric absorption band over the range $\lambda = 9.3-10.0\,\mu$m (Lord~\cite{lord92}) can make the visibilities unreliable in this domain, as the photometric calibration is made more difficult by the lower flux level.

\begin{table}
\caption{Log of the MIDI observations of $\ell$\,Car and RS\,Pup. $\phi$ is the phase of the observation for the Cepheids, relative to the maximum light in $V$ (see Sect.~Ê\ref{intro} for $P$ and $T_0$ values), $B$ is the projected baseline length, PA is the position angle of the projected baseline (counted East of North\,$=0^\circ$), and AM is the airmass.}
\label{midi_log}
\begin{tabular}{lcclccc}
\hline \hline
\# & MJD & $\phi$ & Target & $B$ (m) & $PA$ ($^\circ$) & AM \\
\hline
\noalign{\smallskip}
A$^{*}$ & 53104.060 & 0.90 & $\ell$\,Car & 40.82 & 48.7 & 1.27 \\
B$^{*}$ & 53104.076 & & HD\,67582 & 38.86 & 58.9 & 1.20 \\
C$^{*}$ & 53104.090 & 0.90 & $\ell$\,Car &  39.18 & 56.4 & 1.28 \\
D$^{*}$ & 53104.104 & & HD\,67582 & 35.85 & 63.5 & 1.31 \\
\hline
\noalign{\smallskip}
E & 53\,842.082 & 0.66 & $\ell$\,Car &  117.52 & 74.4 & 1.30 \\
F & 53\,842.099 & & HD\,67582 & 122.80 & 70.1 & 1.21 \\
\hline
\noalign{\smallskip}
G$^{**}$ & 54\,045.290 & 0.90 & RS\,Pup &  121.25 & 31.1 & 1.29 \\
\hline
\noalign{\smallskip}
H & 54\,046.275 & 0.92 & RS\,Pup &  119.80 & 27.5 & 1.36 \\
I & 54\,046.312 & & HD\,70555 & 122.87 & 37.2 & 1.20 \\
J$^{***}$ & 54\,046.343 & 0.41 & $\ell$\,Car &  130.10 & 26.5 & 1.52 \\
K & 54\,046.372 & & HD\,94510 & 130.11 & 20.7 & 1.57 \\
\hline
\noalign{\smallskip}
L & 54\,047.267 & 0.95 & RS\,Pup &  119.10 & 25.7 & 1.40 \\
M & 54\,047.286 & & HD\,70555 & 119.62 & 30.3 & 1.31 \\
\hline
\noalign{\smallskip}
N & 54\,048.244 & & HD\,70555 & 114.88 & 17.3 & 1.63 \\
O & 54\,048.270 & 0.97 & RS\,Pup &  119.79 & 27.5 & 1.36 \\
P & 54\,048.347 & 0.46 & $\ell$\,Car &  130.03 & 29.3 & 1.48 \\
Q & 54\,048.366 & & HD\,94510 & 130.10 & 20.4 & 1.57 \\
\hline
\end{tabular}
\begin{list}{}{}
\item[$^{*}$] $\ell$\,Car observations obtained in 2004 (see Sect.~\ref{midi_obs}).
\item[$^{**}$] Observation \#G of RS\,Pup was not calibrated.
\item[$^{***}$] Observation \#J of $\ell$\,Car has a low interferometric flux.
\end{list}
\end{table}

\subsection{Fourier analysis technique \label{fourier-technique}}

In Sections~\ref{spitzer-analysis}, \ref{visir-analysis} and \ref{midi-analysis} we present the analysis of the spatially resolved emission around $\ell$\,Car and RS\,Pup as observed by \emph{Spitzer}, VISIR and MIDI. The spatial scale sampled by these three instruments covers a broad range from arcminutes with \emph{Spitzer} to arcseconds with VISIR and milliarcseconds with MIDI. This coverage allows us to probe the different scales of the infrared emission around the two Cepheids, both angularly and spectrally.

In the present work, we search for spatially extended emission in the \emph{Spitzer} and VISIR images using a Fourier based technique, similar in its principle to the calibration technique used for interferometry. For this purpose, we divide the Fourier transform modulus of the image of the Cepheid ($I_{\rm Cepheid}$) by that of the calibrator image ($I_{\rm Calibrator}$):
\begin{equation}
\Psi(\nu_x, \nu_y) = \left| \frac{\mathrm{FFT}\left[ I_{\rm Cepheid}(x,y)\right]}{\mathrm{FFT}\left[ I_{\rm Calibrator}(x,y)\right]} \right| \label{fouriercal}
\end{equation}
where $(x,y)$ and $(\nu_x, \nu_y)$ are conjuguate Fourier variables (sky coordinates and angular spatial frequencies for instance).
This Fourier technique has three major advantages over direct image analysis techniques (e.g. PSF subtraction).
Firstly, it is similar in its principle to a deconvolution of the image, but the absence of an inverse Fourier transform step after the division in Eq.~\ref{fouriercal} avoids the appearance of uncontrolled phase noise. Secondly, the background noise is rejected more efficiently, as it is well separated from the image itself by its spatial frequency spectrum. It is therefore more sensitive than direct image analysis techniques. Thirdly, interferometric observations provide measurements of the Fourier transform of the intensity distribution of the observed object. The Fourier analysis applied to images allows us to analyze them within exactly the same framework as interferometric observations. The derived CSE properties (angular size, flux contribution) are therefore strictly comparable, which is particularly interesting for the present multi-instrument study.

We compute the ring median of $\Psi$ (i.e. median for a given spatial frequency radius $\nu$ over all azimuth directions in the Fourier space) to increase the SNR. This operation provides us with a quantity $\Psi(\nu)$ that is mathematically equivalent to the squared visibility of interferometric fringes. In the present Fourier-based analysis, we will always consider ring median properties in order to obtain the global properties of the CSEs. The zero frequency, corresponding to the average ratio of the Cepheid and calibrator images, is normalized to unity in all cases.
Our estimation of the error bars on $\Psi$ takes into account both calibration and statistical uncertainties. The calibration uncertainties are derived from the relative dispersion of the PSF calibrator's Fourier modulus over the night. The statistical uncertainty is taken as the rms dispersion of the calibrated $\Psi$ function over all azimuth directions, for each spatial frequency. Both error sources are added quadratically to get the total uncertainty. Our choice to consider azimuth-averaged $\Psi$ values implies that deviations from central symmetry of the CSEs will be undetectable. However, the error bars associated with the derived $\Psi$ values include the dispersion that may arise from asymmetric flux contributors.

From the Van Cittert-Zernike theorem, the $\Psi(\nu)$ functions are the Fourier transforms of the spatial flux distribution of the CSEs. It is therefore possible to retrieve the CSE intensity distribution from the shape of these curves, by adjusting the Fourier transform of an approximate model of the intensity distribution. We thus define the simple model of a point-like star surrounded by a Gaussian shaped CSE with a FWHM of $\rho_\lambda$ and a relative flux contribution of
\begin{equation}
\alpha_\lambda = \frac{f_{\mathrm{CSE}}(\lambda)}{f_\star(\lambda)},
\end{equation}
where $f_\mathrm{CSE}$ is the photometric flux from the Gaussian envelope and $f_\star$ the flux from the unresolved component.
This model is suitable for the \emph{Spitzer}, VISIR and MIDI UT2-UT3 observations, for which the FWHM of the stellar photosphere is significantly smaller than the diffraction limit $\lambda/D$ of the instrument (with $D$ the aperture). For the longer baseline UT1-UT4 observations with MIDI, the photospheric extension must be taken into account (see Sect.~\ref{midi-analysis}).
This type of model was already used by Kervella et al.~(\cite{kervella06}) and Kervella \& Domiciano de Souza~(\cite{kervella06b}). The $\Psi(\nu)$ function of this model is:
\begin{equation} \label{Gaussian-model}
\Psi(\nu, \rho, \alpha_\lambda)=\frac{1} { 1 + \alpha_\lambda}\ \left[ 1 + \alpha_\lambda\ \exp\left( -\frac{\left( \pi\, \rho\, \nu \right)^2}{4\,\ln 2} \right) \right].
\end{equation}
%

\subsection{Fourier technique validation}

\begin{figure}[]
\centering
\includegraphics[bb=12 8 363 281, width=8.7cm, angle=0]{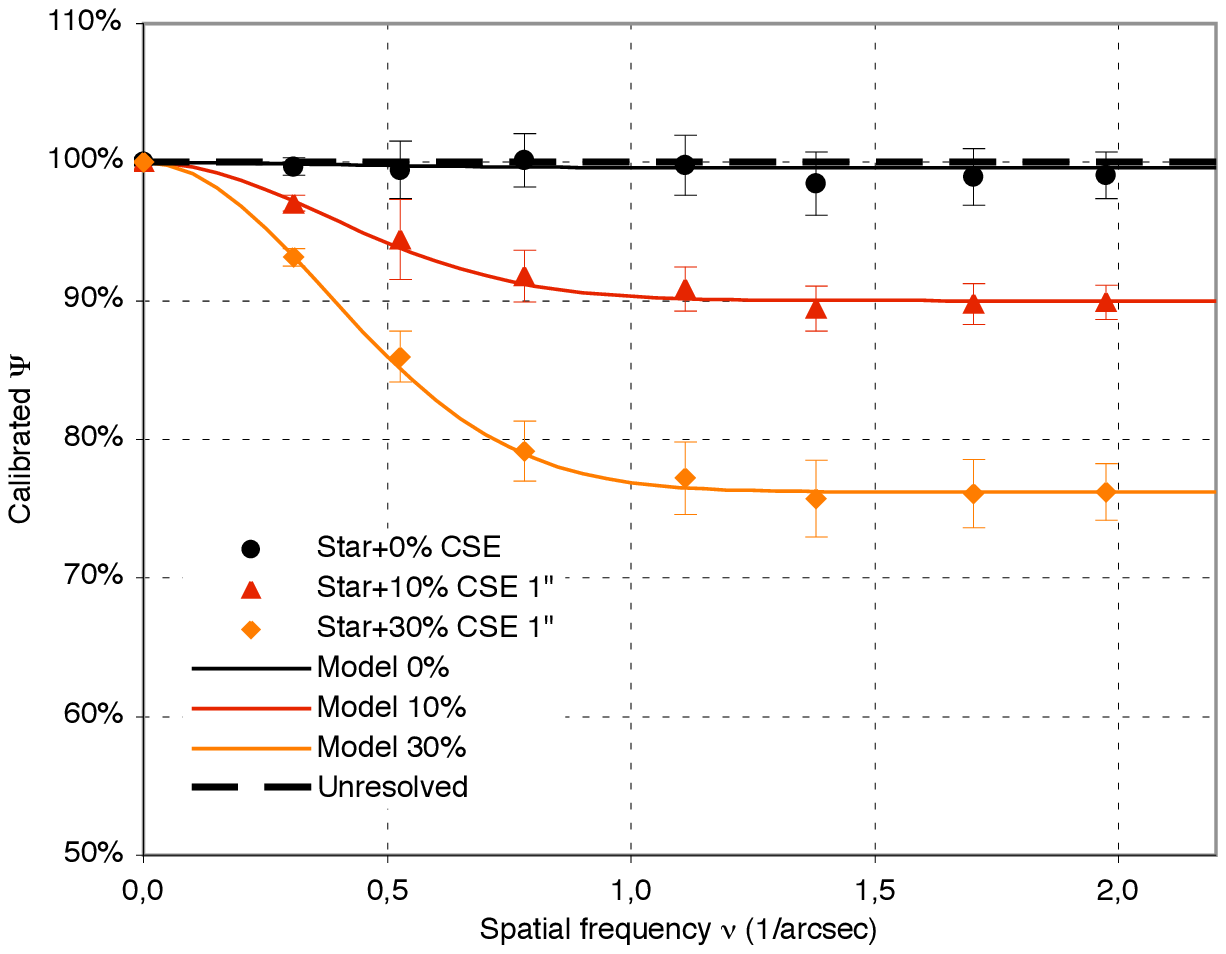}
\caption{$\Psi$ function in the PAH1 filter resulting from the Fourier analysis of HD\,89682 considered as the scientific target, with HD\,98118 and HD\,124294 as PSF calibrators. To validate our Fourier analysis method, we artificially introduced two Gaussian CSEs with fluxes of 10 and 30\% of the stellar flux and a FWHM of 1\arcsec, as discussed in Sect.~\ref{fourier-technique}.\label{visir-calib-psi}}
\end{figure}

As a validation test for our Fourier analysis technique, we applied it to the VISIR observations of \object{HD 89682} obtained in the PAH1 filter, using it as the scientific target (observations \#g and \#q in Table~\ref{visir_log}). We used \object{HD 98118} and \object{HD 124294} as PSF calibrators (observations \#i and \#s). As shown in Fig.~\ref{visir-calib-psi} (upper curve), the resulting average $\Psi$ function of \object{HD 89682} is statistically identical to 100\%, as expected for an unresolved target. The derived flux contribution for the extended emission is $\alpha = 0.4 \pm 0.5\%$, while the value of $\rho$ is not constrained.

To check the sensitivity of the Fourier technique to the presence of a resolved Gaussian CSE component, we created artificial images using a point-like central source surrounded by an extended Gaussian emission with a fixed FWHM of $1\arcsec$ and two values of its flux contribution $\alpha$ relative to the central star: 10\% and 30\%. These model images were then convolved with the \#g and \#i images of \object{HD 89682} separately, to obtain realistic images (separately for each observing night) of the model as observed with VISIR. We then processed these synthetic images separately for each night using the Fourier technique. We finally took the average of the $\Psi$ functions of the two nights, that are shown in Fig.~\ref{visir-calib-psi} (lower two curves). By fitting the Gaussian model presented in Eq.~\ref{Gaussian-model}, we finally derive the following $\alpha$ and $\rho$ parameters for the star+10\% CSE model:
\begin{equation}
\rho(\mathrm{Model\,10\%}) = 0.99 \pm 0.12\,\mathrm{arcsec},
\end{equation}
\begin{equation}
\alpha(\mathrm{Model\,10\%}) = 11.1 \pm 1.0\,\%,
\end{equation}
and for the star+30\% CSE model:
\begin{equation}
\rho(\mathrm{Model\,30\%}) = 1.00 \pm 0.04\,\mathrm{arcsec},
\end{equation}
\begin{equation}
\alpha(\mathrm{Model\,30\%}) = 31.2 \pm 1.2\,\%.
\end{equation}
The derived relative fluxes and FWHMs are in agreement within $1\sigma$ with the input model values, thus validating our Fourier-based approach.
In the following Section, we detail the application of this analysis approach to the \emph{Spitzer}, VISIR and MIDI instruments, by order of increasing resolution.
 
\section{Spitzer data analysis\label{spitzer-analysis}}

In this Section, we present several analysis methods applied to the \emph{Spitzer} images of $\ell$\,Car and RS\,Pup. Table~\ref{spitzer-summary} summarizes the results of these different techniques.

\begin{table}
\caption{Summary of the results of the analysis of the \emph{Spitzer} images.} 
\label{spitzer-summary}
\begin{tabular}{llll}
\hline \hline
Technique & IRAC & MIPS 24\,$\mu$m & MIPS 70\,$\mu$m \\
\hline
\noalign{\smallskip}
{\bf $\ell$\,Car} \\
\hline
Image & saturated & unresolved & unresolved\\
\hline
 Fourier & saturated & unresolved & unresolved\\
\hline  
\hline
\noalign{\smallskip}
{\bf RS\,Pup}\\
\hline
Image & unresolved & $\rho \approx 26\arcsec$ & $\rho \approx 118\arcsec$\\
 & & $\alpha \approx 100$\% & (star invisible) \\
\hline
 Fourier & unresolved & $\rho = 60 \pm 3\arcsec$ & not applicable\\
 & & $\alpha = 98.6 \pm 1.3$\% & (star invisible) \\
\hline
\end{tabular}
\end{table}

\subsection{$\ell$\,Car images\label{lcar-spitzer}}

$\ell$\,Car was observed with both the IRAC and MIPS instruments of \emph{Spitzer}, in all the available spectral channels of these instruments:  3.8 to 8.0\,$\mu$m for IRAC, and 24 to $160\,\mu$m for MIPS. The 160\,$\mu$m observation is not useable, and the IRAC images of $\ell$\,Car are all saturated close to the star position, and therefore are not suitable for small-scale CSE searches. On a larger spatial scale, the IRAC images do not present an obvious extended emission that would indicate the presence of a very extended nebula ($\gtrsim 30\arcsec$ from the star). 

In order to check for the presence of spatially extended emission from the immediate surroundings of $\ell$\,Car in the MIPS images, we used as references the experimental point response functions (PRFs) provided in the MOPEX calibration package that are elaborated from calibration observations of the Trapezium in Orion. We also checked synthetic PRFs from the \emph{sTinyTim} software package\footnote{http://ssc.spitzer.caltech.edu/archanaly/contributed/stinytim/}, but they do not provide as clean subtraction residuals as the MOPEX experimental PRFs.

\begin{figure}[]
\centering
\includegraphics[width=8.9cm]{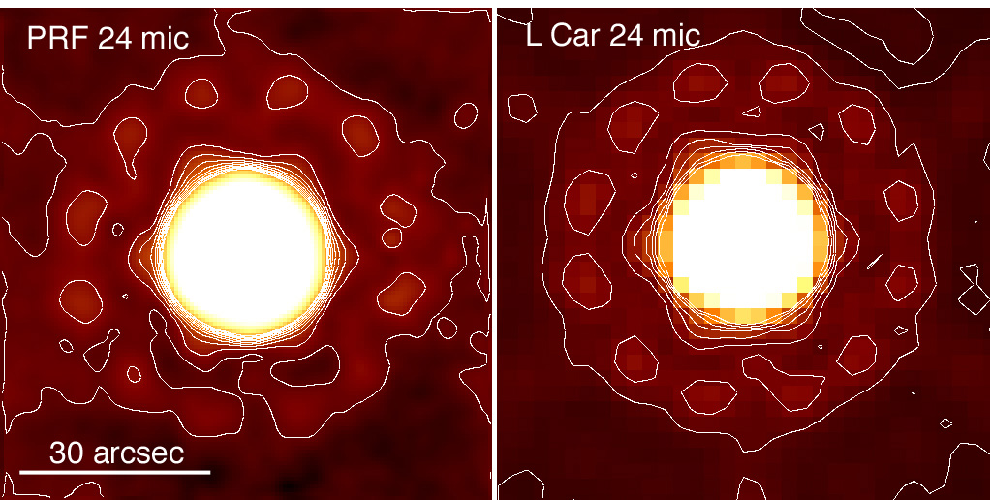}
\includegraphics[width=8.9cm]{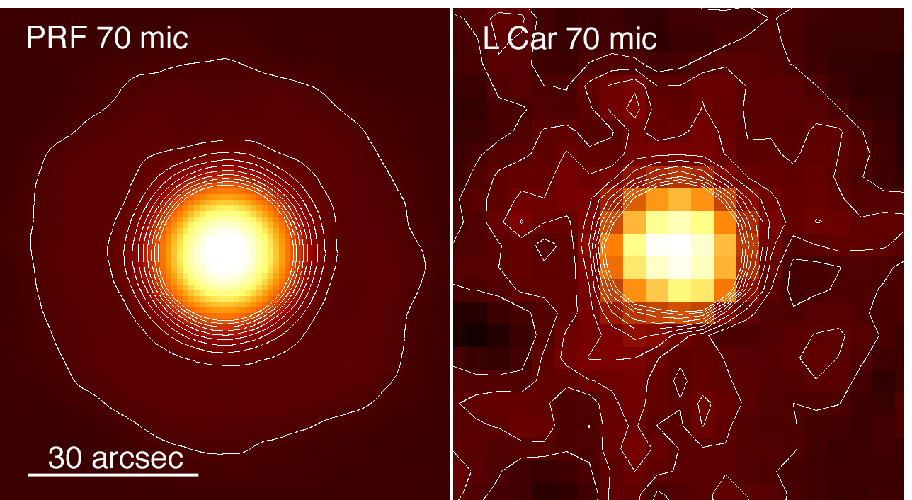}
\caption{\emph{Spitzer} images of the reference PRF and $\ell$\,Car at 24 and 70\,$\mu$m.
\label{lcar-images}}
\end{figure}

The MIPS images of the PRF reference and $\ell$\,Car at 24 and 70\,$\mu$m are shown in Fig.~\ref{lcar-images}. The two images on the left show the oversampled PRF of \emph{Spitzer} ($0.6125\arcsec$/pix) and the 24\,$\mu$m image of $\ell$\,Car ($2.45\arcsec$/pix). The two images on the right show the same data for 70\,$\mu$m (pixel scales of 1 and $4\,\arcsec$/pix). The square-root color scale at 24\,$\mu$m has been chosen to enhance the faint extensions of the PRF, hence it cuts the inner core of the star images. The 70\,$\mu$m color scale is linear, with contours corresponding to 3 to 30\% of the peak level (steps of 3\%).

\subsection{RS\,Pup images\label{rspup-spitzer}}

RS\,Pup was observed both with IRAC's sub-array and MIPS (in all channels except at 160\,$\mu$m).
We searched for extended emission in the IRAC images using the Fourier transform technique presented in Sect.~\ref{fourier-technique}. We used for this purpose a PSF calibrator, \object{BD+68 1022}, that was observed on 20 June 2008 specifically for the calibration of the instrumental PRF. RS\,Pup appears unresolved in the four channels of IRAC. From the resolution of the IRAC observations, we can set an upper limit on the extension of the CSE to $5\arcsec$ in the IRAC bands.

\begin{figure*}[ht]
\centering
\includegraphics[height=6.7cm]{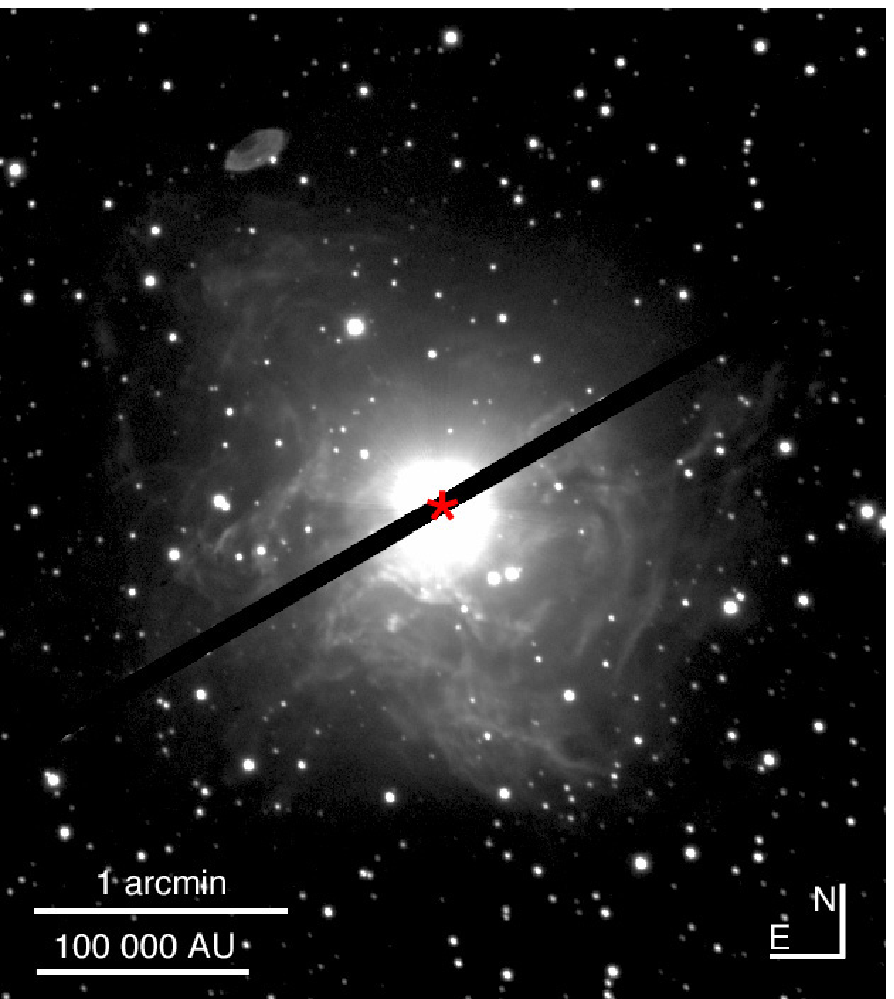}
\includegraphics[height=6.7cm]{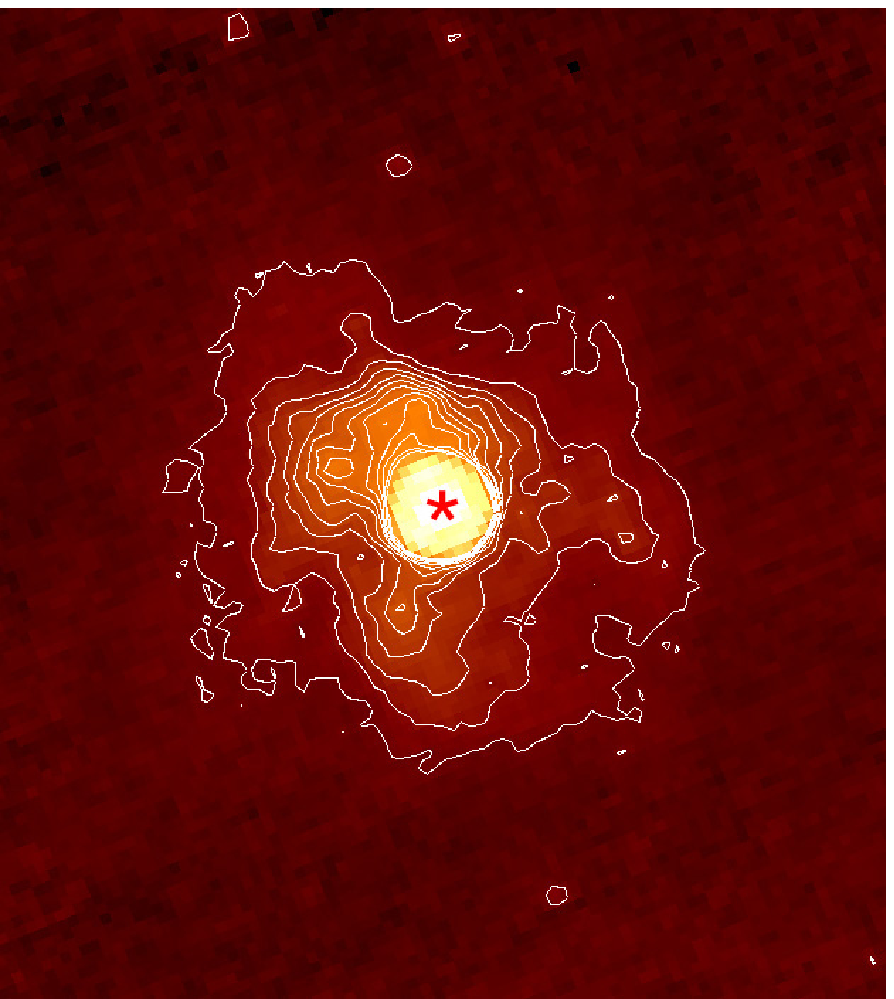} 
\includegraphics[height=6.7cm]{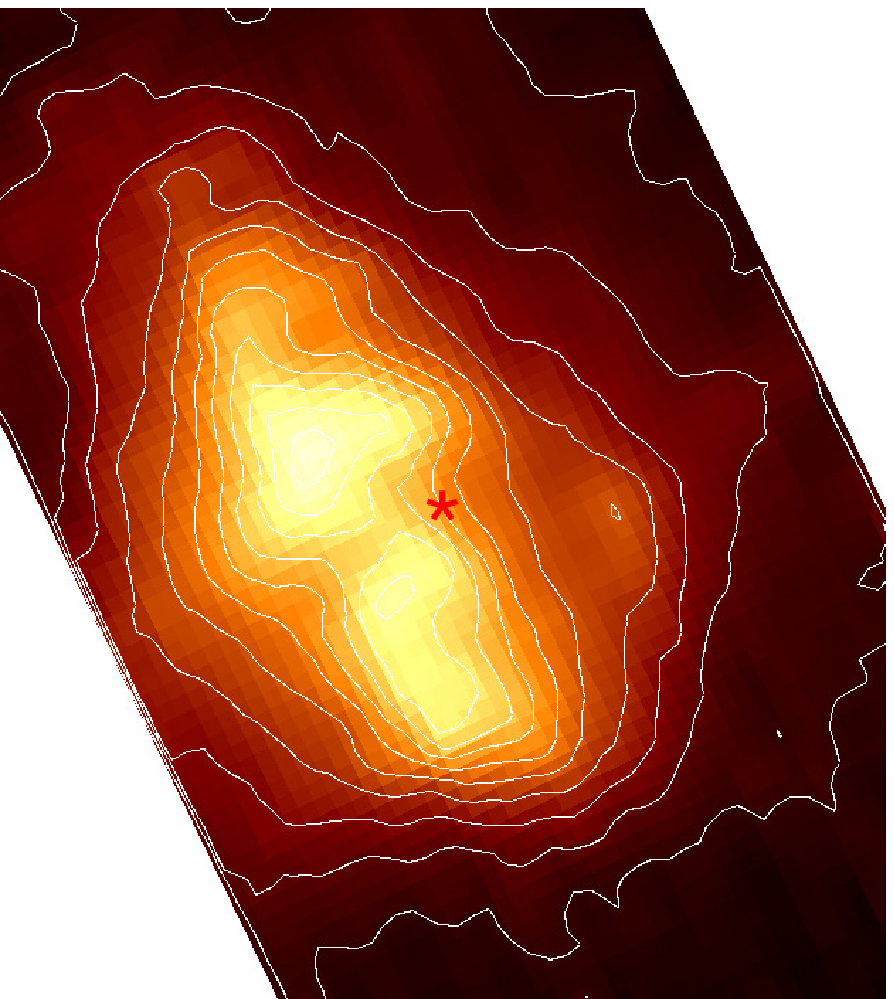} 
\caption{Images of the nebula surrounding RS\,Pup in the $V$ band (left, pulsation phase $\phi=0.877$), at 24\,$\mu$m (middle) and at 70\,$\mu$m (right), to the same spatial scale. The images are not corrected for the background emission. The position of the star is marked with the $\ast$ symbol. The contours are 576 to 705\,$\mu$Jy.arcsec$^{-2}$ (step 5.9) in the 24\,$\mu$m image, and 627 to 5012\,$\mu$Jy.arcsec$^{-2}$ (step 627) in the 70\,$\mu$m image. The linear scale is given for a distance of 1\,992\,pc (Kervella et al.~\cite{kervella08}). \label{rspup-images}}
\end{figure*}

The left part of Figure~\ref{rspup-images} shows the circumstellar nebula of RS\,Pup in the $V$ band, from observations with the NTT/EMMI instrument by Kervella et al.~(\cite{kervella08}). In this wavelength range, we observe the scattered light from the central star.
The middle and right images are the \emph{Spitzer}/MIPS images at 24 and 70\,$\mu$m. They show the thermal emission from the circumstellar dust (the three images have the same spatial scale and orientation). The nebula appears elongated along a north-east axis both in scattered light and IR emission. Interestingly, RS\,Pup itself is offset from the brightest part of the nebula, particularly at 70\,$\mu$m. It also remains invisible at this long wavelength as the image is fully dominated by the thermal emission from the envelope.

\subsection{Radial CSE intensity distribution \label{cse-sb}}

\begin{figure}[]
\centering
\includegraphics[bb=12 8 363 281, width=8.7cm, angle=0]{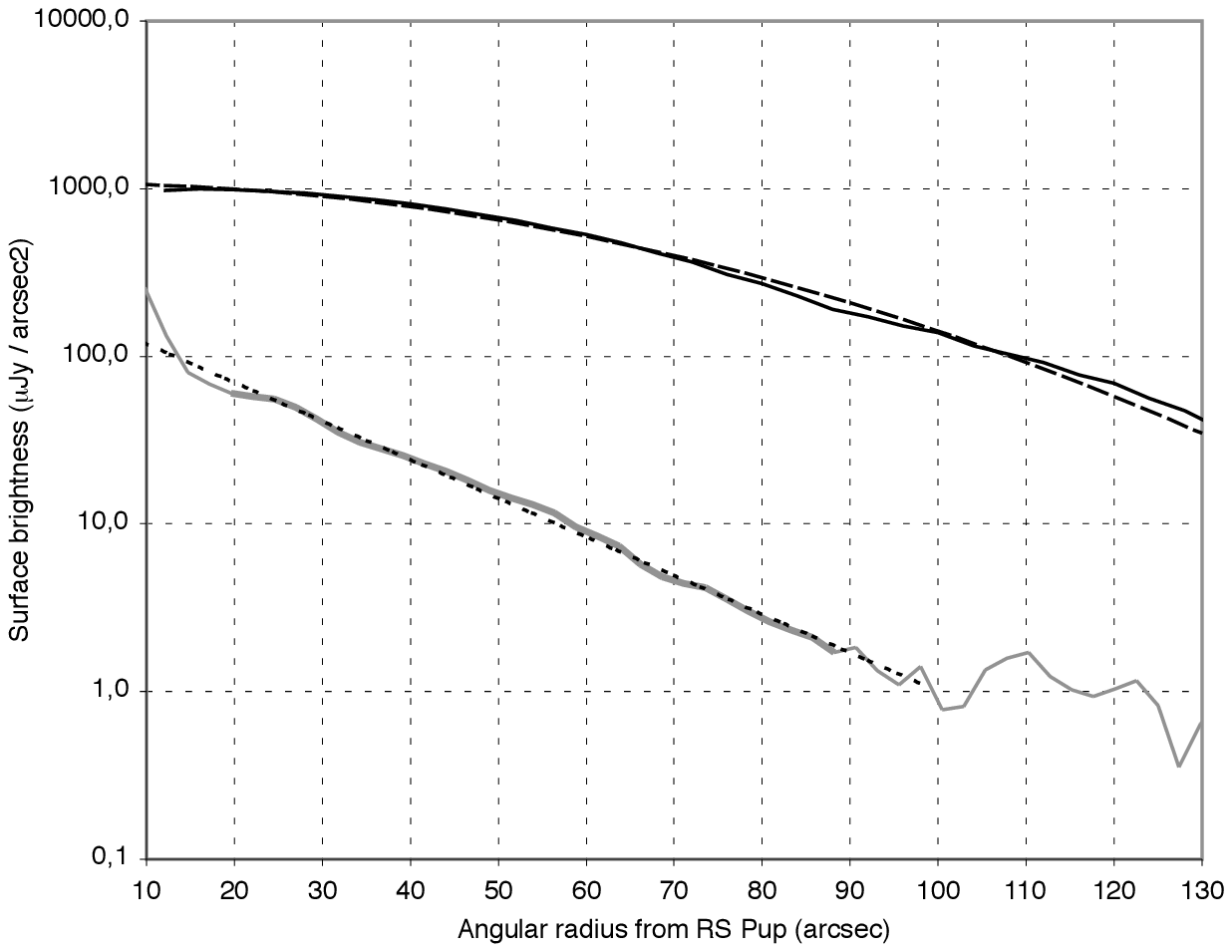}
\caption{Surface brightness of the CSE of RS\,Pup at 70\,$\mu$m (top, black curve) and 24\,$\mu$m (bottom, grey curve) as a function of the angular radius to the star. The result of an exponential fit to the $24\,\mu$m curve between 20 and 90\,arcsec is shown as a dotted line. The dot-dashed line is a Gaussian fit to the 70\,$\mu$m profile (see text for details). \label{rspup-sb}}
\end{figure}

The average surface brightness at 24 and 70\,$\mu$m of RS\,Pup's CSE as a function of the distance from the central star is shown in Fig.~\ref{rspup-sb}. The maximum surface brightness of the CSE of RS\,Pup reaches 0.1\,mJy.arcsec$^{-2}$ at 24\,$\mu$m (outside the star image) and more than 4\,mJy.arcsec$^{-2}$ at 70\,$\mu$m. While the CSE is inhomogeneous locally, its overall surface brightness follows closely an exponential law as a function of the angular separation from the star. An {\it ad hoc} fit from 20 to $90\,\arcsec$ from the star at 24\,$\mu$m  gives:
\begin{equation}
f_{24\,\mu{\rm m}}(\mathrm{RS\,Pup\ CSE}) = 201.6 \, \exp{\left(-\frac{\theta}{18.9}\right)}
\end{equation}
with $f_{24\,\mu{\rm m}}$ in $\mu$Jy/arcsec$^2$ and $\theta$ in arcsec.
This corresponds to a full width at half maximum of $26\,\arcsec$ at 24\,$\mu$m (but the distribution is not strictly Gaussian).
As the wings of the PSF of RS\,Pup are included in this fit, they tend to create a more peaked light distribution than the nebular emission alone.

The distribution of flux in the nebula at 70\,$\mu$m does not show this effect, as the stellar PSF is negligible compared to the envelope emission. It closely follows a Gaussian law of the form:
\begin{equation}
f_{70\,\mu{\rm m}}(\mathrm{RS\,Pup\ CSE}) = 1081 \, \exp{\left[-\left(\frac{\theta}{49.6}\right)^2\right]} \label{Gaussian-70mic}
\end{equation}
with $f_{70\,\mu{\rm m}}$ in $\mu$Jy/arcsec$^2$ and $\theta$ in arcsec. This corresponds to a full width at half maximum of $118\,\arcsec$. 

\subsection{Fourier analysis \label{spatial-spitzer}}

An alternative method to analyse the IRAC and MIPS images is through a Fourier analysis, using the model defined in Sect.~\ref{fourier-technique}.
For the calibration of the $\Psi$ function of the RS\,Pup images obtained with IRAC, we used the archive images of \object{BD+68 1022} that was observed on 20 June 2008 specifically for the calibration of the instrumental PRF. The resulting $\Psi$ functions for the four IRAC channel images show that RS\,Pup appears unresolved. We cannot apply the Fourier analysis technique to the $\ell$\,Car images as they are saturated close to the star.

\begin{figure}[]
\centering
\includegraphics[bb=12 8 363 281, width=9cm, angle=0]{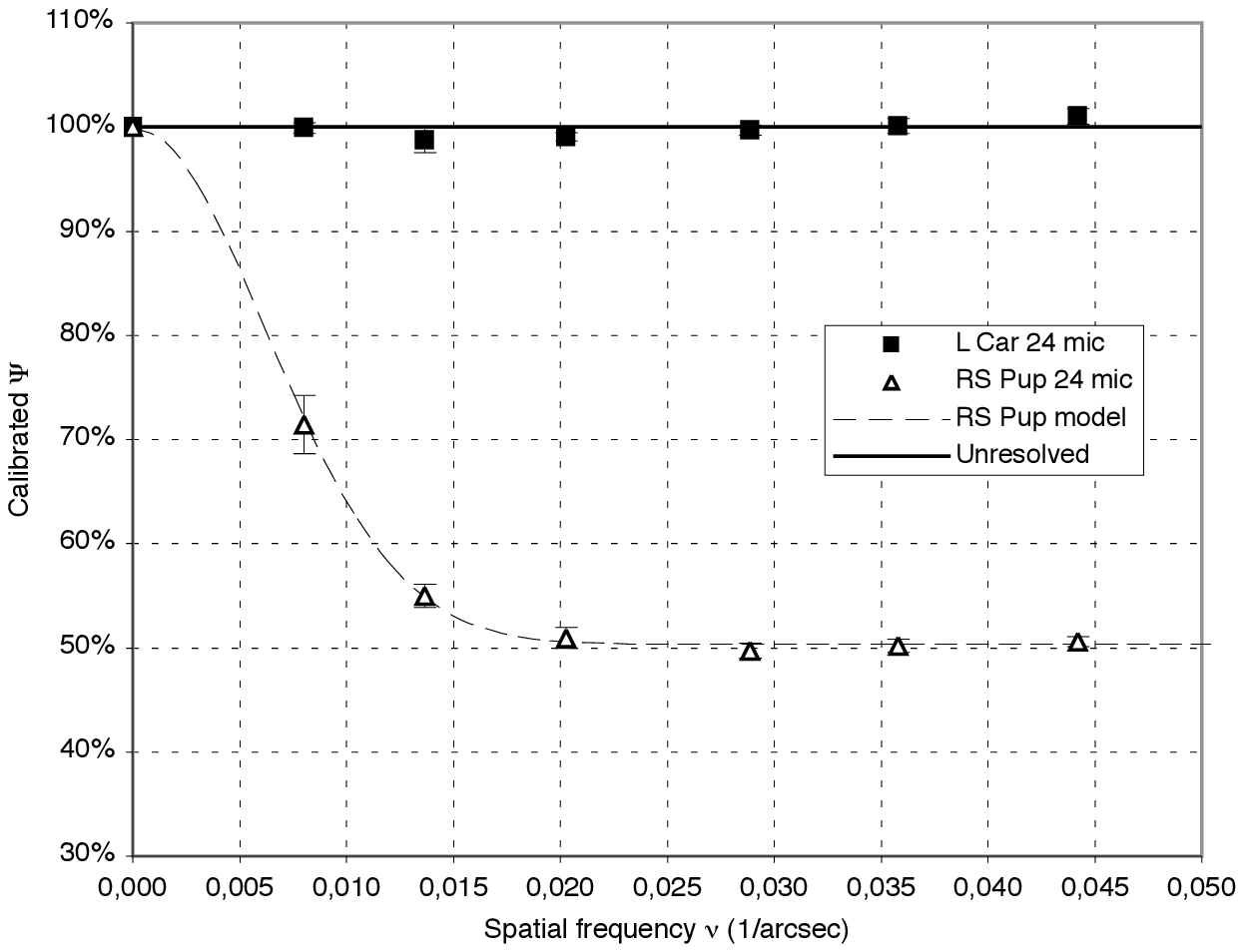}
\caption{Calibrated $\Psi$ function of the MIPS images of $\ell$\,Car and RS\,Pup.\label{spitzer-mips-psi}}
\end{figure}

We analyzed the MIPS images using the theoretical PRF reference presented in Fig.~\ref{lcar-images}. The result of the $\Psi$ function calibration is presented in Fig.~\ref{spitzer-mips-psi}. $\ell$\,Car appears unresolved in the MIPS images at 24 and 70\,$\mu$m, with a $\Psi$ function close to 100\% at all spatial frequencies. As expected from the visual aspect of RS\,Pup in the 24\,$\mu$m MIPS band (Fig.~\ref{rspup-images}), the $\Psi$ function shows a steep decline around a spatial frequency of $\nu = 0.015$\,arcsec$^{-1}$. This frequency corresponds to the typical $\approx 1\arcmin$ angular radius of the nebula, as shown in Fig.~\ref{rspup-images}. In this case, the adjustment of our Gaussian CSE model using a classical $\chi^2$ minimization gives:
\begin{equation}
\rho_{24\,\mu\mathrm{m}}(\mathrm{RS\,Pup}) = 60 \pm 3\,\mathrm{arcsec},
\end{equation}
\begin{equation}
\alpha_{24\,\mu\mathrm{m}}(\mathrm{RS\,Pup}) = 98.6 \pm 1.3\,\%.
\end{equation}
We here retrieve a larger FWHM than from the direct exponential model-fitting to the image shown in Fig.~\ref{rspup-sb}. This is due to the fact that the Fourier technique cleanly separates the unresolved stellar emission, including the wings of the stellar PSF, and the resolved nebular emission.

The analysis of the 70\,$\mu$m MIPS image of the nebula is better done directly on the images using a simple Gaussian fit (as done in Sect.~\ref{cse-sb}), because the star is invisible and the nebula is well-resolved in the image.

\subsection{Aperture photometry \label{spitzer-phot}}

We explored two routes to obtain the photometry of $\ell$\,Car and RS\,Pup from \emph{Spitzer} images: using the APEX single-frame tool included in MOPEX to obtain PRF-fitting fluxes, and from classical aperture photometry using self-developed routines. The results of the two approaches are in excellent agreement for $\ell$\,Car, but it appeared that due to the presence of the irregularly shaped CSE around RS\,Pup in the MIPS images, the PRF-fitting was not robust on this star. For this reason, we adopted the aperture photometry technique for both stars at MIPS wavelengths. For IRAC images, we adopted a slightly different measurement technique for $\ell$\,Car and RS\,Pup due to the saturation of the core of $\ell$\,Car's images (discussed below). The PRF-fitting method is in any case not recommended by the \emph{Spitzer} Science Center for IRAC images.
The systematic uncertainty in the conversion of the IRAC instrumental fluxes to absolutely calibrated units is estimated by Reach et al.~(\cite{reach05}) to be less than 3\% (value retained here). The same values for MIPS fluxes are found by Engelbracht et al.~(\cite{engelbracht07}) to be 4\% at 24\,$\mu$m, and by Gordon et al.~(\cite{gordon07}) to be 5\% at 70\,$\mu$m. As we observe bright stars, we neglect the statistical uncertainties on our photometry and adopt these values as the total uncertainties of our measurements. The measured flux densities on the two stars are summarized in Table~\ref{phot_table}, while the details of the photometry are given in the following two paragraphs.

\subsubsection{IRAC}

Due to the heavy saturation of the core of the IRAC images of $\ell$\,Car, we chose to compare the fluxes of the theoretical extended PRF reference\footnote{http://ssc.spitzer.caltech.edu/irac/psf.html} and $\ell$\,Car over an unsaturated ring between angular distances of 10 and 20\,pixels (1.22\,$\arcsec$/pixel) from the star.
This angular separation range allows us to stay in the linear part of the detector's sensitivity curve: we have a sufficient SNR for the measurement, but avoid a bias from the saturated core.
For an example of core-saturated images obtained with IRAC, the reader is referred to Marengo et al.~(\cite{marengo06}).
We corrected the 10-20\,pixel flux of $\ell$\,Car for the sky background using an annulus of 20-30\,pixels, in the four IRAC channels. We measured the relevant aperture corrections $\gamma$ on the extended PRF image for this specific combination, according to the IRAC Data Handbook version 3.0\footnote{http://ssc.spitzer.caltech.edu/irac/dh/}. We then obtained the total flux of the star from a multiplication by $\gamma$.

For RS\,Pup, the photometric analysis is easier than at longer wavelengths as the star appears unresolved (Sect.~\ref{rspup-spitzer}), and we obtained classical aperture photometry. Due to the small field of view, we used a small aperture of 5 pixels, and a background annulus between 5 and 10 pixels. The aperture corrections listed by the SSC for these apertures are 1.061, 1.064, 1.067 and 1.089 respectively for the 3.8, 4.5, 5.8 and 8.0\,$\mu$m channels.

\subsubsection{MIPS}

The aperture photometry of $\ell$\,Car is simplified by the fact that it appears as a bright (but unsaturated) isolated point source within a relatively large field void of other bright sources. Taking advantage of this configuration, we obtained the aperture photometry of $\ell$\,Car at 24\,$\mu$m in a classical way, using a 35$\arcsec$ aperture radius, and a sky background annulus of 40-50\arcsec. This configuration corresponds to an aperture correction of 1.082 according to the SSC\footnote{http://ssc.spitzer.caltech.edu/mips/apercorr/}. For the $70\,\mu$m image, we used a $30\arcsec$ aperture, with a sky annulus of 40-60\arcsec, for an aperture correction of 1.295.

The CSE of RS\,Pup has a significant extended contribution at 24\,$\mu$m and it fully dominates the 70\,$\mu$m \emph{Spitzer} image. The goal of the present work being to establish the \emph{total} infrared excess of the star, the circumstellar and photospheric emissions are considered together in the photometry. For this reason, we used a large photometric aperture of $3.0\arcmin$ in radius at 24\,$\mu$m to include the full CSE emission.
At 70\,$\mu$m, a difficulty is that the nebula appears larger than the field of view of the MIPS mosaic along its narrower direction, as shown in Fig.~\ref{rspup-images}. As a consequence, there is a slight truncation of the CSE faint extensions. For this reason we chose to integrate the average Gaussian profile given in Eq.~\ref{Gaussian-70mic} from zero to infinite radius to derive the total nebular flux density, and obtain $f_{70\,\mu{\rm m}}(\mathrm{RS\,Pup\ CSE}) = 16.9$\,Jy.

We can estimate the compact emission at 24\,$\mu$m by extrapolating the flux over a small aperture that includes essentially the star and little of the extended nebula. Using an aperture of $13\arcsec$ and an aperture correction of 1.167, we obtain a flux of $0.39$\,Jy for the core of RS\,Pup, or about half of the total flux of the star and its CSE. An analysis of the radial flux distribution in the nebula is presented in Sect.~\ref{spatial-spitzer}.

\section{VISIR data analysis\label{visir-analysis}}

In this Section, we present several analysis methods applied to the VISIR observations of $\ell$\,Car and RS\,Pup. Table~\ref{visir-summary} summarizes the corresponding results.

\begin{table}
\caption{Summary of the results of the analysis of the VISIR images.} 
\label{visir-summary}
\begin{tabular}{llll}
\hline \hline
Technique & PAH1 & PAH2 & SiC \\
\hline
\noalign{\smallskip}
{\bf $\ell$\,Car} \\
\hline
Image & resolved & resolved & unresolved\\
\hline
 PSF sub. & no residual & no residual & unresolved \\
\hline
 Fourier & $\rho = 0.75 \pm 0.15\arcsec$ & $\rho = 1.17 \pm 0.18\arcsec$ & unresolved \\
 & $\alpha = 4.0 \pm 1.0$\,\% & $\alpha = 10.6 \pm 1.2$\,\% & \\
\hline  
\hline
\noalign{\smallskip}
{\bf RS\,Pup}\\
\hline
Image & resolved & resolved & too faint \\
\hline
 PSF sub. & faint residual & residual $\approx 10\%$ & too faint \\
\hline
 Fourier & $\rho = 0.8^{+0.8}_{-0.5}$\,arcsec & $\rho = 1.04 \pm 0.21\arcsec$ & too faint \\
 & $\alpha = 2.3 \pm 1.2$\,\%& $\alpha = 23.5 \pm 2.0$\,\% & \\
\hline
\end{tabular}
\end{table}

\subsection{Cepheid and calibrator images\label{visir-images}}

\begin{figure*}[]
\centering
\includegraphics[height=6cm]{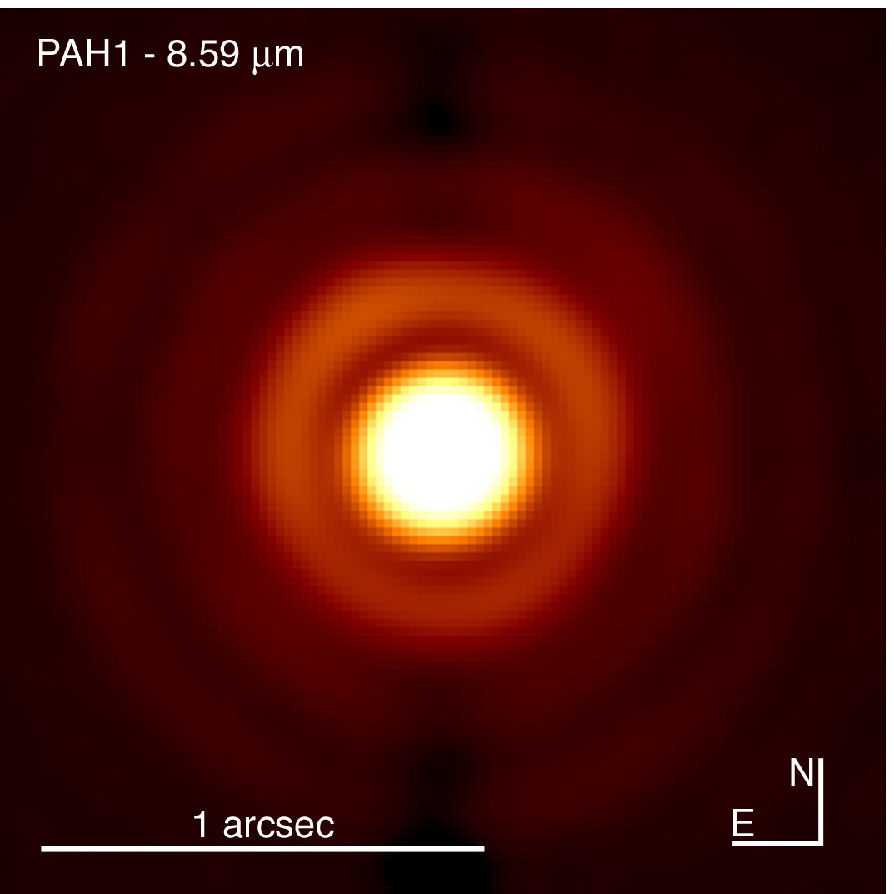}
\includegraphics[bb=0 12 357 369, height=6cm, angle=0]{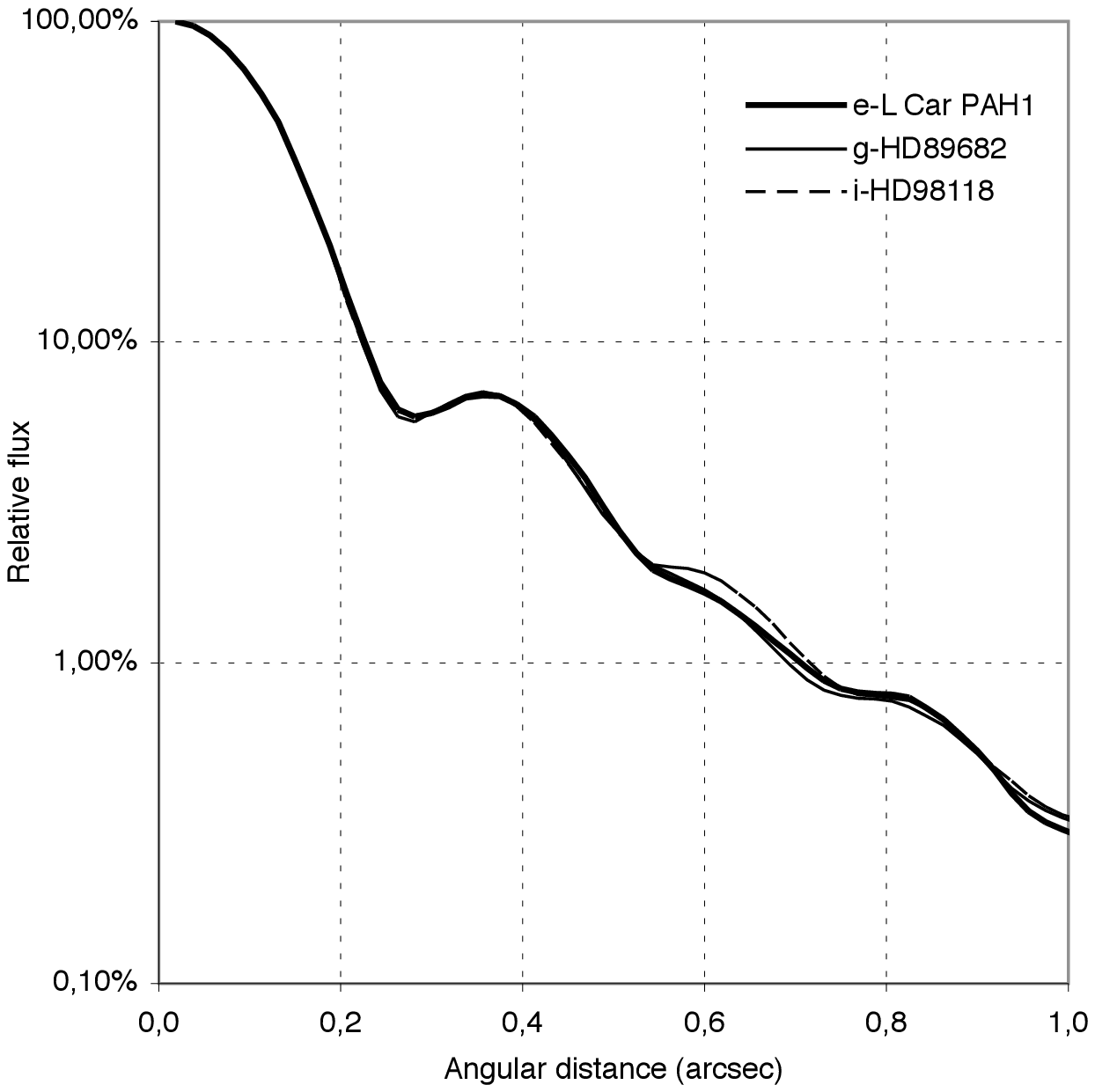}
\includegraphics[bb=0 12 357 369, height=6cm, angle=0]{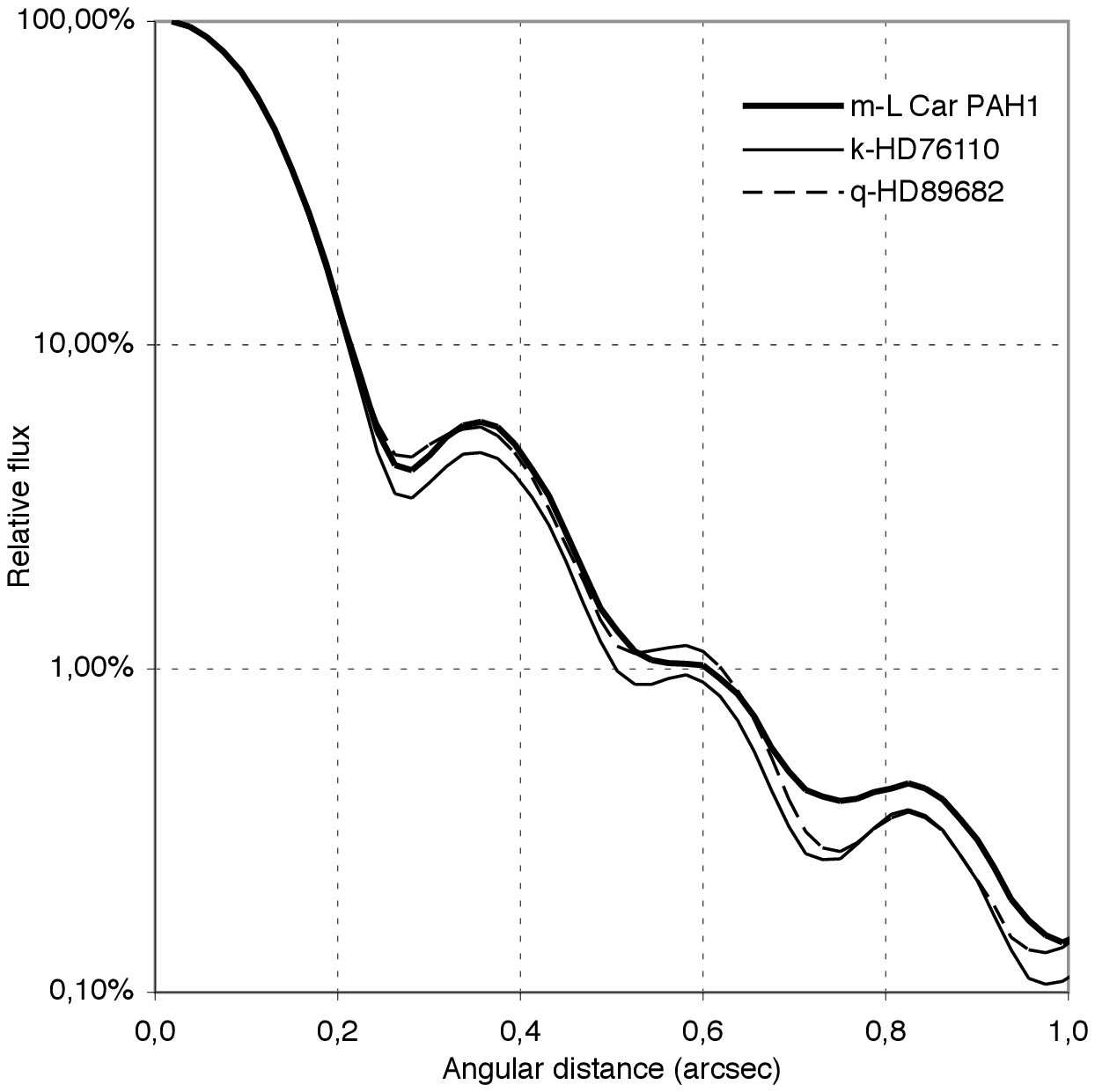}
\caption{VISIR \#i image of $\ell$\,Car in the PAH1 filter (left) and median radial profiles (normalized to unity at maximum) for the two observations \#i and \#m (see Table~\ref{visir_log}), with the associated PSF references.\label{lcar-visir1}}
\end{figure*}

\begin{figure}[]
\centering
\includegraphics[height=4.4cm]{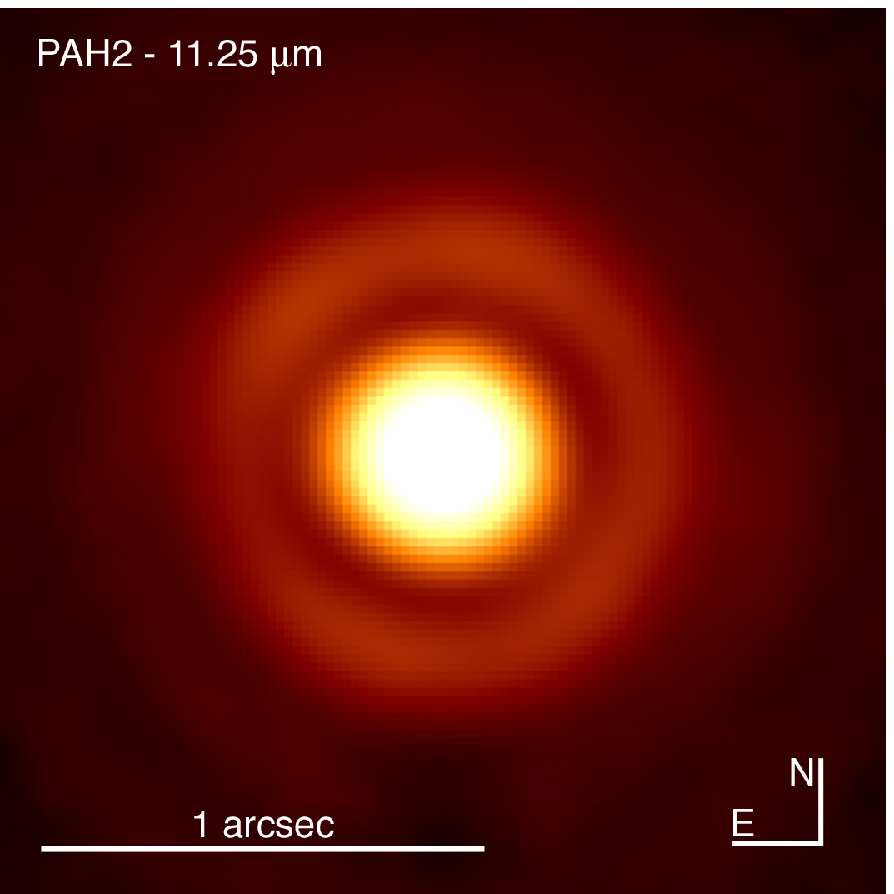}
\includegraphics[bb=0 12 357 369, height=4.4cm, angle=0]{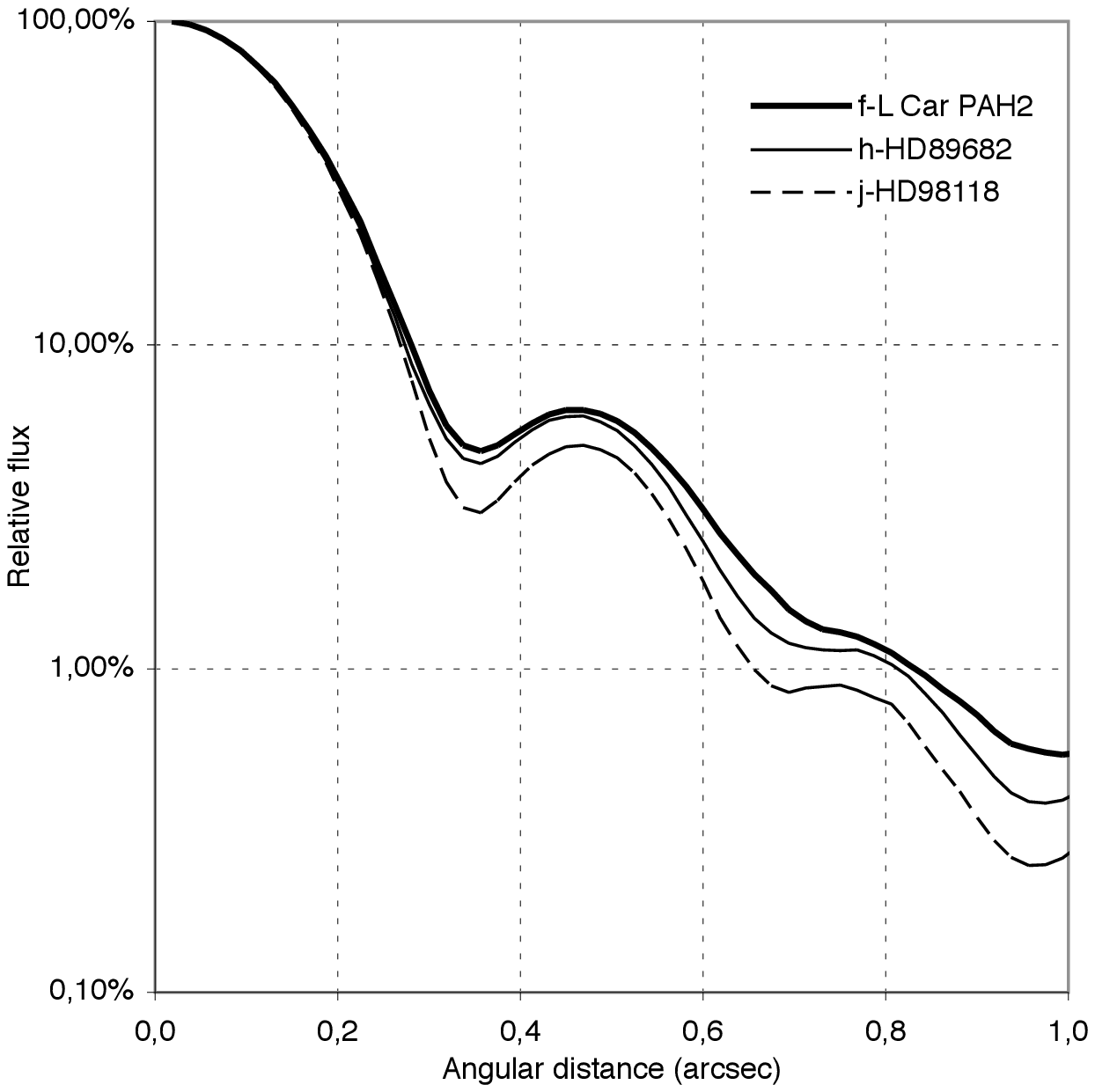}
\includegraphics[height=4.4cm]{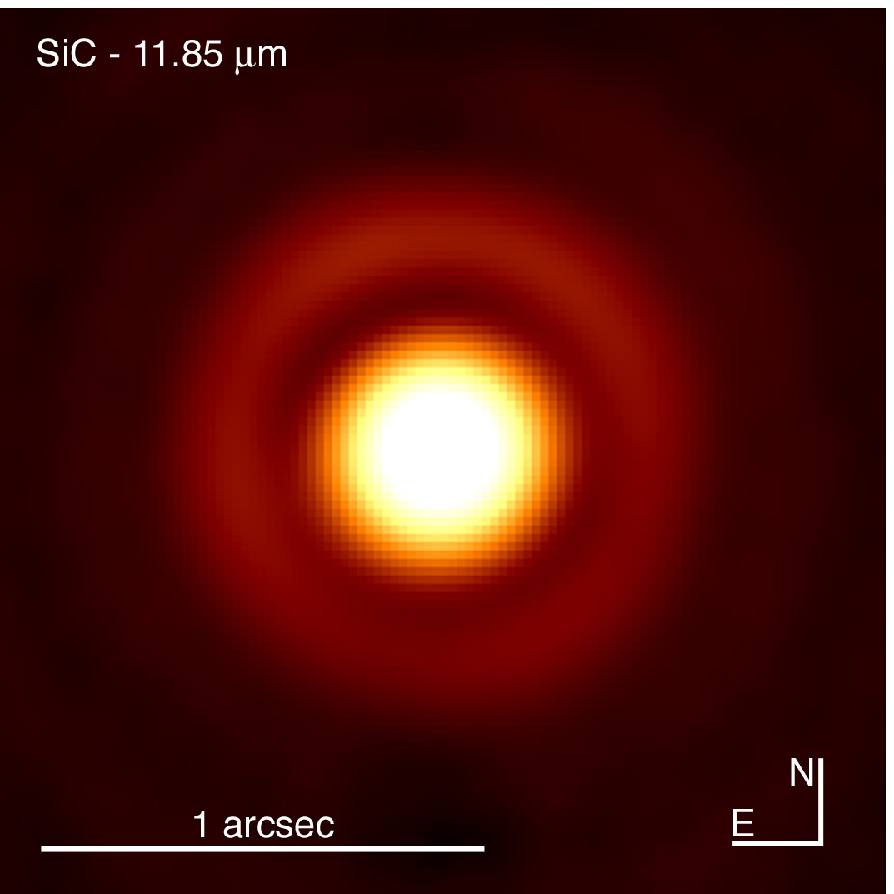}
\includegraphics[bb=0 12 357 369, height=4.4cm, angle=0]{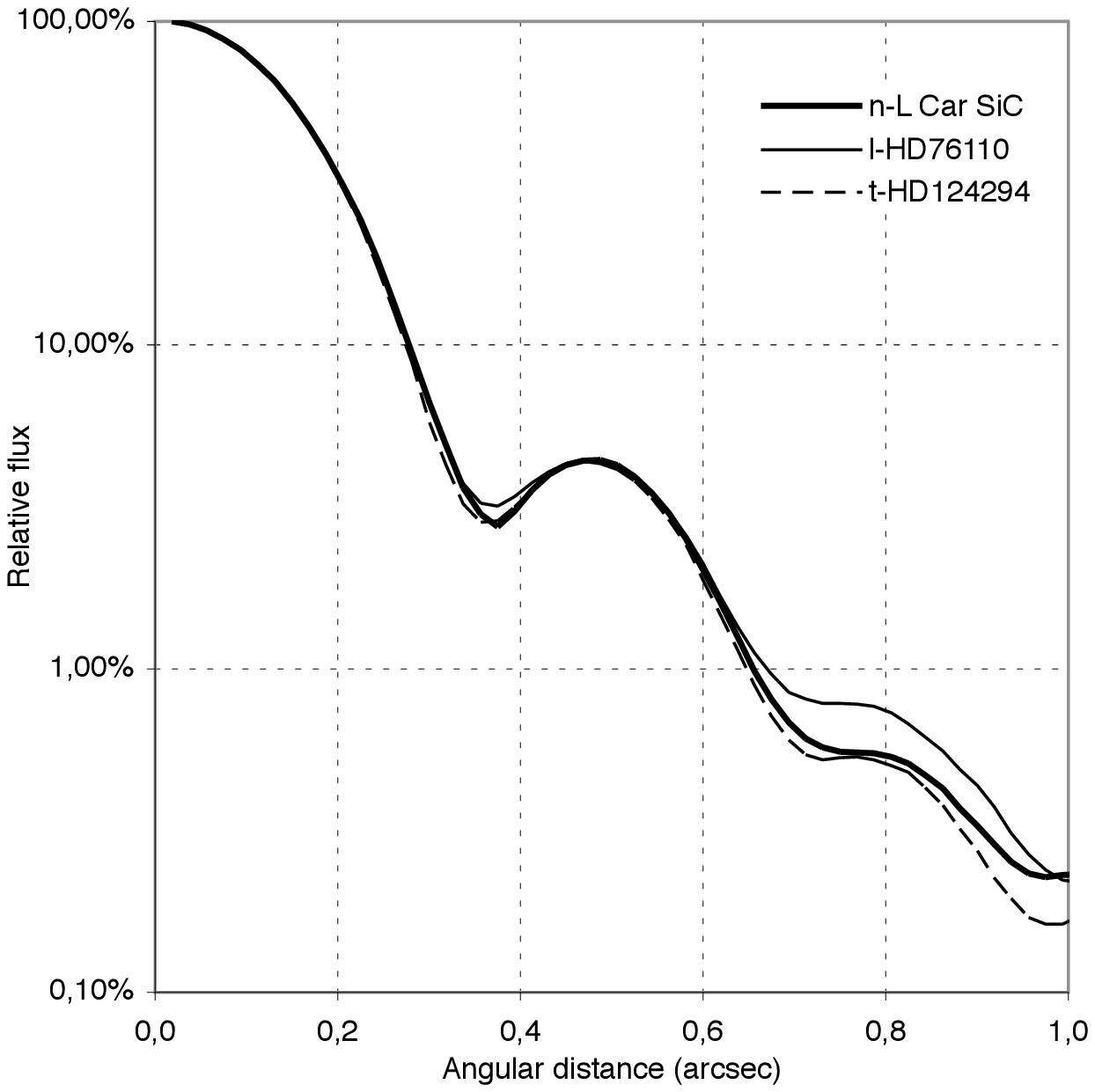}
\caption{VISIR \#f and \#n images of $\ell$\,Car in the PAH2 and SiC filters (normalized to unity at maximum) and the corresponding median radial profiles (with the PSF calibrators).\label{lcar-visir2}}
\end{figure}

The VISIR images of $\ell$\,Car in the PAH1, PAH2 and SiC filters are shown in Figs.~\ref{lcar-visir1} and \ref{lcar-visir2}, together with their median radial profiles. These $I(\theta)$ profiles were obtained by computing the median of the pixels located within an annulus with internal radius $\theta$ and external radius $\theta+1$\,pixel (resampled scale of 19\,mas/pixel). 

\begin{figure}[]
\centering
\includegraphics[height=4.4cm]{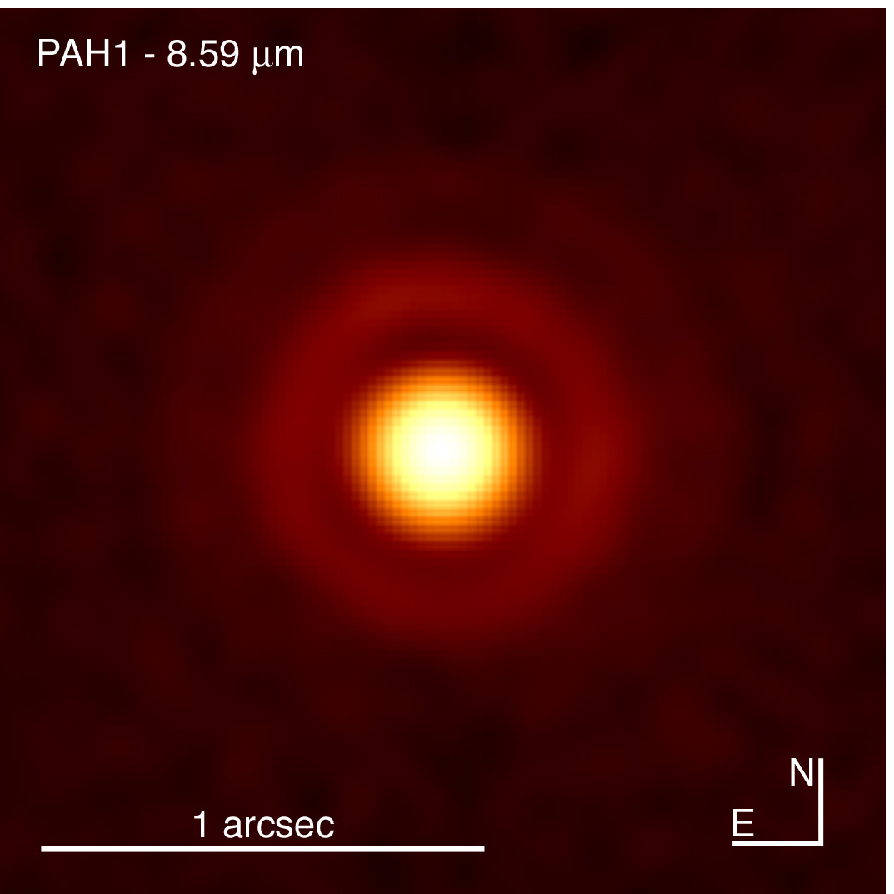}
\includegraphics[bb=0 12 357 369, height=4.4cm, angle=0]{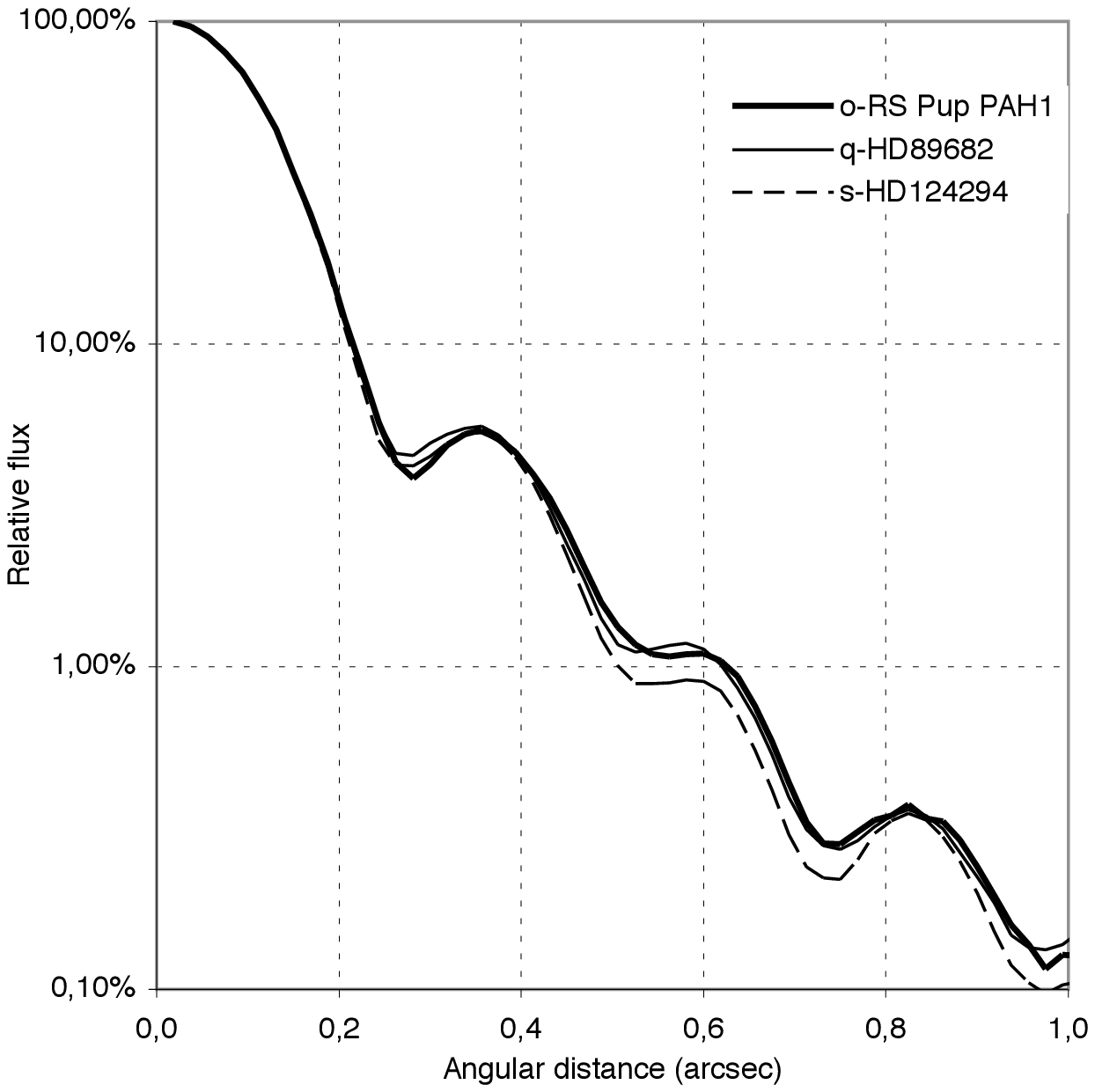}
\includegraphics[height=4.4cm]{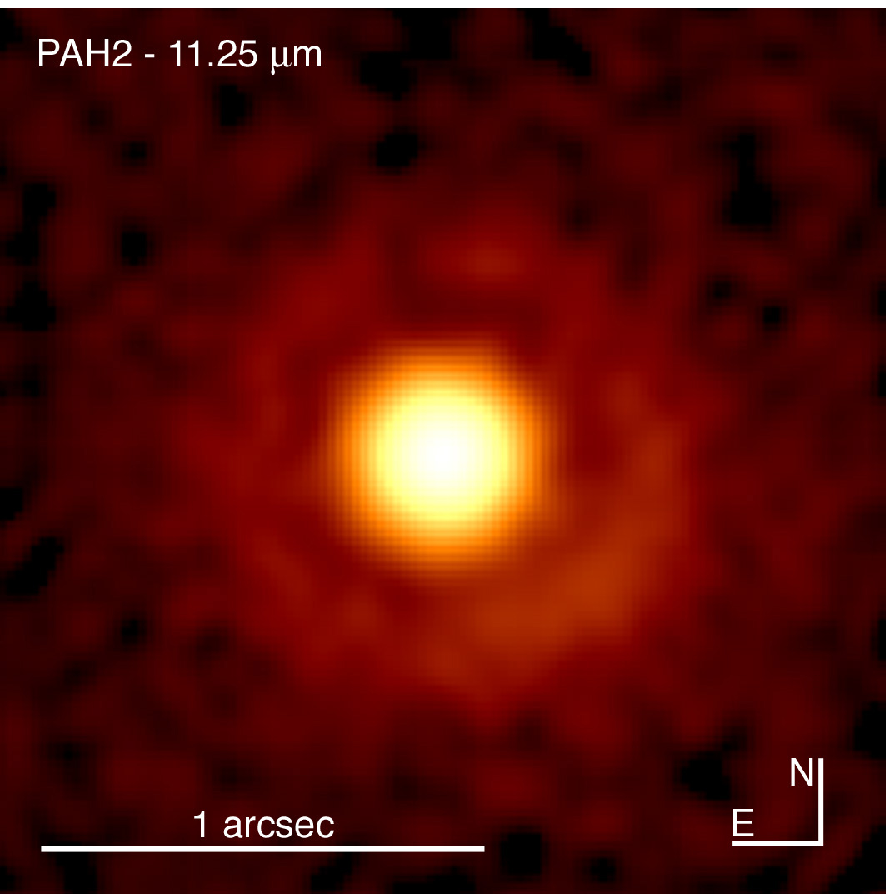}
\includegraphics[bb=0 12 357 369, height=4.4cm, angle=0]{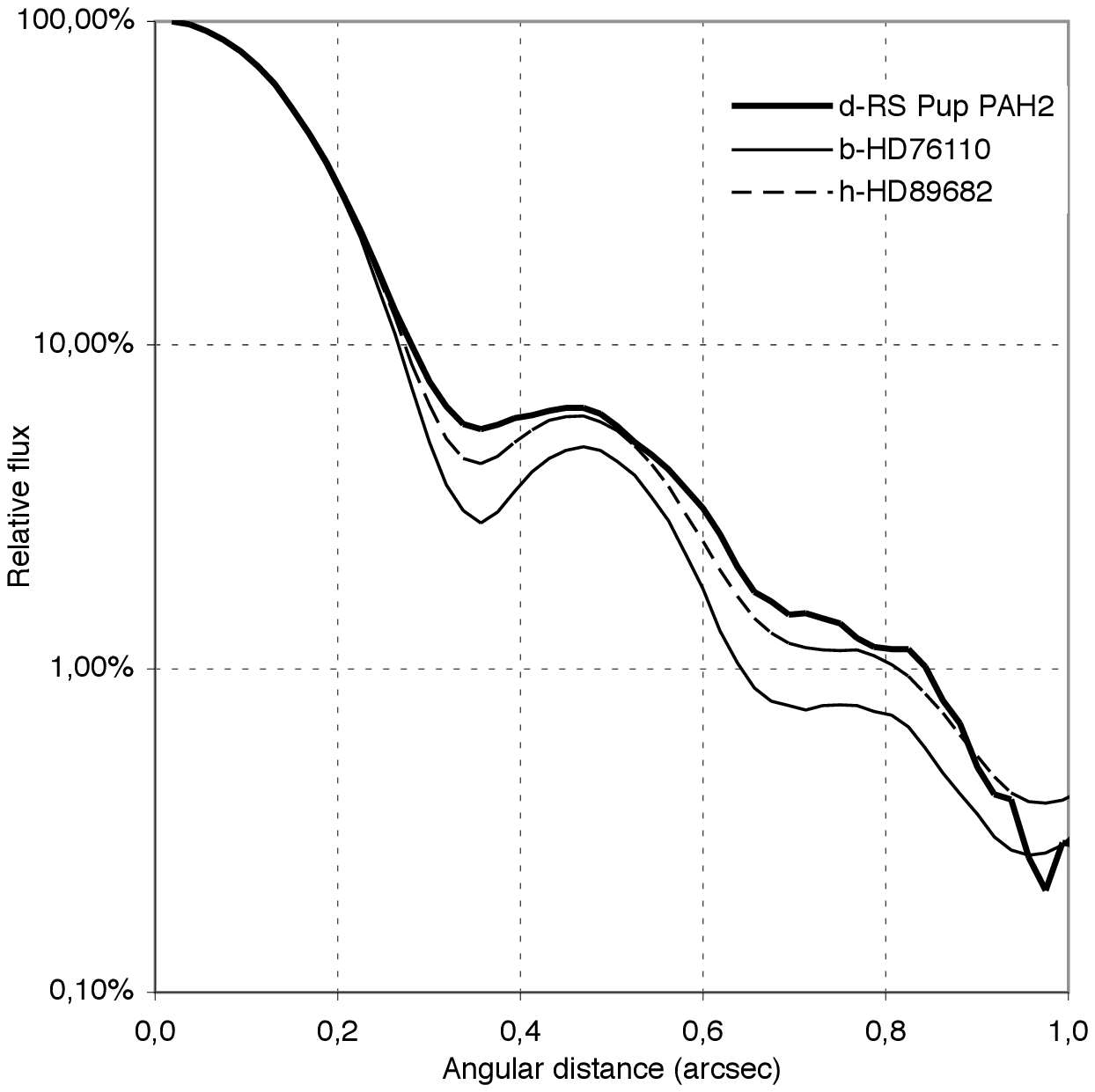}
\caption{VISIR image \#o and \#d of RS\,Pup in the PAH1 and PAH2 filters (left) and the corresponding median radial profiles compared to the PSF calibrators.\label{rspup-visir}}
\end{figure}

The images of RS\,Pup in the PAH1 and PAH2 bands are presented in Fig.~\ref{rspup-visir}. The observation \#n (Table~\ref{visir_log}) of RS\,Pup in the SiC filter did not give a satisfactory image due to the limited sensitivity of the instrument in this band.

\subsection{PSF subtracted images\label{psf-subtraction}}

\begin{figure}[]
\centering
\includegraphics[height=4.4cm]{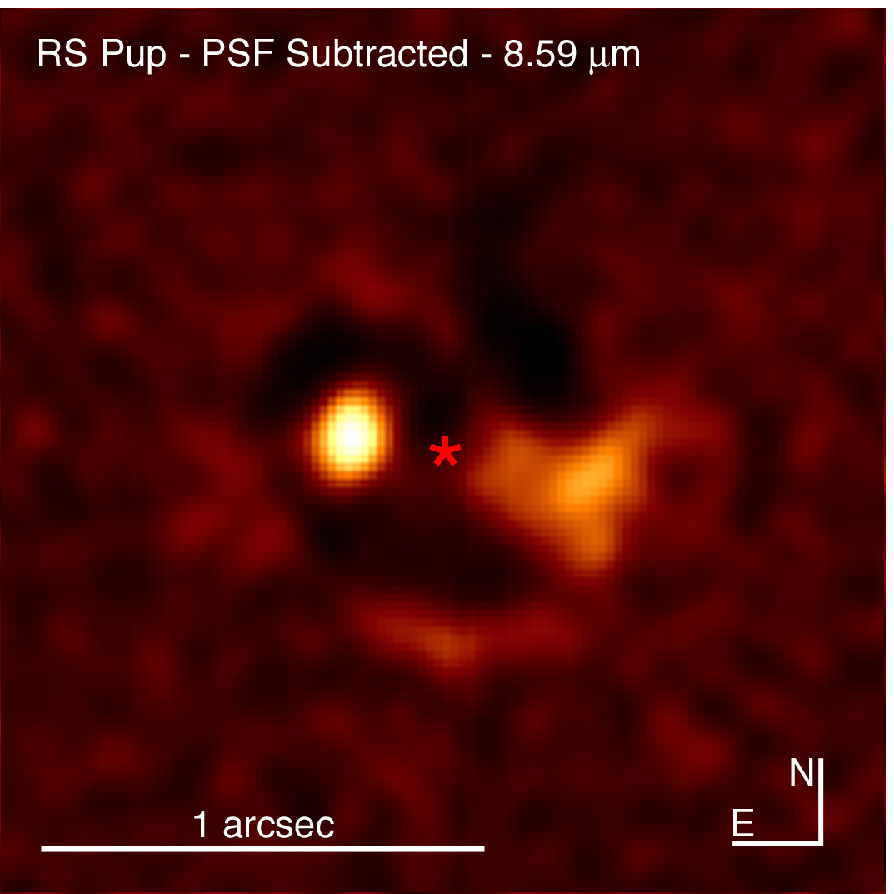}
\includegraphics[height=4.4cm]{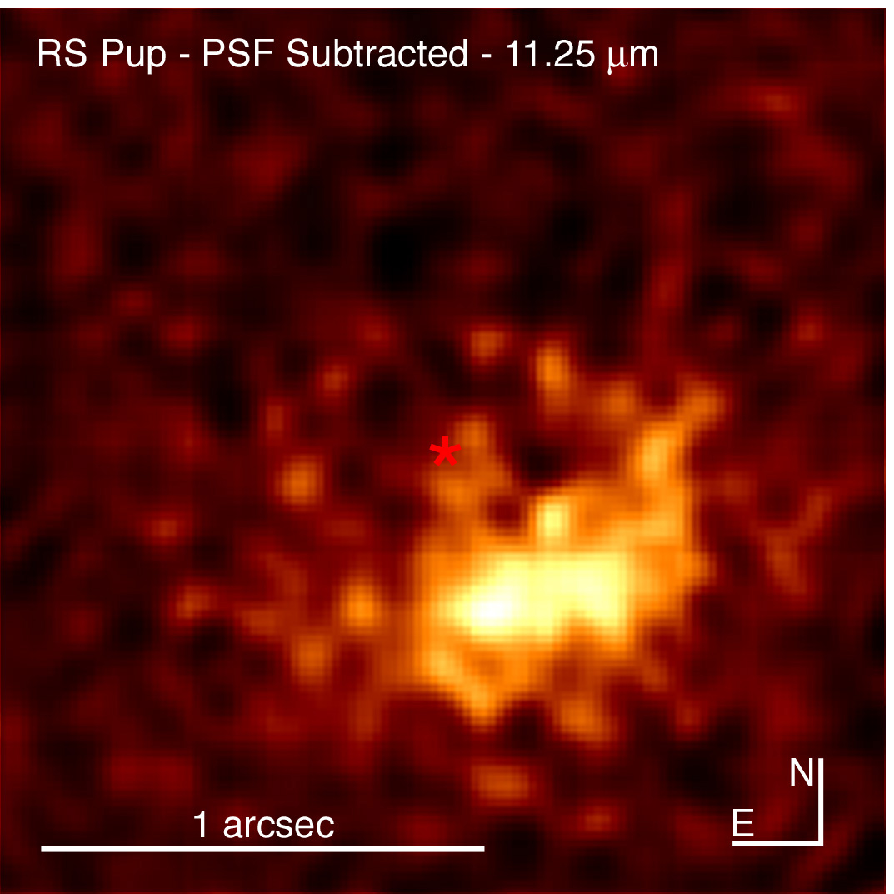}
\caption{PSF subtracted VISIR images of RS\,Pup in the PAH1 (left) and PAH2 (right) bands. The position of the star is marked with ``$\star$".\label{visir-rspup-psfsub}}
\end{figure}

The RS\,Pup image in the PAH2 band shows some asymmetry. An additional emission appears in the south-western quadrant relative to the star.
In order to remove the contribution from the star itself, we subtracted the average image of the PSF calibrators, scaled using a classical $\chi^2$  minimization of the residuals. The resulting image is shown in Fig.~\ref{visir-rspup-psfsub} (right). This emission is also apparent in the original image (Fig.~\ref{rspup-visir}), specifically as an asymmetry of the first Airy ring that appears brighter in its south-western section. The integrated flux density of this additional component is $2.8\ 10^{-15}$\,W/m$^2/\mu$m, or $\approx 10\%$ of the total emission of RS\,Pup in this band. Its position is 0.4\arcsec from the star, along an azimuth of 220$^\circ$ and its FWHM angular extension is approximately 0.5\arcsec.

Although it appears slightly more extended than a point source (for which the FWHM would be 0.3\arcsec), it is not excluded that this additional component is stellar in nature. Its position on the first Airy ring of RS\,Pup could have distorted its shape. However, it is more likely a particularly dense clump of dust in the close-in CSE of RS\,Pup. Interestingly, its position places it almost aligned with the major axis of the nebula at 70\,$\mu$m as shown in Fig.~\ref{rspup-images}.

The PSF-subtracted PAH1 image (Fig.~\ref{visir-rspup-psfsub}, left) also shows a small extended emission in the same quadrant as in PAH2, but its detection is more uncertain. A compact flux contributor is also present to the East of RS\,Pup, but its detection is very uncertain due the high PSF residual noise level so close to the star.

We applied the same processing to the images of $\ell$\,Car, but they do not reveal any significant subtraction residual.

\subsection{Fourier analysis\label{visir-fourier}}

\begin{figure*}[]
\centering
\includegraphics[bb=8 8 363 281, width=8.9cm, angle=0]{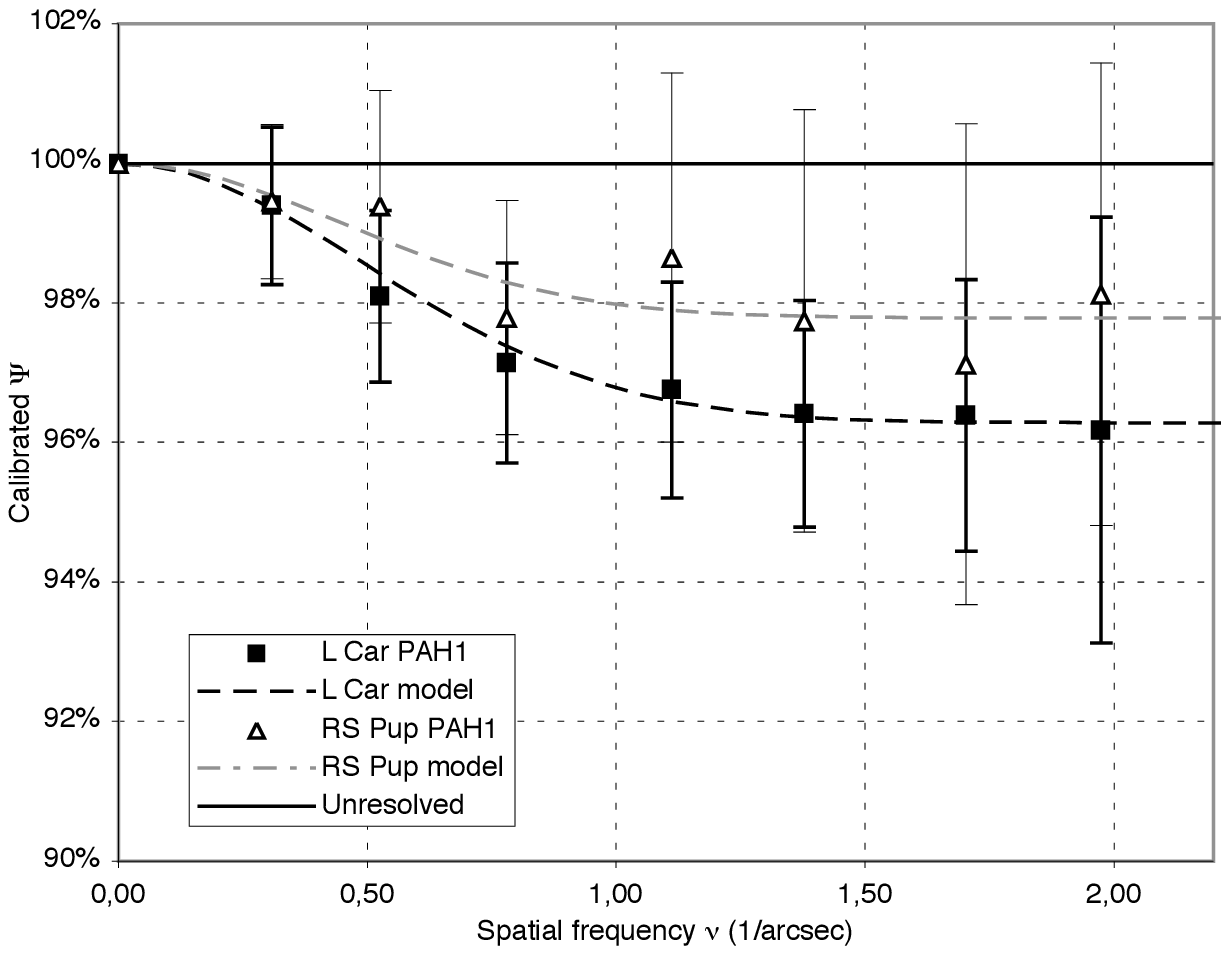} \hspace{2mm}
\includegraphics[bb=8 8 363 281, width=8.9cm, angle=0]{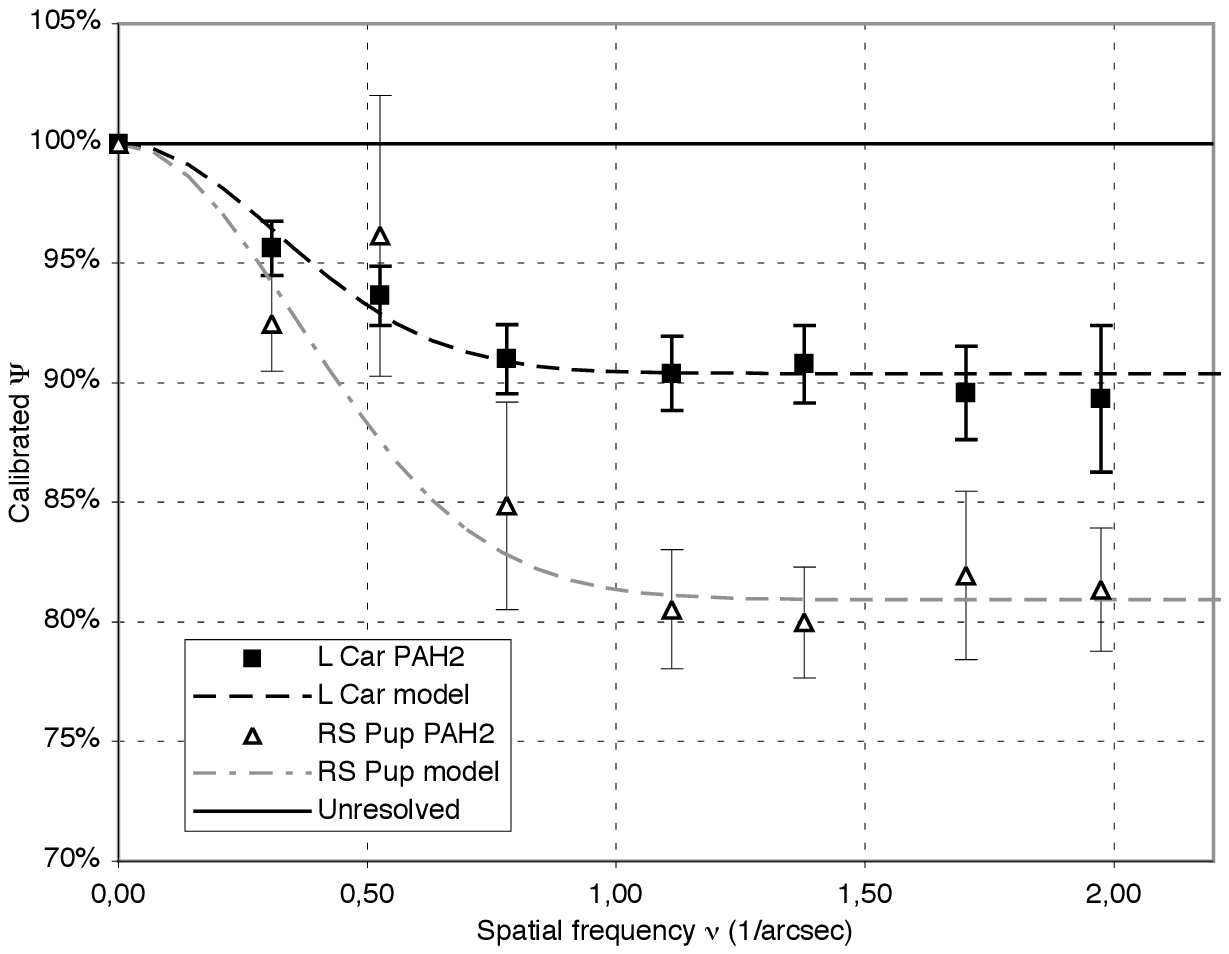}
\caption{Fit of a Gaussian CSE + unresolved point source model to the $\Psi$ functions of $\ell$\,Car and RS\,Pup in the VISIR PAH1 ($\lambda = 8.59\,\mu$m, left) and PAH2 ($\lambda = 11.25\,\mu$m, right) bands.\label{visir-gaussfit}}
\end{figure*}

We search here for extended emission in the VISIR images using the same Fourier transform technique as for the \emph{Spitzer} data.
A spatially extended emission appears as a deficit in the $\Psi$ function, starting at the spatial frequency corresponding to the angular scale of the emission. As for \emph{Spitzer}, we consider here ring median values of $\Psi(\nu)$ over all azimuth directions. It is important to note that the derived $\alpha$ contributions correspond to the extended emission contained within the full field of view of the VISIR images ($3.2\arcsec \times 3.2\arcsec$) and extrapolated outside this field as a Gaussian. In the PAH1 band (Fig.~\ref{visir-gaussfit}, left), $\ell$\,Car and RS\,Pup present a similar level of resolution, with:
\begin{equation}
\rho_{8.59\,\mu\mathrm{m}}(\ell\,{\rm Car}) = 0.75 \pm 0.15\,\mathrm{arcsec},
\end{equation}
\begin{equation}
\alpha_{8.59\,\mu\mathrm{m}}(\ell\,{\rm Car}) = 4.0 \pm 1.0\,\%,
\end{equation}
\begin{equation}
\rho_{8.59\,\mu\mathrm{m}}({\rm RS\,Pup}) = 0.8^{+0.8}_{-0.5}\,\mathrm{arcsec},
\end{equation}
\begin{equation}
\alpha_{8.59\,\mu\mathrm{m}}({\rm RS\,Pup}) = 2.3 \pm 1.2\,\%.
\end{equation}

In the PAH2 band (Fig.~\ref{visir-gaussfit}, right), the $\Psi$ functions also show that the two stars present a spatially resolved flux contributor.
The best-fit parameters for $\ell$\,Car are:
\begin{equation}
\rho_{11.25\,\mu\mathrm{m}}(\ell\,{\rm Car}) = 1.17 \pm 0.18\,\mathrm{arcsec},
\end{equation}
\begin{equation}
\alpha_{11.25\,\mu\mathrm{m}}(\ell\,{\rm Car}) = 10.6 \pm 1.2\,\%,
\end{equation}
%
The $\Psi$ function of RS\,Pup in the PAH2 filter shows a ``bump" for spatial frequencies around 0.5\arcsec. This behavior could be due to the compact emission source present in the field of view of VISIR (Sect.~\ref{psf-subtraction}). The Gaussian model fitting gives the following properties for RS\,Pup:
\begin{equation}
\rho_{11.25\,\mu\mathrm{m}}({\rm RS\,Pup}) = 1.04 \pm 0.21\,\mathrm{arcsec},
\end{equation}
\begin{equation}
\alpha_{11.25\,\mu\mathrm{m}}({\rm RS\,Pup}) = 23.5 \pm 2.0\,\%.
\end{equation}

The SiC image of $\ell$\,Car does not show a resolved component. The broad filter bandpass ($40\%$) degrades the contrast of the Airy diffraction pattern, as different wavelengths corresponding to different spatial resolutions are superimposed. As the detected CSE in PAH1 and PAH2 is compact, this likely prevents its detection.

\subsection{Aperture photometry \label{visir-photometry}}

We derive here the photometric flux density of $\ell$\,Car and RS\,Pup in the VISIR filters through classical aperture photometry.
In order to absolutely calibrate the flux density, we used as references the spectrophotometric templates from Cohen et al.~(\cite{cohen99}). The irradiance of the calibrators was read from the templates taking into account the transmission curve of each filter\footnote{http://www.eso.org/sci/facilities/paranal/instruments/visir/}.
The integrated fluxes are affected by differential atmospheric absorption due to the different airmass of each observation (listed in Table.~\ref{visir_log}). To take this effect into account, we used the empirical formula by Sch\"utz \& Sterzik~(\cite{schutz05}), that gives the multiplicative correction $C(\lambda,{\rm AM})$ to apply to the raw flux to calibrate the atmospheric absorption:
\begin{equation}
C(\lambda,{\rm AM}) = 1 + \left[ 0.220 - \frac{0.104}{3}\,\left(\lambda - 8.6\,\mu{\rm m}\right) \right]\,({\rm AM} - 1).
\end{equation}
As we used the same apertures for the photometry of the Cepheids and the flux calibrators ($1.3\arcsec$ in angular radius), no aperture correction is required.

The measured flux densities on $\ell$\,Car and RS\,Pup in the VISIR filters are summarized in Table~\ref{phot_table}. The angular radius of the effective aperture over which the VISIR photometry was obtained is $1.31\arcsec$ in all filters.
The associated uncertainties include the statistical dispersion of the aperture photometry, the dispersion of the calibrator flux densities over the night, and the absolute calibration uncertainty ($\approx 1.8\%$ for our calibrators) as listed by Cohen et al.~(\cite{cohen99}).

\section{MIDI data analysis\label{midi-analysis}}

\subsection{Differential visibilities}

The absolute calibration of the MIDI visibilities has a typical accuracy of 10-15\% in the {\tt HIGH\_SENS} mode. The {\tt SCI\_PHOT} mode allows for a better accuracy, but is only available for very bright objects. As our two Cepheids are relatively faint stars for MIDI standards, the {\tt HIGH\_SENS} mode was the only available choice. Its large calibration uncertainty would completely mask the small contributions of the CSE in the absolute visibility spectrum. However, the \emph{shape} of the spectrum is known with a much better accuracy, and the relative values of the visibility between different wavelengths are accurate to a few percent (Chesneau~\cite{chesneau07}).
In the following, we therefore adopt a differential approach for the interpretation of the MIDI visibilities. We define the differential visibility as:
\begin{equation}
\Delta V(\lambda, B) = \frac{V(\lambda, B)}{V(8\,\mu\mathrm{m}, B)} - 1 \label{differential-def}
\end{equation}
with $B$ the projected baseline of the measurement, and $\lambda$ the wavelength. By definition, $\Delta V(8\,\mu\mathrm{m}, B) = 0$. In practice, we estimated the value of $V(8\,\mu\mathrm{m}, B)$ from a linear fit to the whole visibility spectrum. To estimate the uncertainties, we considered the standard deviation of the residuals of this fit to $\Delta V$, uniform with wavelength.
This differential visibility approach gives us a very much improved sensitivity to color-dependent variations of the CSE properties. However, it comes with a price, as the signature of any achromatic component is obliterated (Sect.~\ref{sensitivity-domain}).
In the following, we compare the resolved flux component measured with VISIR in the PAH1 and PAH2 bands to the MIDI differential visibilities, taking into account the expected differential visibility contribution from the stellar photosphere, as discussed in the next paragraph.

\subsection{Visibility model\label{extended-compact-model}}

The MIDI differential visibility measurements $\Delta V(\lambda)$ characterize physically the ratio of the spatially correlated flux $f_\mathrm{corr}(\lambda, B)$ (measured from the contrast of the interference fringes) to the total flux $f_\mathrm{tot}(\lambda)$, at the spatial frequency $\lambda/B$ sampled by the interferometer, relative to this ratio at $8\,\mu$m. The total flux considered here is the flux enclosed within the aperture stop of the instrument. For MIDI , this aperture has an angular radius of approximately 0.4\,arcsec (see also Sect.~\ref{midi-phot}). Within this aperture, the flux from the resolved component of the CSE around the Cepheids contributes to increase $f_\mathrm{tot}$ up to the edge of the aperture stop, even beyond the core of the point spread function (PSF) of the telescopes. This is different for instance from fibered interferometric beam combiners (e.g. VINCI; Kervella et al.~\cite{kervella04b}), whose field of view closely matches the PSF.

The goal of this analysis is to search for a compact component in the CSE, in addition to the extended component resolved with VISIR. Our approach is the following: we present hereafter a model of the known differential visibility contributors, and compare its prediction with the observed values. In this process, we will initially suppose that the CSE flux density at 8\,$\mu$m is negligible, i.e. $\alpha(8\,\mu\mathrm{m})=0$. This hypothesis is reasonable in view of the small photometric infrared excess measured at this wavelength with \emph{Spitzer} (Sect~\ref{spitzer-phot}).

The known contributors to the visibility measured with MIDI are the stellar photosphere visibility $V_\star(\lambda, B)$, and the contribution from the extended CSE resolved with VISIR $V_L(\lambda,B)$. As this last component is already resolved spatially by a single telescope, it is also fully resolved by the interferometer, and its visibility $V_L(\lambda,B)=0$. In other words, the CSE detected with VISIR contributes only spatially incoherent flux to the MIDI observations.
These two contributors create a visibility dependence of the form:
\begin{equation}
V_\mathrm{model}(\lambda, B) =  \frac{\mathrm{Coherent\ flux\ density}}{\mathrm{Total\ flux\ density}} = \frac{f_\star(\lambda)\ V_\star(\lambda,B)}{f_\mathrm{\star}(\lambda) + f_L(\lambda)}
\end{equation}
where $f_\star(\lambda)$ is the stellar flux density, $V_\star(\lambda,B)$ the visibility of the photosphere, and $f_L(\lambda)$ the flux density of the large CSE component.
This can be rewritten as:
\begin{equation}\label{vis-model-1}
V_\mathrm{model}(\lambda, B) =  \frac{V_\star(\lambda,B)}{1 + \alpha(\lambda)}
\end{equation}
where $\alpha(\lambda) = f_L(\lambda)/f_\star(\lambda)$ is the relative contribution of the large component. This parameter was directly measured from the VISIR observations in the PAH1 and PAH2 filters (Sect.\,\ref{visir-analysis}). As we assume that $\alpha(8\,\mu\mathrm{m})=0$, this gives us the differential visibility expression:
\begin{equation}\label{vis-model-2}
\Delta V_\mathrm{model}(\lambda, B) = \frac{\Delta V_\star(\lambda,B) - \alpha(\lambda)}{1 + \alpha(\lambda)}.
\end{equation}
In the next paragraphs of this Section, we compare the measured value of $\Delta V_\mathrm{obs}(\lambda, B)$ in the PAH1 and PAH2 bands to the predicted $\Delta V_\mathrm{model}(\lambda, B)$, for the two stars.

\subsection{Sensitivity domain\label{sensitivity-domain}}

Now let us discuss the sensitivity of our approach for the detection of a compact CSE component, within the angular dimension and photometric contribution parameter space.
We now consider, in addition to the large CSE component discussed precedently, a compact CSE contributing a fraction $\beta(\lambda)$ of the stellar flux density, with a non-zero visibility $V_C(\lambda, B)$. This additional component represents the perturbing signal that we seek in our MIDI observations, once the extended VISIR CSE has been taken into account. As this new component adds a partially coherent flux contribution, its presence will modify Eq.~\ref{vis-model-1} towards:
\begin{equation}\label{vis-model-3}
V_\mathrm{model}(\lambda, B) =  \frac{V_\star(\lambda,B) + \beta(\lambda)\,V_C(\lambda, B)}{1 + \alpha(\lambda) + \beta(\lambda)},
\end{equation}
where $\beta(\lambda)\,V_C(\lambda, B)$ is the coherent part of the total flux density of the compact CSE, and $\beta(\lambda)$ its total flux density. The sensitivity of our study to the presence of a compact CSE can be divided in two domains, depending on its typical angular size $\rho_C$:

\begin{itemize}
\item $\rho_C < \lambda/B$: the compact component visibility $V_C(\lambda, B)$ will be significantly different from zero. In this case, the MIDI differential visibility measurement will show the presence of the compact CSE as an additional slope (to first order) on the differential visibility spectrum compared to the model expected from the VISIR extended CSE alone. However, if the coherent flux contribution $\beta(\lambda) V_C(\lambda,B)$ is uniform with wavelength, this signature will disappear in the computation of the differential quantity through Eq.~\ref{differential-def}.

\item $\lambda/B < \rho_C < \lambda/D$: in this case its interferometric visibility is negligible, and it will remain undetectable in the differential visibility spectrum (no effect on the differential visibility $\Delta V_\mathrm{model}$ computed from Eq.~\ref{vis-model-3}).
\end{itemize}

The angular resolution $\lambda/B$ of the interferometer in the $N$ band is typically $\approx 55$\,mas on the $40$\,m UT2-UT3 baseline, and $\approx 17$\,mas on the $130$\,m UT1-UT4 baseline, while the resolution of a single Unit Telescope is $\lambda/D \approx 500$\,mas.

It thus appears that the differential visibility technique is sensitive only to a restricted kind of compact CSE. We can only detect the \emph{color excess} of the \emph{partially resolved part} of a compact CSE component. This color excess is measured over the $N$ band, relative to $\lambda = 8\,\mu$m. We are not sensitive to its absolute flux level. The contribution from the envelope resolved with VISIR is taken into account for this comparison. Within these limitations, and as shown in the following numbers, our sensitivity expressed as a fraction of the stellar photosphere flux is between 1.6 and 7.5\%.

\subsection{$\ell$\,Car differential visibility spectrum\label{lcar-midi}}

We observed $\ell$\,Car with MIDI on the UT2-UT3 baseline ($B = 40.0$\,m) in 2004 (\#A and \#C, Table~\ref{midi_log}), and on the UT1-UT4 baseline in 2006 (\#E and \#P). Observation \#J was rejected due to its very low interferometric flux. The average projected baseline length of the two UT1-UT4 observations is $B = 123.8$\,m.

\begin{figure}[]
\centering
\includegraphics[bb=12 8 363 281, width=8.7cm, angle=0]{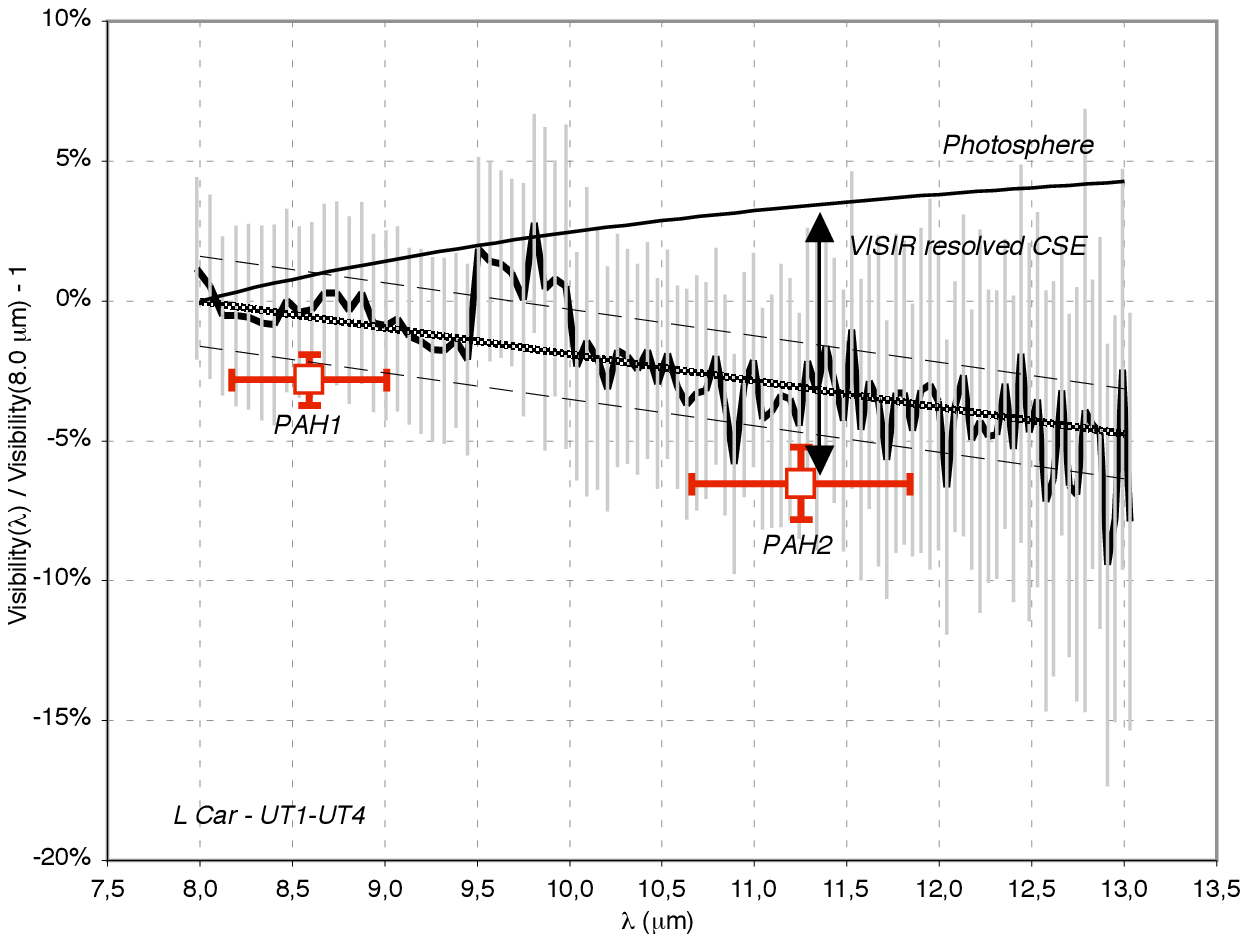}
\includegraphics[bb=12 8 363 281, width=8.7cm, angle=0]{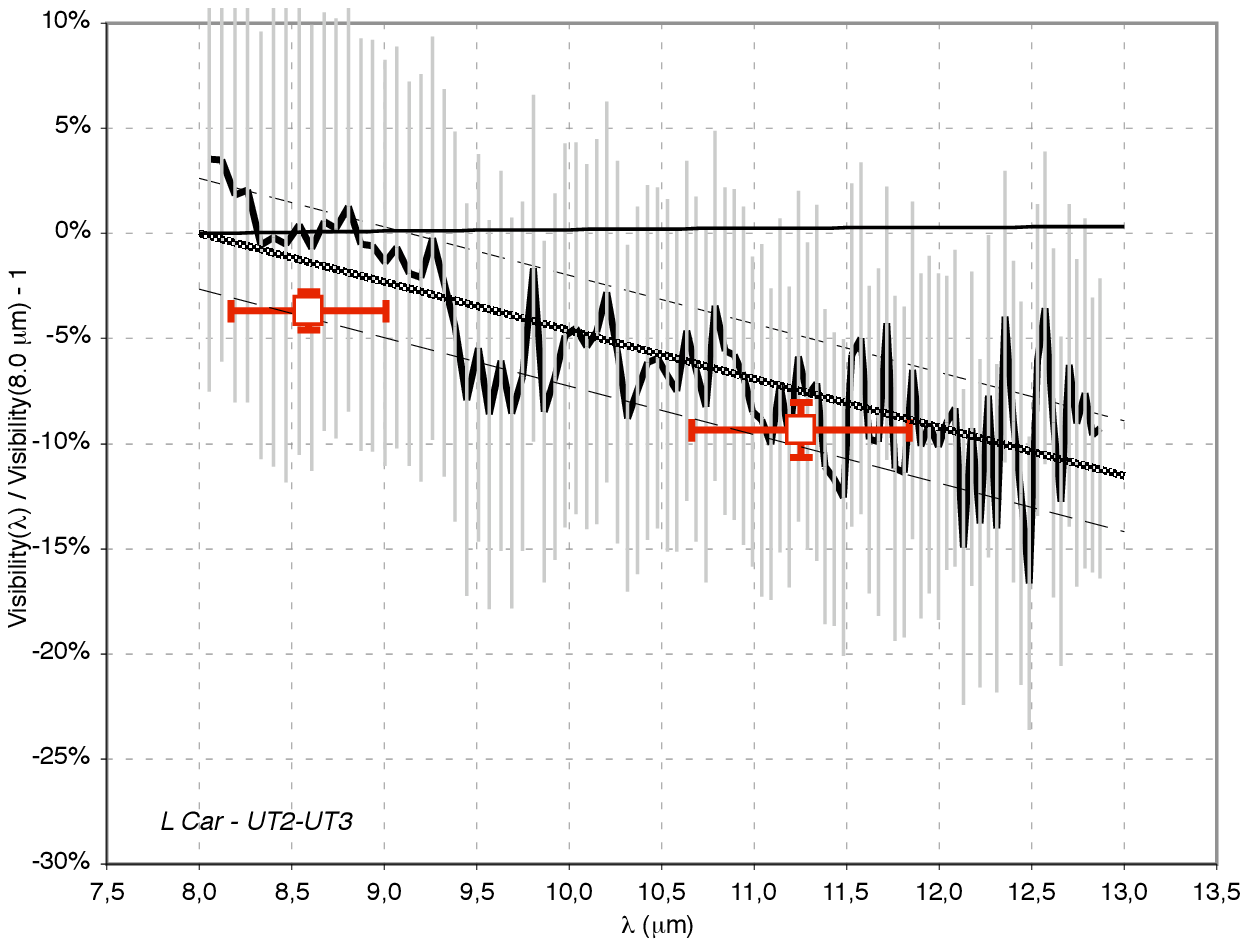}
\caption{Differential visibility of $\ell$\,Car on the UT1-UT4 (top) and UT2-UT3 (bottom) baselines. The details of the different curves are discussed in Sect.~\ref{lcar-midi}.\label{lcar-visib}}
\end{figure}

The differential visibility spectra are presented in Fig.~\ref{lcar-visib}. The MIDI average spectra for 2004 and 2006 are shown as thick black curves, with the original error bars on the absolute visibilities produced by the EWS software. The thin rising curve represents the expected differential visibility of the photosphere of the star without CSE. The open squares show the expected differential visibility in PAH1 and PAH2 of the stellar photosphere plus the extended component resolved with VISIR (as defined in Sect.~\ref{extended-compact-model}). The width of the PAH1 and PAH2 filters bandpass are shown as horizontal error bars. In the upper plot of Fig.~\ref{lcar-visib}, the contribution of the CSE to the decrease of the differential visibility in indicated for the PAH2 band. The thick grey curve is a linear fit to the MIDI visibility spectrum, scaled so that it equals zero for $\lambda = 8\,\mu$m. The thin dashed curves represent the rms dispersion domain of the differential visibilities around the linear fit model, that equal 1.6\% and 2.6\% respectively for the UT1-UT4 and UT2-UT3 baselines.

The differential visibility of the photosphere $\Delta V_\star(\lambda,B)$ was computed from the expected limb darkened angular diameter of the star. For the 2004 epoch, $\ell$\,Car was at its minimum diameter phase ($\phi = 0.90$), for which Kervella et al.~(\cite{kervella04a}) find a photospheric angular diameter of $\theta_{\rm LD} = 2.69$\,mas. The 2006 epochs correspond to phases of $\phi=0.66$ and $\phi=0.46$. The average angular diameter of $\ell$\,Car for these two phases is $\theta_{\rm LD} = 3.09$\,mas. We can consider these two observations together, as the variation of the photospheric angular diameter of the star between the two phases is only $\delta \theta_{\rm LD} = \pm 0.10$\,mas. This corresponds on the 130\,m UT1-UT4 baseline to a change in visibility of 1.7\% at 8.0\,$\mu$m, below the typical uncertainty on the MIDI differential visibilities.

The UT2-UT3 differential visibility spectrum obtained with the present reduction differs in slope from the $V^2$ curve presented in Fig.~6 of Kervella et al.~(\cite{kervella06}). The reason for this difference is likely to be the different definition of the detector masks used to extract the source spectrum and estimate the background. This step is critical for the correct analysis of the MIDI data, and the computation method of the EWS software used for the present work (mask adapted to the observed object) appears more robust than that of the Paris Observatory package (fixed mask position on the detector). But although the spectrum slopes are different, the two calibrated visibility spectra agree well within their statistical uncertainties.
From an integration of the measured $\Delta V(\lambda, B)$, weighted by the PAH1 and PAH2 filter transmissions, we derive for the UT1-UT4 baseline:
\begin{equation}
\Delta V_\mathrm{obs}(\mathrm{PAH1})= - 0.2 \pm 1.6\,\%,
\end{equation}
\begin{equation}
\Delta V_\mathrm{obs}(\mathrm{PAH2})= - 3.3 \pm 1.6\,\%,
\end{equation}
and we compute model values of:
\begin{equation}
\Delta V_\mathrm{model}(\mathrm{PAH1})= - 2.8 \pm 0.9\,\%,
\end{equation}
\begin{equation}
\Delta V_\mathrm{model}(\mathrm{PAH2})= - 6.5 \pm 1.3\,\%.
\end{equation}
Similarly, on the UT2-UT3 baseline, we measure:
\begin{equation}
\Delta V_\mathrm{obs}(\mathrm{PAH1})= + 0.1 \pm 2.6\,\%,
\end{equation}
\begin{equation}
\Delta V_\mathrm{obs}(\mathrm{PAH2})= - 8.6 \pm 2.6\,\%.
\end{equation}
and we obtain model values of:
\begin{equation}
\Delta V_\mathrm{model}(\mathrm{PAH1})= - 3.7 \pm 0.9\,\%,
\end{equation}
\begin{equation}
\Delta V_\mathrm{model}(\mathrm{PAH2})= - 9.4 \pm 1.3\,\%.
\end{equation}

As shown in Fig.~\ref{lcar-visib} and the comparison of the $\Delta V_\mathrm{obs}$ and $\Delta V_\mathrm{model}$ values above, the model differential visibilities $\Delta V_\mathrm{model}$ are smaller than the observed $\Delta V_\mathrm{obs}$ by $\approx 3 \pm 1.5\%$ in PAH1 and PAH2. This indicates that the relative CSE flux contribution at 8.0\,$\mu$m is probably of this order, as the effect of the presence of extended emission at this wavelength is a global negative shift of the $\Delta V_\mathrm{obs}$ curve relative to the photospheric visibility.
This good agreement between the observed and model differential visibilities imply that the deficit observed between the expected photospheric visibility and the measured MIDI visibility is caused by the CSE component resolved with VISIR alone. We thus do not find a significant additional compact component.

\subsection{RS\,Pup differential visibility spectrum\label{rspup-midi}}

\begin{figure}[]
\centering
\includegraphics[bb=12 8 363 281, width=8.7cm, angle=0]{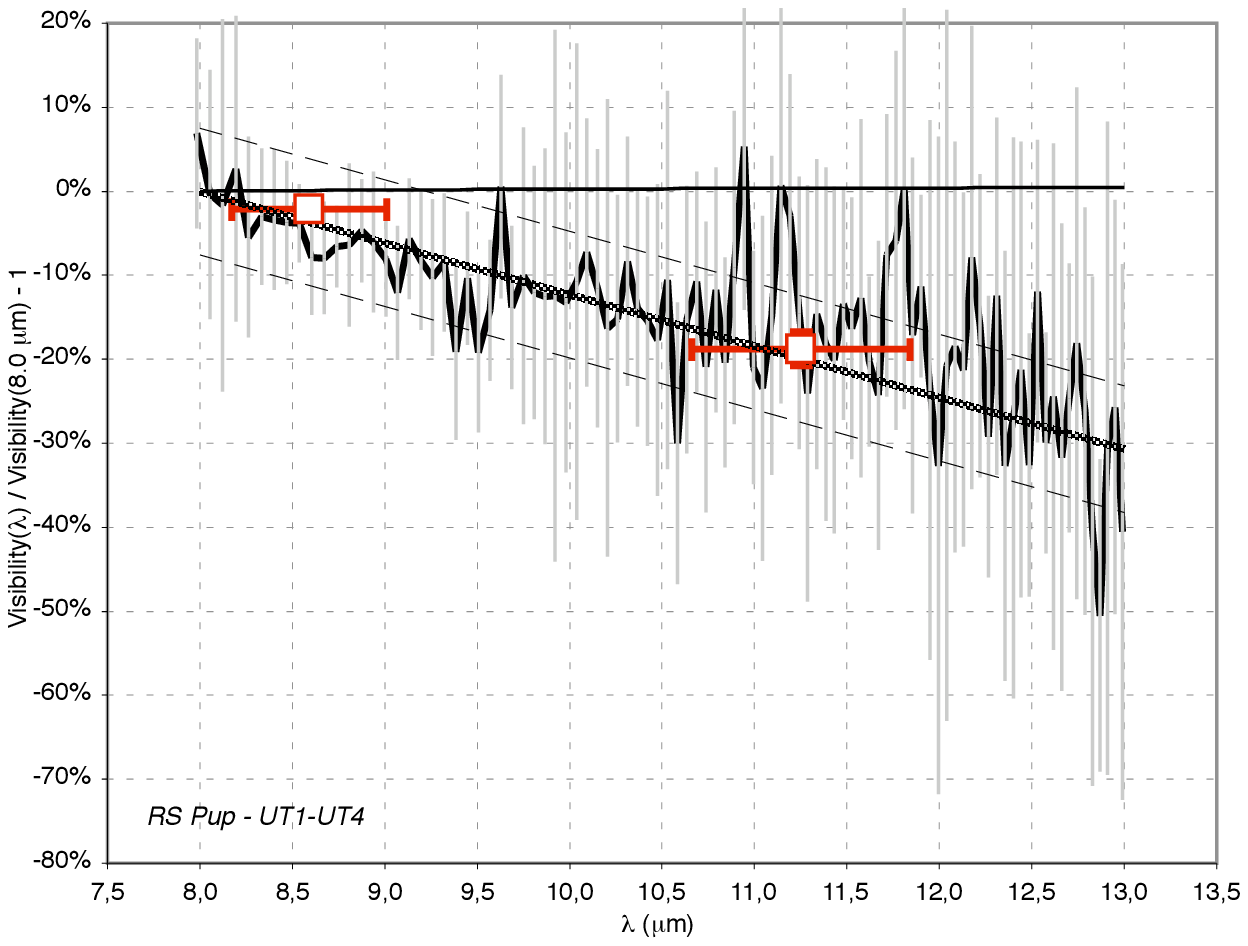}
\caption{Differential visibility of RS\,Pup on the UT1-UT4 baseline. The symbols are the same as in Fig.~\ref{lcar-visib}, but the vertical scale is different.\label{rspup-visib}}
\end{figure}

The differential visibility spectrum of RS\,Pup is shown in Fig.~\ref{rspup-visib}. This curve is the average of the three observations of the star labeled \#H, \#L and \#O in Table~\ref{midi_log}, obtained between phases $\phi = 0.92$ and $\phi = 0.97$. The variation in projected baseline length is negligible between the three observations. As observed for $\ell$\,Car, the visibility decreases significantly from 8 to 13\,$\mu$m, and the residual dispersion of the differential visibilities around the linear fit shown in Fig.~\ref{rspup-visib} as a thick grey line is 7.5\%.
The measured differential visibilities are well reproduced by our single-component extended emission model, and we obtain:
\begin{equation}
\Delta V_\mathrm{obs}(\mathrm{PAH1})= - 5.6 \pm 7.5\,\%,
\end{equation}
\begin{equation}
\Delta V_\mathrm{obs}(\mathrm{PAH2})= - 14.2 \pm 7.5\,\%,
\end{equation}
\begin{equation}
\Delta V_\mathrm{model}(\mathrm{PAH1})= - 2.1 \pm 1.2\,\%,
\end{equation}
\begin{equation}
\Delta V_\mathrm{model}(\mathrm{PAH2})= - 18.7 \pm 2.0\,\%.
\end{equation}
The observed and modeled values are in agreement well within $1\sigma$. As for $\ell$\,Car, we thus do not need to invoke a compact CSE component in addition to the VISIR resolved flux to explain the measured differential visibilities. Moreover, we do not detect a significant extended emission at 8.0\,$\mu$m at a level of 7\% of the stellar flux, as we do not detect a global negative shift of the model visibilities as in the case of $\ell$\,Car.

\subsection{$N$ band spectroscopy \label{midi-phot}}

The data acquisition of MIDI requires the observation of a visibility calibrator that is usually chosen to be also a photometric standard. The absolutely calibrated spectrum of the target can be obtained by dividing its average MIDI spectrum by that of its calibrator, and then multiplying the result by the absolutely calibrated template spectrum of the latter, for instance from Cohen et al.~(\cite{cohen99}).

Two of the 2006 observations of $\ell$\,Car were obtained with HD\,94510 as calibrator. This K1III giant is not in the Cohen et al.~(\cite{cohen99}) catalogue, and we derived its thermal infrared spectrum using its {\tt CalVin}\footnote{http://www.eso.org/observing/etc/}Ê limb darkened angular diameter ($\theta_{\rm LD} = 2.23 \pm 0.01$\,mas) and the theoretical spectrum by Castelli \& Kurucz~(\cite{castelli03}) for an effective temperature of 5\,000\,K, solar metallicity and $\log g = 3.0$ (parameters from Cayrel de Strobel et al.~\cite{cayrel01}). We checked {\it a posteriori} that the photometry compiled by Ducati~(\cite{ducati02}) is well reproduced by the resulting model spectrum, in the $B$ to $L$ bands. This gave us the template spectrum of HD\,94510 used for the photometric calibration of the $\ell$\,Car spectra \#J and \#P in Table~\ref{midi_log}.

\begin{figure}[]
\centering
\includegraphics[bb=12 8 363 281, width=8.9cm, angle=0]{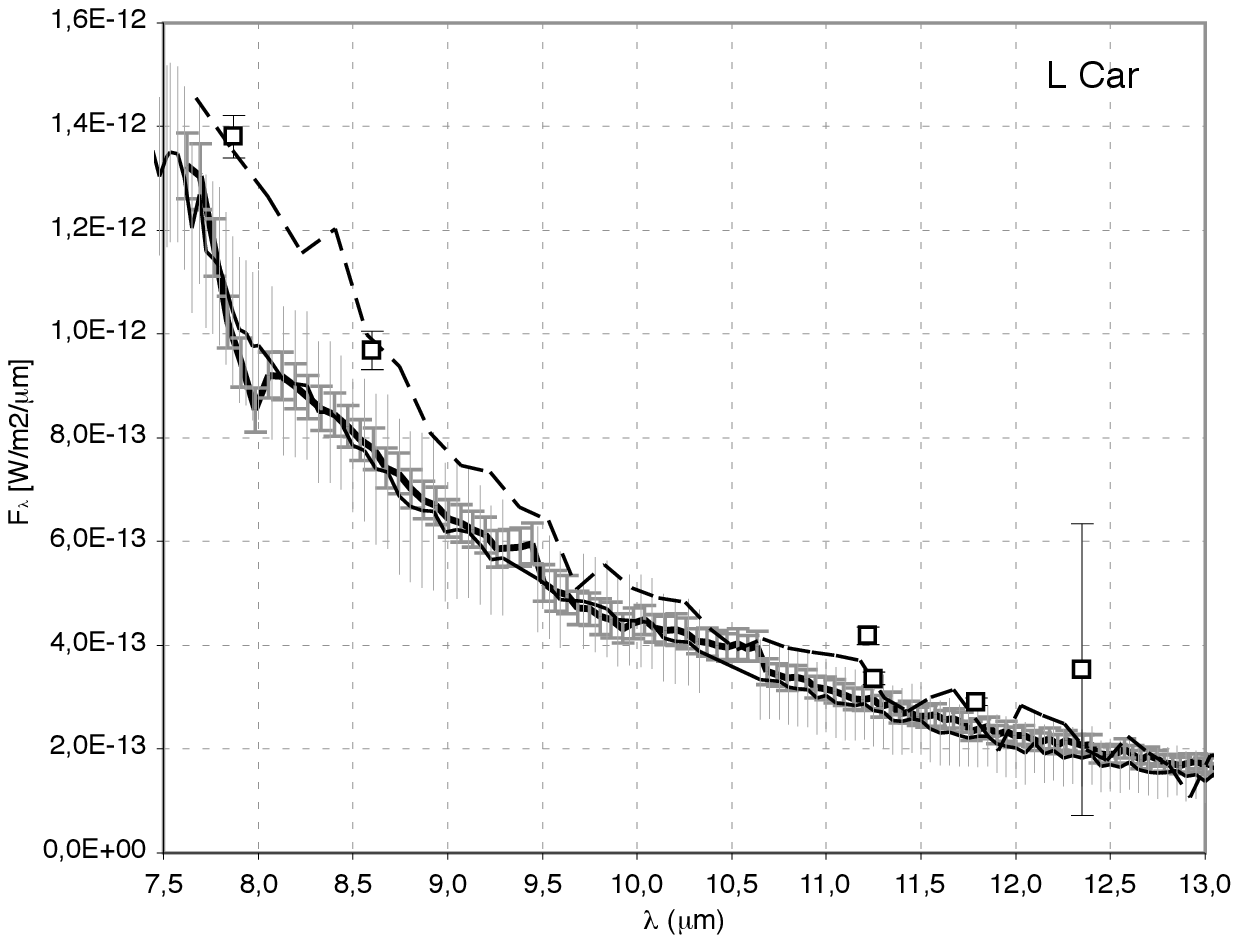}
\includegraphics[bb=12 8 363 281, width=8.9cm, angle=0]{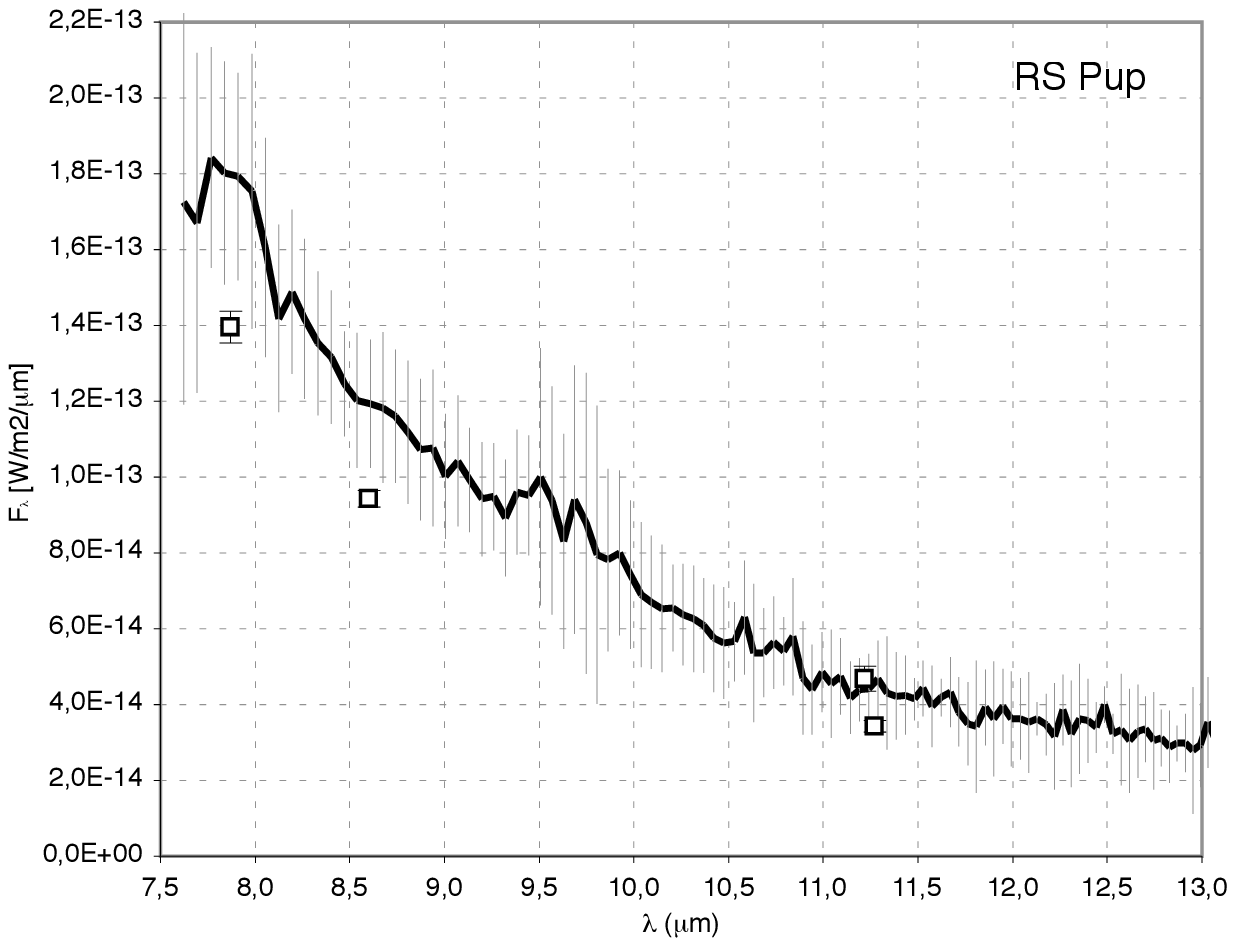}
\caption{Thermal infrared spectrum of $\ell$\,Car (top) and RS\,Pup (bottom) from MIDI. The thin curve in the top figure corresponds to the 2004 observation, while the thick curve is the 2006 data. The dashed curve is the IRAS spectrum of $\ell$\,Car, and the open square symbols in both plots represent the photometric measurements listed in Table~\ref{phot_table}, excluding the extinction correction.\label{midi-spectro}}
\end{figure}

The calibrated $N$ band spectra of $\ell$\,Car is presented in Fig.~\ref{midi-spectro} (top). The agreement with the IRAS spectrum from Volk \& Cohen~(\cite{volk89}) is qualitatively satisfactory. The slightly greater flux in the IRAS data is due to the larger aperture over which the photometry has been obtained. The extended component resolved with VISIR (Sect.~\ref{visir-fourier}) was excluded from the measurement, causing an underestimation of the total flux. There is no statistically significant change in the spectral energy distribution between the 2004 ($\phi = 0.90$) and 2006 ($\phi = 0.56$) epochs. 
The spectrum of RS\,Pup in the bottom part of Fig.~\ref{midi-spectro} is the first of this star in this wavelength domain. It was obtained close to the phase of maximum light in the visible ($\phi = 0.9-1.0$). Fig.~\ref{midi-spectro} shows that the agreement of the MIDI spectra with the broadband photometry listed in Table~\ref{phot_table} is satisfactory at the $\approx 1\,\sigma$ level for RS\,Pup (bottom curve).

An important aspect of the MIDI spectrophotometry is that it is obtained with a very narrow aperture. The field of view of the MIDI cold stop is approximately $1.5\arcsec$ in radius, over which is superimposed a broad slit of $0.6\arcsec \times 2.0\arcsec$, that rotates with the Earth rotation (due to the absence of field derotator). In the MIDI data processing software, the spectrum is extracted from the detector frames using a Gaussian weighting perpendicularly to the dispersion direction, that effectively reduces the field to $\approx 0.4\arcsec$ FWHM. At this stage, the background is subtracted from two lateral bands parallel to the dispersion direction within the field of view, thus subtracting also the contribution of any extended emission located beyond a radius of 0.2 to $0.3\arcsec$ from the star. For an example of aperture photometry with MIDI on the extended CSE of $\eta$\,Car, the interested reader is refered to Chesneau et al.~(\cite{chesneau05}).

Due to these difficulties, we did not use the MIDI spectra for the analysis of the CSE spectral properties in Sect.~\ref{discussion}, and restrict their use to the search for spectral features that are absent from the $N$ band spectra of the two stars. 

\section{VINCI data\label{vinci-2006}}

To complete our new observations, we also consider here the observations of $\ell$\,Car obtained with the VINCI instrument in the $K$ band by Kervella et al.~(\cite{kervella06}). They are well reproduced by the stellar photosphere and a Gaussian CSE with an angular radius at half maximum of
\begin{equation}
\rho_{2.2\,\mu\mathrm{m}}(\ell\,\mathrm{Car}) = 2.9 \pm 2.3\,\mathrm{mas}
\end{equation}
and a relative flux contribution $\alpha$ of
\begin{equation}
\alpha_{2.2\,\mu\mathrm{m}}(\ell\,\mathrm{Car}) = 4.2 \pm 0.2\%.
\end{equation}
This model is of the same nature as those we adjusted to the other data presented here. We can therefore directly include these parameters in our analysis.

\section{Properties of the CSEs of $\ell$\,Car and RS\,Pup\label{properties-sect}}


\subsection{Spectral energy distribution\label{sed-section}}

\begin{table*}
\caption{Overview of the infrared photometry of $\ell$\,Car and RS\,Pup from the present work and bibliography. All flux densities are corrected for interstellar extinction, except longwards of 30\,$\mu$m, where it is considered negligible. The ``$\bullet$" symbols in the ``Fit" column indicate the photometry used for the synthetic spectrum fitting. $\lambda_0$ is the average of the half maximum extreme wavelengths of each instrument/filter combination, and $\Delta \lambda$ the half-bandwidth. $\lambda_{\rm eff}$ is the effective wavelength of the measurement considering the spectral energy distribution of the star. The ``Ap." column lists the effective radius of the aperture over which the flux density was measured, when this parameter was available. In the ``Ref." column, the references are the following: F07 = Fouqu\'e et al.~(\cite{fouque07}), I86 = IPAC~(\cite{ipac86}), H88 = Helou \& Walker~(\cite{helou88}), K08 = present work, S04 = Smith et al.~(\cite{smith04}). ``$\alpha$" is the measured flux density excess, expressed in percentage of the photospheric flux (see Sect.~\ref{sed-section} for details), except for RS\,Pup's three longest wavelengths, where it is expressed in stellar flux units, followed by ``$\times$". The last column ``$N\sigma$" gives the detection level in number of times the RMS uncertainty.} 
\label{phot_table}
\begin{tabular}{lcclccclllr}
\hline \hline
\noalign{\smallskip}
Inst.\,/ & Band  & Fit & $\lambda_0 \pm \Delta \lambda$ & $\lambda_{\rm eff}$ & Ap. & Flux density & Flux density & Ref. & $\alpha \pm \sigma(\alpha)$ & $N\sigma$\\
{\it System} & &  & [$\mu$m] & [$\mu$m]  & [\arcsec] & [${\rm W/m}^{2}/\mu{\rm m}$] & [Jy] & & [\%] & \\
\noalign{\smallskip}
\hline
\noalign{\smallskip}
 $\mathbf{\ell}$\,{\bf Car} \\
{\it Johnson} & $B$ & $\bullet$ & $0.44 \pm 0.05$ & 0.44 & $-$ & $1.13 \pm 0.03\ 10^{-09}$  &  $72.6 \pm 2.0$ & F07 & $-1.5 \pm 2.8$ & -0.5 \\
{\it Johnson} & $V$ & $\bullet$ & $0.55 \pm 0.04$ & 0.56 & $-$  & $1.80 \pm 0.05\ 10^{-09}$  &  $181 \pm 5.1$  & F07 & $1.0 \pm 2.8$ & 0.3 \\
{\it Cousins}  & $I_c$ & $\bullet$ & $0.75 \pm 0.11$ & 0.75 & $-$  & $1.47 \pm 0.04\ 10^{-09}$  &  $274 \pm 7.7$  & F07 & $3.5 \pm 2.9$ & 1.2 \\
{\it 2MASS} & $J$ & $\bullet$ & $1.25 \pm 0.15$ & 1.22 & $-$  & $7.34 \pm 0.35\ 10^{-10}$  &  $382 \pm 18$  &  F07 & $4.3 \pm 4.9$ & 0.9 \\
 {\it 2MASS} & $H$ & $\bullet$ & $1.62 \pm 0.10$ & 1.65 & $-$  & $3.91 \pm 0.18\ 10^{-10}$  &  $342 \pm 16$   &  F07 & $-4.5 \pm 4.7$ & -1.0 \\
 {\it 2MASS} & $K$ & $\bullet$ & $2.20 \pm 0.30$ & 2.16 & $-$  & $1.67 \pm 0.08\ 10^{-10}$  &  $269 \pm 13$ &  F07 & $-1.1 \pm 4.7$ & -0.2 \\
 COBE$^{*}$ & 3.5\,$\mu$m & & $3.57 \pm 0.49$ & 3.55 & $\infty$ & $3.03 \pm 0.41\ 10^{-11}$  &  $127 \pm 17$ & S04 & $8.4 \pm 15$ & 0.6 \\
 IRAC & I1 & & $3.54 \pm 0.39$ & 3.55 & 24 & $2.78 \pm 0.08\ 10^{-11}$  &  $117 \pm 4$ & K08 & $3.1 \pm 3.1$ & 1.0 \\
 IRAC & I2 & & $4.51 \pm 0.51$ & 4.49  & 24 & $1.01 \pm 0.03\ 10^{-11}$  &  $68 \pm 2$ & K08 & $-0.3 \pm 3.0$ & -0.1 \\
COBE$^{*}$ & 4.9\,$\mu$m & & $4.93 \pm 0.37$ & 4.90 & $\infty$ & $7.47 \pm 1.22\ 10^{-12}$  &  $60 \pm 10$ & S04 & $11 \pm 18$ & 0.6 \\
 IRAC & I3 & & $5.73 \pm 0.70$ & 5.73  & 24 & $4.23 \pm 0.13\ 10^{-12}$  &  $46.3 \pm 1.4$  & K08 & $4.3 \pm 3.1$ & 1.4 \\
 IRAC & I4 & & $7.89 \pm 1.44$ & 7.87 & 24 & $1.41 \pm 0.04\ 10^{-12}$  &  $29.1 \pm 0.9$ & K08 & $8.6 \pm 3.3$ & 2.6 \\
 VISIR & PAH1 & & $8.59 \pm 0.42$ & 8.60 & 1.3 &  $9.96 \pm 0.38\ 10^{-13}$  &  $24.6 \pm 0.9$ & K08 & $14.9 \pm 4.4$ & 3.4 \\
 IRAS$^{**}$  & 12\,$\mu$m & & $11.5 \pm 3.5$ & 11.22 & 25 & $4.32 \pm 0.17\ 10^{-13}$  &  $18.1 \pm 0.7$ & I86 & $17.1 \pm 4.7$ & 3.6 \\
 VISIR & PAH2 & & $11.25  \pm 0.59$ & 11.25 & 1.3 &  $3.46 \pm 0.12\ 10^{-13}$  &  $14.6 \pm 0.5$ & K08 & $16.1 \pm 4.2$ & 3.9 \\
 VISIR & SiC & & $11.85  \pm 2.34$ & 11.79 & 1.3 & $3.00 \pm 0.07\ 10^{-13}$  &  $13.9 \pm 0.3$ & K08 & $16.7 \pm 2.9$ & 5.8 \\
COBE$^{*}$ & 12\,$\mu$m & & $12.71 \pm 4.08$ & 12.35 & $\infty$ & $3.64 \pm 2.89\ 10^{-13}$  &  $19 \pm 15$ & S04 & $18.6 \pm 94$ & 0.2 \\
 IRAS$^{**}$  & 25\,$\mu$m & & $24 \pm 6$ & 23.34 & 25 &  $2.37 \pm 0.14\ 10^{-14}$  &  $4.31 \pm 0.26$ & I86 & $25.3 \pm 7.5$ & 3.4 \\	
 MIPS & M1 & & $23.5 \pm 2.7$ & 23.68 & 35 & $2.00 \pm 0.08\ 10^{-14}$  &  $3.73 \pm 0.15$ & K08 & $23.4 \pm 4.9$ & 4.7 \\
 IRAS$^{**}$  & 60\,$\mu$m & & $62 \pm 17$ & 59.32 & 60 & $7.63 \pm 0.69\ 10^{-16}$  &  $0.90 \pm 0.08$ & I86 & $51.3 \pm 14$ & 3.8 \\	
 MIPS & M2 & & $70.4 \pm 9.54$ & 71.42 & 30 & $2.63 \pm 0.13\ 10^{-16}$  &  $0.45 \pm 0.02$ & K08 & $30.5 \pm 6.5$ & 4.7 \\
\hline
\noalign{\smallskip}
{\bf RS\,Pup} \\
 {\it Johnson} & $B$ & $\bullet$ & $0.44 \pm 0.05$ & 0.44 & $-$ & $1.56 \pm 0.04\ 10^{-10}$  &  $10.1 \pm 0.3$ & F07 & $1.5 \pm 2.8$ & 0.5 \\
 {\it Johnson} & $V$ & $\bullet$ & $0.55 \pm 0.04$ & 0.56 & $-$  & $2.21 \pm 0.06\ 10^{-10}$  &  $22.3 \pm 0.6$  & F07 & $-0.5 \pm 2.8$ & -0.2 \\
 {\it Cousins}  & $R_c$ & $\bullet$ & $0.65 \pm 0.15$ & 0.65 & $-$  & $2.30 \pm 0.06\ 10^{-10}$  &  $32.1 \pm 0.9$  & F07 & $7.7 \pm 3.0$ & 2.6 \\
 {\it Cousins}  & $I_c$ & $\bullet$ & $0.75 \pm 0.11$ & 0.75 & $-$  & $1.76 \pm 0.05\ 10^{-10}$  &  $32.8 \pm 0.9$  & F07 & $4.6 \pm 2.9$ & 1.6 \\
 {\it 2MASS} & $J$ & $\bullet$ & $1.25 \pm 0.15$ & 1.22 & $-$  & $8.40 \pm 0.40\ 10^{-11}$  &  $43.7 \pm 2.1$ & F07 & $6.3 \pm 5.0$ & 1.3 \\
 {\it 2MASS} & $H$ & $\bullet$ & $1.62 \pm 0.10$ & 1.65 & $-$  & $4.28 \pm 0.20\ 10^{-11}$  &  $37.4 \pm 1.8$  & F07 & $-3.0 \pm 4.6$ & -0.7 \\
 {\it 2MASS} & $K$ & $\bullet$ & $2.20 \pm0.30$ & 2.16 & $-$  & $1.82 \pm 0.09\ 10^{-11}$  &  $29.4 \pm 1.4$ & F07 & $0.9 \pm 4.8$ & 0.2 \\
  IRAC & I1 & & $3.54 \pm 0.39$ & 3.55 & 6.1 &  $3.41 \pm 0.10\ 10^{-12}$ &  $14.3 \pm 0.4$ & K08 & $19.2 \pm 3.6$ & 5.4 \\
 IRAC & I2 & & $4.51 \pm 0.51$ & 4.49  & 6.1 &  $1.31 \pm 0.04\ 10^{-12}$  &  $8.82 \pm 0.26$ & K08 & $20.8 \pm 3.6$ & 5.7 \\
 IRAC & I3 & & $5.73 \pm 0.70$ & 5.73 & 6.1 &  $5.16 \pm 0.16\ 10^{-13}$  &  $5.22 \pm 0.17$ & K08 & $18.7 \pm 3.6$ & 5.3 \\
 IRAC & I4 & & $7.89 \pm 1.44$ & 7.87 & 6.1 &  $1.49 \pm 0.05\ 10^{-13}$  &  $3.08 \pm 0.09$ & K08 & $8.2 \pm 3.2$ & 2.5 \\
 VISIR & PAH1 & & $8.59  \pm 0.42$ & 8.60 & 1.3 &  $1.03 \pm 0.02\ 10^{-13}$  &  $2.53 \pm 0.06$ & K08 & $11.4 \pm 2.5$ & 4.5 \\
 IRAS$^{**}$  & 12\,$\mu$m & & $11.5 \pm3.5$ & 11.22 & 25 & $5.15 \pm 0.36\ 10^{-14}$  &  $2.16 \pm 0.15$ & I86 & $31 \pm 9.2$ & 3.4 \\	
 VISIR & PAH2 & & $11.25 \pm 0.59$ & 11.27 & 1.3 &  $3.76 \pm 0.16\ 10^{-14}$  &  $1.59 \pm 0.16$ & K08 & $18.8 \pm 5.2$ & 3.6 \\
  IRAS$^{**}$  & 25\,$\mu$m & & $24 \pm 6$ & 23.34 & 25 &  $3.96 \pm 0.28\ 10^{-15}$  & $0.72 \pm 0.05$ & I86 & $96 \pm 14$ & 7.0 \\	
 MIPS & M1 & & $23.5 \pm 2.7$ & 23.68 & 180 & $4.21 \pm 0.17\ 10^{-15}$  &  $0.79 \pm 0.03$ & K08 & $145 \pm 9.8$ & 15 \\
 IRAS  & 60\,$\mu$m & & $62 \pm 17$ & 59.32 & $\infty$ &  $1.19 \pm 0.26\ 10^{-14}$  &  $13.9 \pm 3.1$ & H88 & $220 \pm 49\times$ & 4.5 \\
 MIPS$^{***}$ & M2 & & $70.4 \pm 9.5$ & 71.42 & $\infty$ & $9.94 \pm 0.50\ 10^{-15}$  &  $16.9 \pm 0.9$ & K08 & $463 \pm 23\times$ & 20 \\
 IRAS  & 100\,$\mu$m & & $103 \pm 18$ & 100.3 & $\infty$ & $7.87 \pm 1.57\ 10^{-15}$  &  $26.4 \pm 5.3$ & H88 & $1\,430 \pm 290\times$ & 5 \\	
\hline
\end{tabular}
\begin{list}{}{}
\item[$^{*}$] The angular resolution of COBE/DIRBE is $\approx 0.7^\circ$, its aperture is thus considered as infinitely large.
\item[$^{**}$] The IRAS Point Source Catalogue (IPAC~\cite{ipac86}) apertures were taken as the 80\% encircled energy from Beichman et al.~(\cite{beichman88}).
\item[$^{***}$] The MIPS/M2 flux density of RS\,Pup's CSE is the integral of the Gaussian profile adjusted in Sect.~\ref{cse-sb}.
\end{list}
\end{table*}

To retrieve the spectral energy distribution (SED) of the CSEs of $\ell$\,Car and RS\,Pup from the total observed flux, we need to estimate the contribution from the stellar photosphere. For this purpose, we used a combination of our new photometry in the near to mid-IR domains with intensity mean photometry assembled by Fouqu\'e et al.~(\cite{fouque07}) in the $BVR_cI_cJHK$ bands from the literature. In the infrared, we also collected the COBE/DIRBE photometry of $\ell$\,Car at 3.5, 4.9 and 12\,$\mu$m (Smith et al.~\cite{smith04}), the photometry of $\ell$\,Car at 12, 25 and $60\,\mu$m and RS\,Pup at 12 and 25\,$\mu$m from the IRAS Point Source Catalogue (IPAC~\cite{ipac86}). We took the photometry of RS\,Pup at 60 and $100\,\mu$m from the IRAS Small Scale Structure Catalogue (Helou \& Walker~\cite{helou88}), as the angular size of the nebula at these wavelengths becomes non-negligible compared to the angular resolution of IRAS.

The $B$ to $K$ photometry from Fouqu\'e et al.~(\cite{fouque07}) is already corrected for interstellar absorption. In order to preserve the consistency of our data set at all wavelengths, we adopted their $E(B-V)$ color excess values, namely $0.147 \pm 0.013$ for $\ell$\,Car and $0.457 \pm 0.009$ for RS\,Pup. We also used the same total-to-selective absorption ratio $R_V = 3.23$ to derive $A_V$ extinctions for the two stars: $A_V(\ell\,\mathrm{Car}) = 0.475$ and $A_V(\mathrm{RS Pup}) = 1.476$. We then used these values and the interstellar extinction curves in the mid-IR by McClure~(\cite{mcclure08}) to correct all our photometric measurements between 3 and 30\,$\mu$m. As a remark, the curves from this author are in excellent agreement with the independent results by Chapman et al.~(\cite{chapman08}), so we do not expect any significant systematics from this operation. Longwards of 30\,$\mu$m, the correction to apply to the flux density is negligible (below 1\% for $\ell$\,Car and 2\% for RS\,Pup), compared to the uncertainties. The complete set of measurements is summarized in Table~\ref{phot_table}.

To model the photospheric emission, we used tabulated stellar atmosphere models obtained with the ATLAS9 simulation code that we retrieved from Castelli \& Kurucz's~(\cite{castelli03}) library. The grid we chose was computed for solar metallicity and a turbulence velocity of 2\,km/s. We finely interpolated this grid in order to compute spectra for any effective temperature and any surface gravity. We multiplied the spectrum by the solid angle of the stellar phosphere, $\pi \theta_\mathrm{LD}^2/4$, where $\theta_\mathrm{LD}$ is the limb darkened angular diameter, in radians, in order to scale the model. The photometric data were adjusted to the model taking into account the spectral response of each instrument listed in Table~\ref{phot_table}. Since we assume that there is no detectable excess (5\% or less) below the 2.2\,$\mu$m $K$ band, we included all the photometric measurements bluer than the $K$ band to estimate the angular diameter and the effective temperature. Note that we did not adjust the surface gravity, since the broadband photometry is mostly insensitive to this parameter.

Although we are not interested here in an unbiased estimate of the angular diameter $\theta_\mathrm{LD}$, the best-fit values of $\theta_{\rm LD}(\ell\,\mathrm{Car}) = 2.92$\,mas is only 2\% smaller than the mean angular diameter of $\theta_{\rm LD}(\ell\,\mathrm{Car}) = 2.99 \pm 0.01$\,mas measured by Kervella et al.~(\cite{kervella04c}) in the $K$ band. For RS\,Pup, the agreement between the present best-fit value of $\theta_{\rm LD}(\mathrm{RS\,Pup}) = 0.93$\,mas and the 1.02\,mas prediction by Moskalik \& Gorynya~Ê(\cite{moskalik06}) is also reasonably good. The best-fit effective temperatures $T_\mathrm{eff}$ of $\ell$\,Car and RS\,Pup are respectively 4860\,K and 5040\,K. One should note however that there is a correlation between $T_\mathrm{eff}$ and the angular diameter in this fit. Following Kervella et al.~(\cite{kervella04c}), we consider an effective gravity of $\log g=1.5$, typical of a long period Cepheid, but changing this parameter by $\pm 1.0$\,dex does not significantly affect the quality of the fit, nor the amount of infrared excess.

\begin{figure}[]
\centering
\includegraphics[bb=12 8 363 281, width=8.9cm, angle=0]{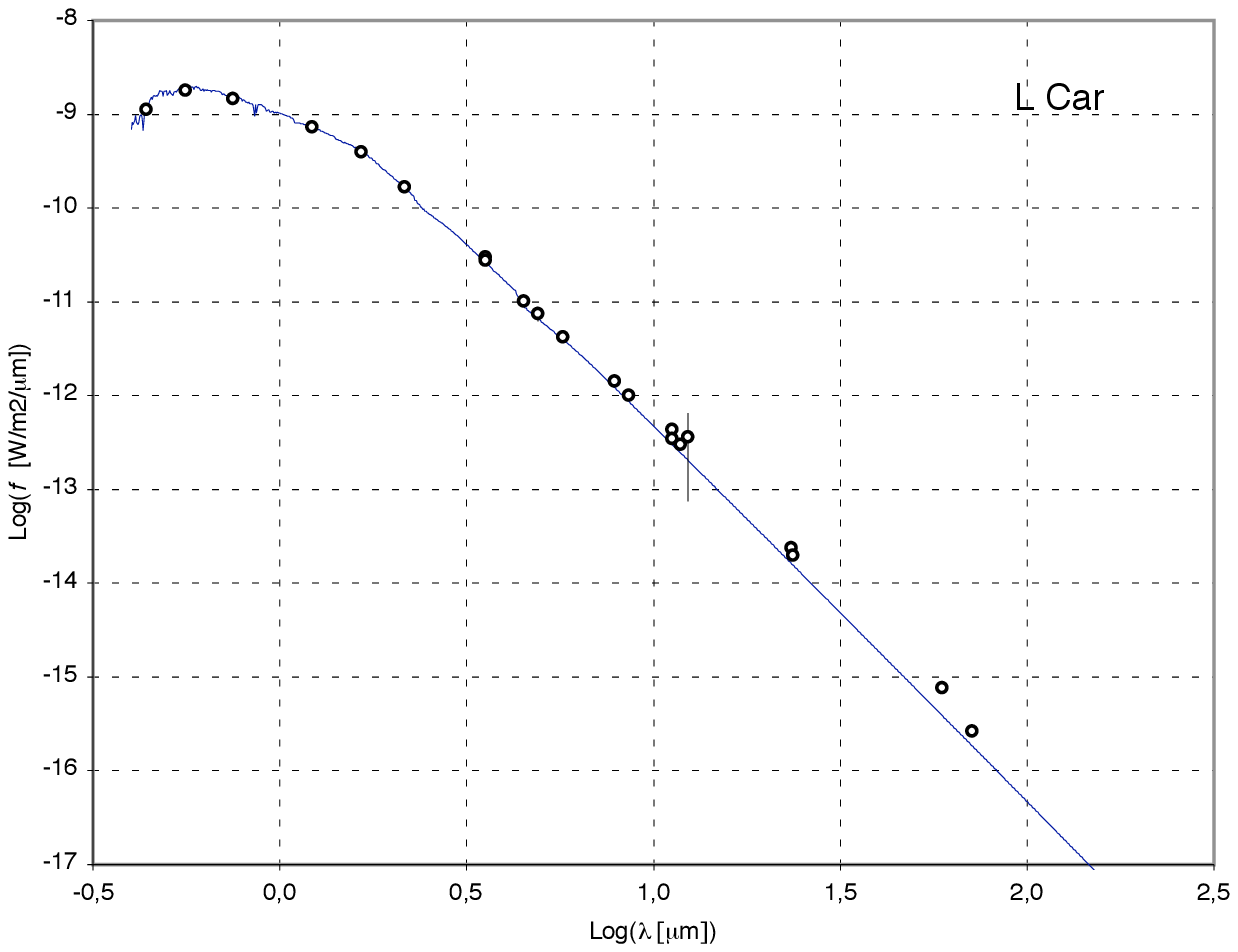}
\includegraphics[bb=12 8 363 281, width=8.9cm, angle=0]{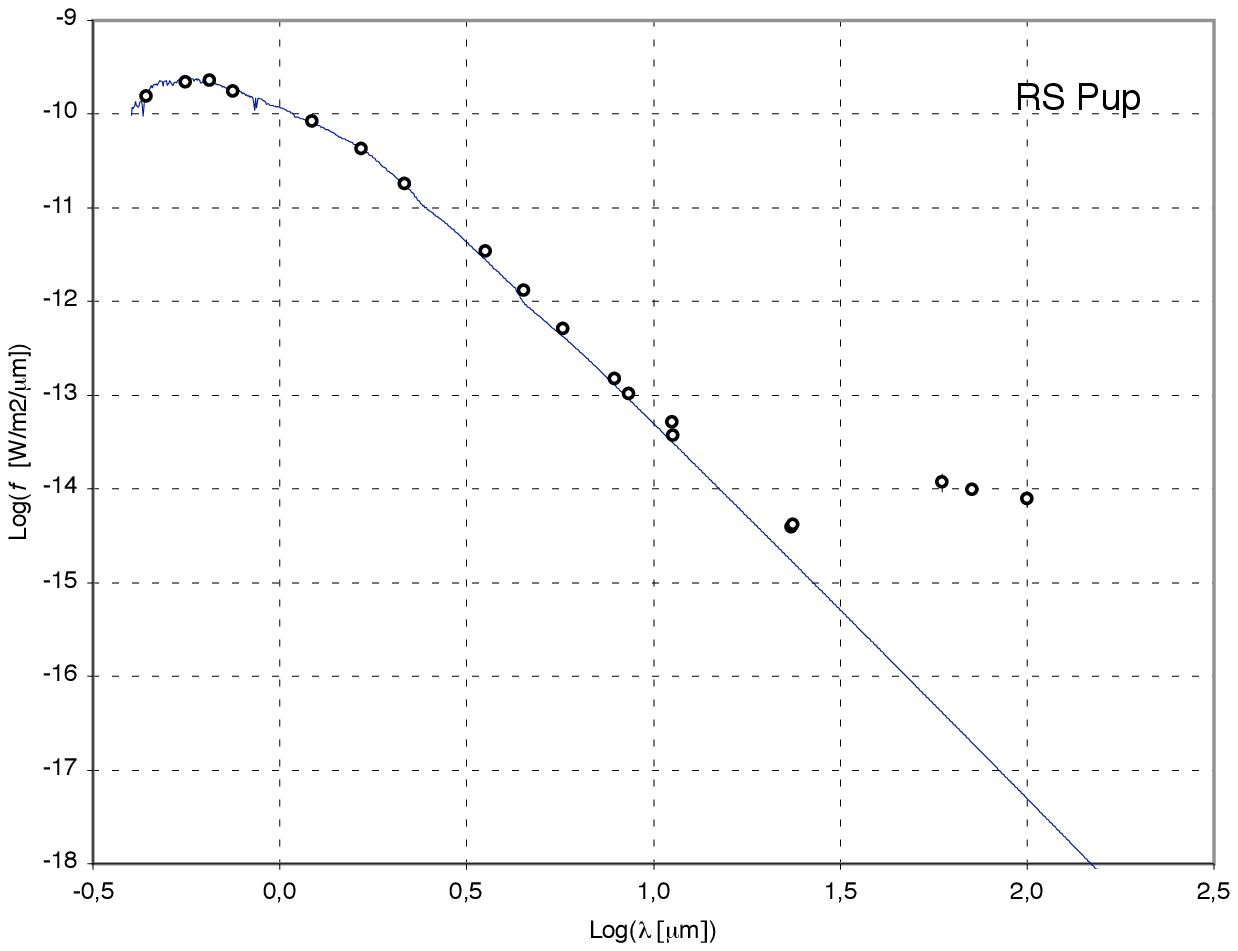}
\caption{Synthetic spectra of $\ell$\,Car and RS\,Pup (solid line) with the photometric measurements listed in Table~\ref{phot_table} (dots). The residuals of the subtraction of the synthetic spectrum from the measurements is shown in panels (c) and (d) of Fig.~\ref{cepheid-excess}. \label{cepheid-spectra}}
\end{figure}

The fitting process produces a good agreement between the model spectrum and the $B$ to $K$ band photometric measurements, with a reduced $\chi^2$ of 0.7 for $\ell$\,Car, and 2.2 for RS\,Pup. It should also be noted that in the case of RS\,Pup, the presence of the nebulosity makes the estimation and correction of the interstellar and circumstellar reddening quite difficult, which might explain in part the slightly worse agreement. The result of this fit is presented in Fig.~\ref{cepheid-spectra} for $\ell$\,Car and RS\,Pup. In both cases, the presence of excess radiation in the thermal infrared domain is clearly detected.

\subsection{Spectral and spatial structure of the infrared excess}

\begin{figure*}[]
\centering
\includegraphics[bb=18 8 369 281, width=8.9cm, angle=0]{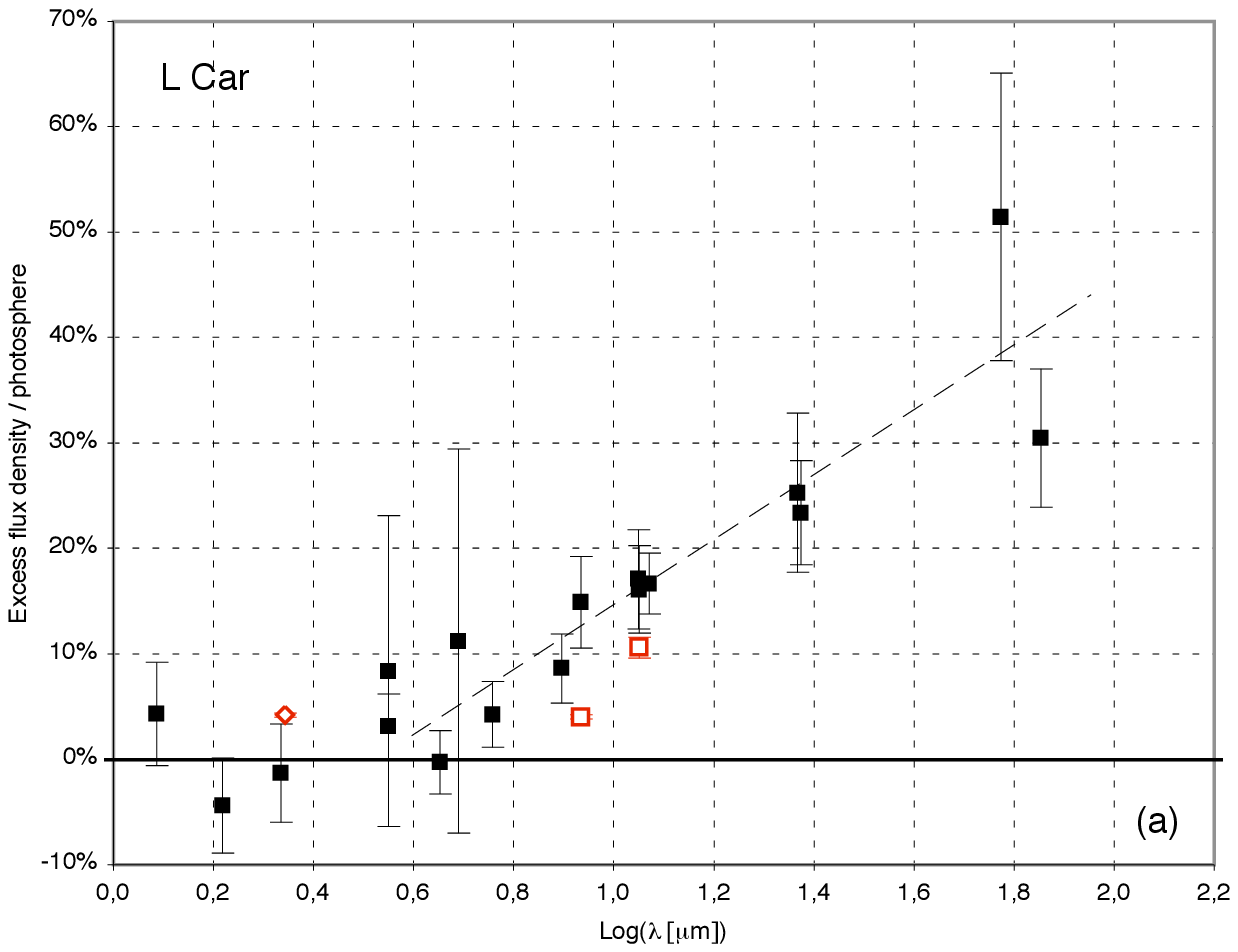}
\includegraphics[bb=12 8 363 281, width=8.9cm, angle=0]{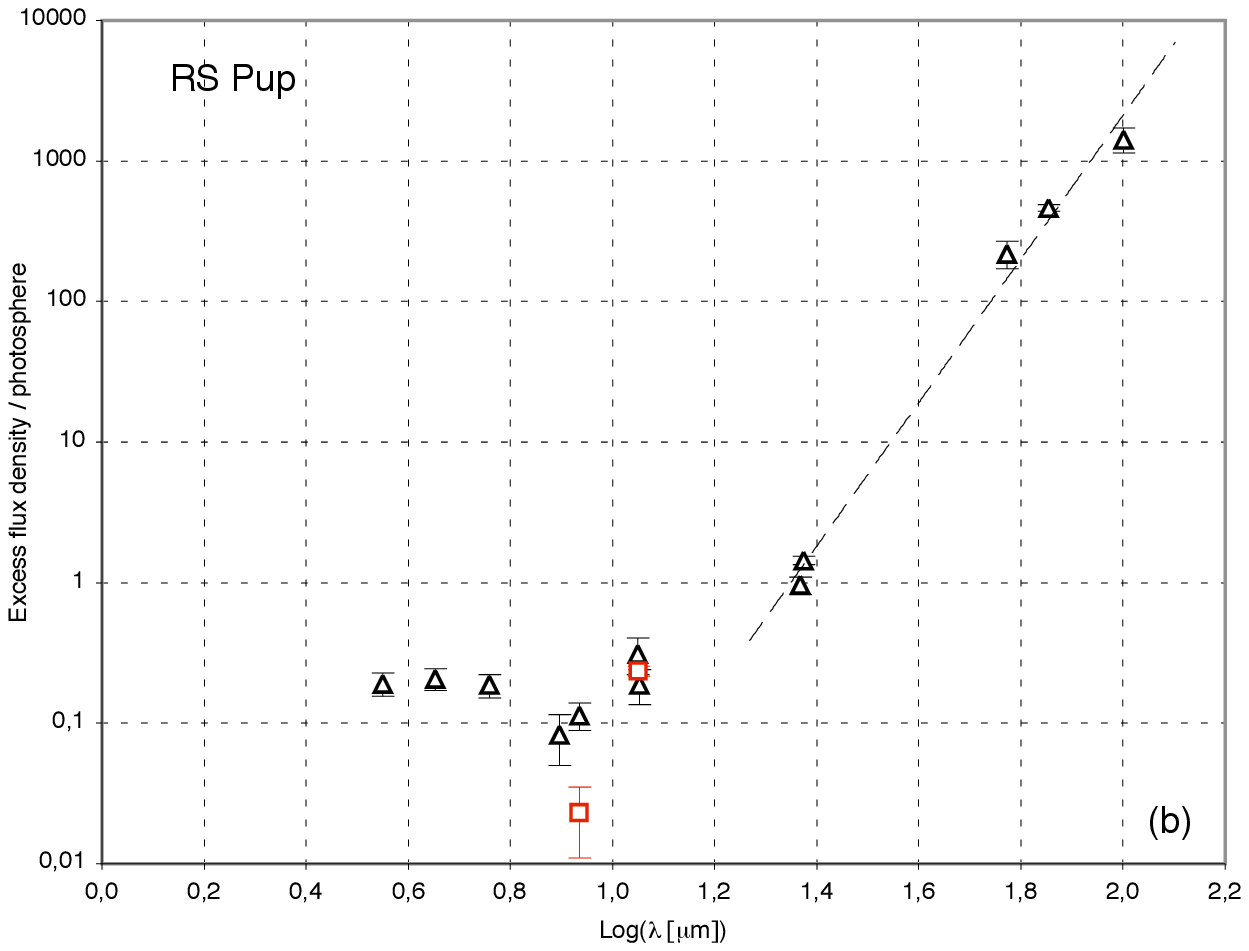}
\includegraphics[bb=18 8 369 281, width=8.9cm, angle=0]{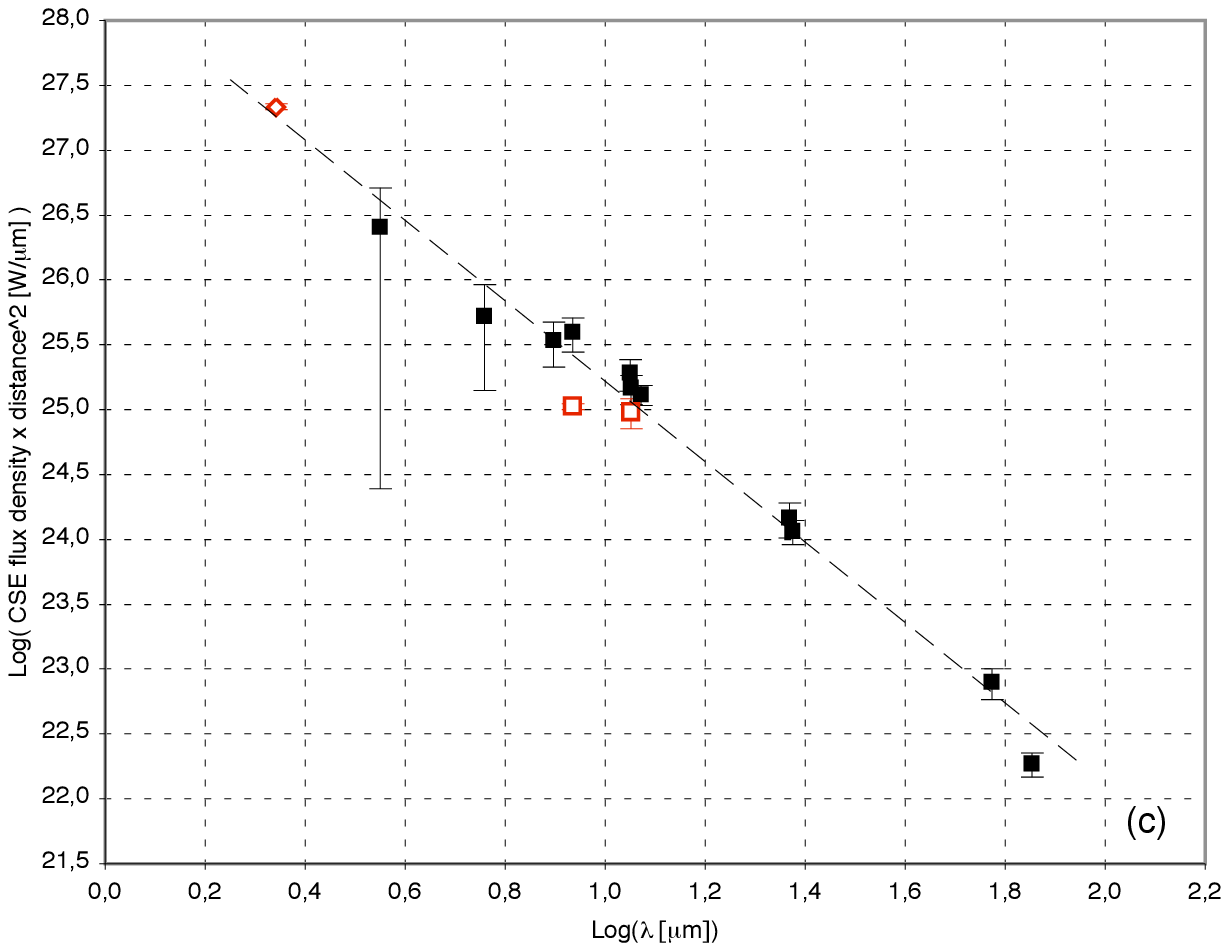}
\includegraphics[bb=12 8 363 281, width=8.9cm, angle=0]{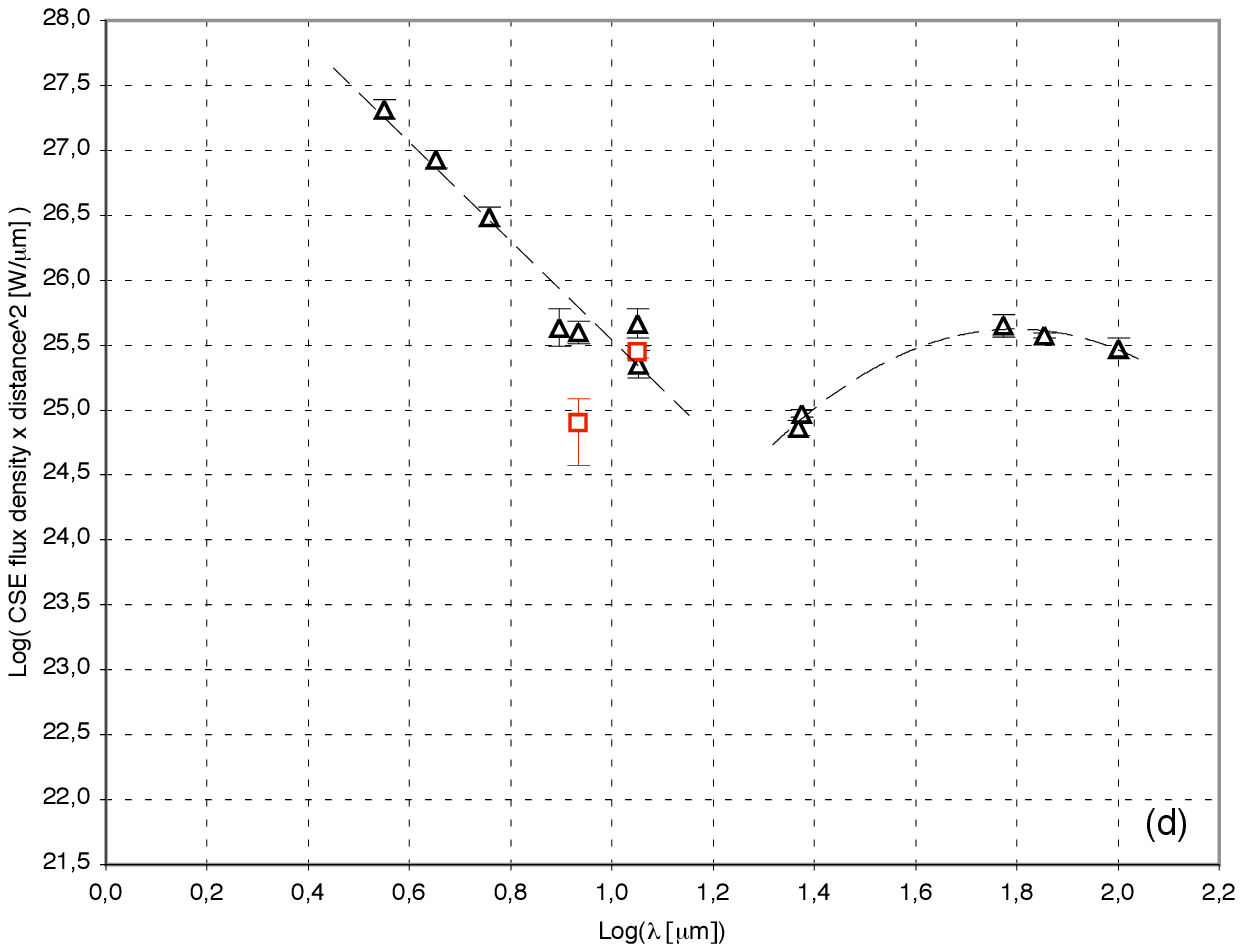}
\includegraphics[bb=18 8 369 281, width=8.9cm, angle=0]{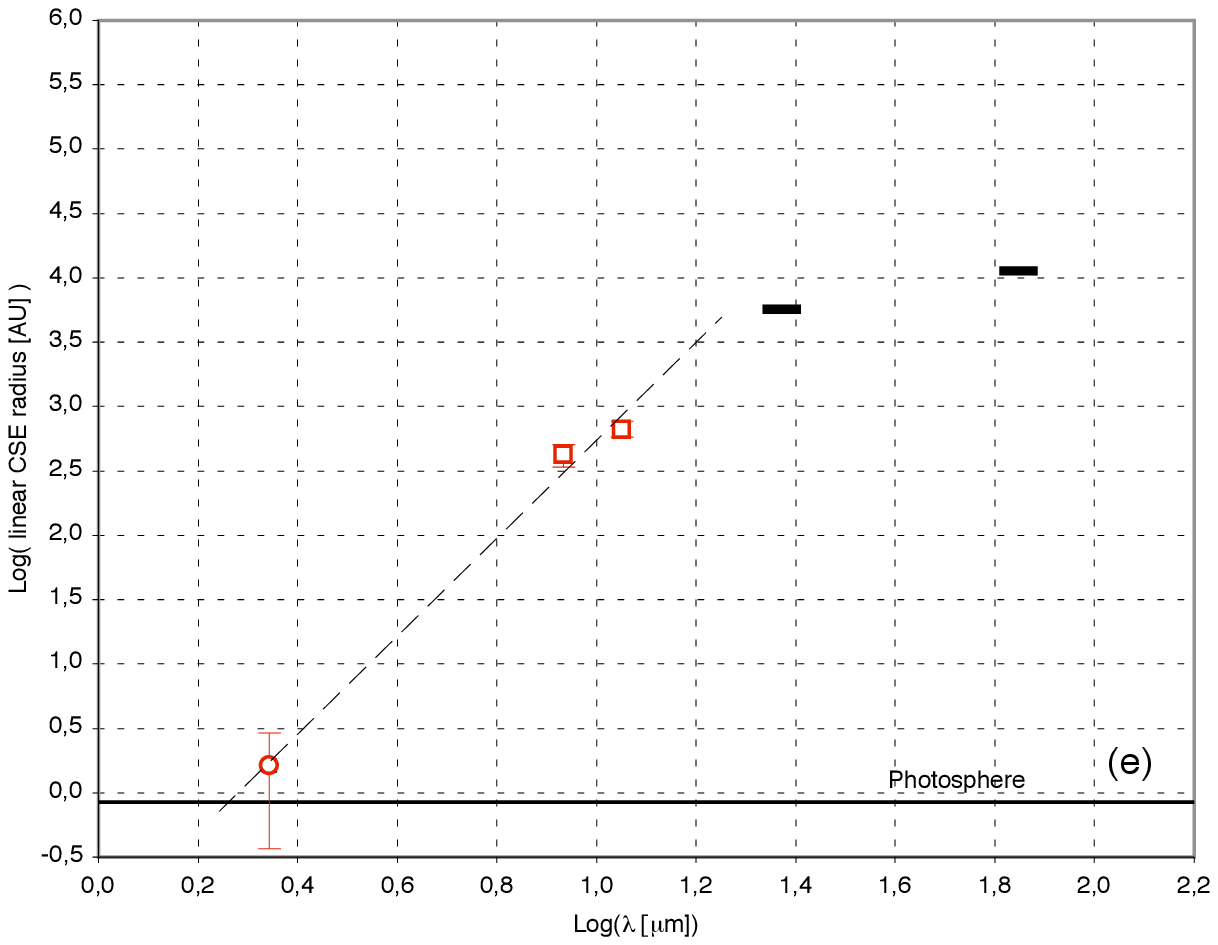}
\includegraphics[bb=12 8 363 281, width=8.9cm, angle=0]{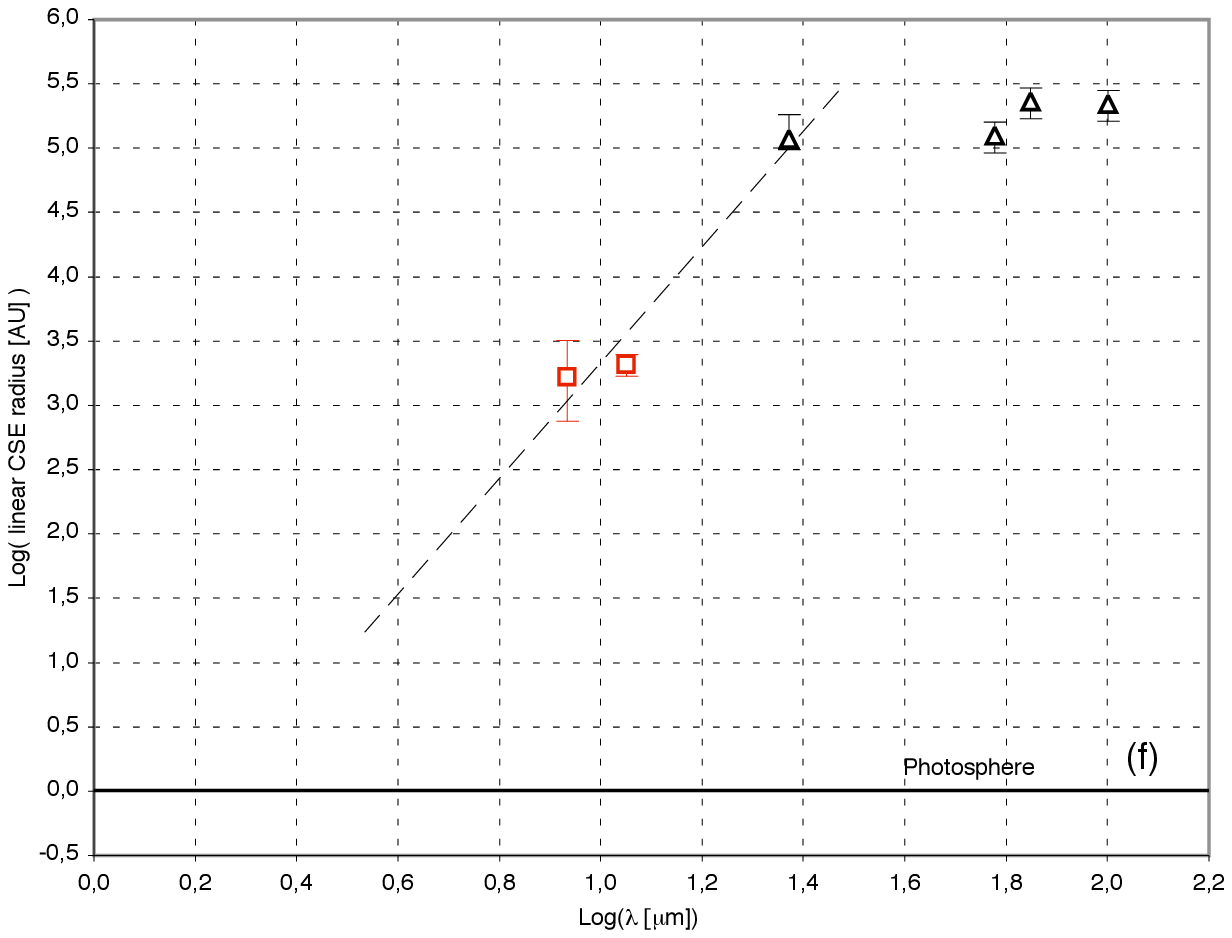}
\caption{Comparison of the spatial and spectral properties of the CSEs of $\ell$\,Car (left) and RS\,Pup (right). In all panels, the open squares represent the CSE resolved with VISIR and the open circles the CSE of $\ell$\,Car in the $K$ band (resolved with VINCI). {\bf (a) and (b):} Excess flux density relatively to the photospheric emission, with a different vertical scale for $\ell$\,Car and RS\,Pup. {\bf (c) and (d):} Flux density from the CSE alone (stellar contribution subtracted), multiplied by the squared distance of each star. {\bf (e) and (f):} Half-maximum linear radius of the envelope as a function of wavelength. The photospheric linear size is shown as a solid line. The \emph{Spitzer} observations of $\ell$\,Car at 24 and 70\,$\mu$m provide upper limits only, as the CSE was not resolved. {\it Ad hoc} models are shown as thin dashed curves to guide the eye on the different plots. \label{cepheid-excess}}
\end{figure*}

We discuss in this paragraph the distribution of the infrared excess as a function of wavelength, both in the spectral energy distribution and as spatially extended emission. The different properties of the CSEs of $\ell$\,Car and RS\,Pup are summarized in Fig.~\ref{cepheid-excess}. While we discuss here the overall properties of the observed circumstellar emission, we postpone the presentation of a detailed model of the Cepheid CSEs to a future publication, that will incorporate the present results and the scattered light observations of RS\,Pup's nebula obtained by Kervella et al.~(\cite{kervella08}).

To convert the angular values determined previously into linear distances from the star, $\ell$\,Car, we used the distance of 566\,pc measured interferometrically by Kervella et al.~(\cite{kervella04a}). While we do not resolve the CSE in the \emph{Spitzer}/MIPS images at 24 and 70\,$\mu$m, the angular resolution of these images allows us to set an upper limit on its extension of $\approx 10\arcsec$ and $20\arcsec$, respectively. The shortest wavelength at which the excess emission of $\ell$\,Car is detected is $\approx 8.0\,\mu$m. For RS\,Pup, we assume a distance of 1\,992\,pc (Kervella et al.~\cite{kervella08}). We included in Fig.~\ref{cepheid-excess} panel (f) the extension of the RS\,Pup nebula measured with IRAS at 60 and 100\,$\mu$m, respectively of 2.1 and 3.7 arcmin in angular diameter (Deasy~\cite{deasy88}).

The CSEs of $\ell$\,Car and RS\,Pup appear different in many respects. The distance corrected flux density of $\ell$\,Car's envelope appears as continuously decreasing with wavelength ( Fig.~\ref{cepheid-excess}, panel (c)), with an almost exponential dependence. The observation of a CSE flux contribution at 2.2\,$\mu$m by interferometry (Kervella et al.~\cite{kervella06}) shows that a very hot and compact component is present in its CSE. From simple radiative equilibrium considerations and its measured linear extension (Fig.~\ref{cepheid-excess}, panel (e)), this implies that the envelope starts within a few stellar radii, and extends continuously up to at least several hundred astronomical units from the star. 

The envelope of RS\,Pup is on average much larger and colder than that of $\ell$\,Car. The most striking difference between $\ell$\,Car and RS\,Pup's CSEs is their brightness around 70\,$\mu$m (Fig.~\ref{cepheid-excess}, panels (c) and (d)). While the two stars exhibit approximately the same brightness around 10\,$\mu$m, RS\,Pup's envelope is $\approx 2\,000$\,times brighter than $\ell$\,Car's at 70\,$\mu$m.
The presence of such a considerable emission at $70\,\mu$m indicates that the temperature of the majority of the dusty material in the CSE of RS\,Pup is around 40\,K, as already established by McAlary \& Welch~(\cite{mcalary86}). However, our VISIR and MIDI observations around 10\,$\mu$m as well as the IRAS photometry at 12\,$\mu$m show that a warm component is also present in the CSE of RS\,Pup. The angular extension of this warm component, as well as its flux contribution, appear to be of the same order of magnitude as that of $\ell$\,Car, as shown in panels (c) to (f) of Fig.~\ref{cepheid-excess}.
The observed properties of RS\,Pup's CSE therefore suggest that it consists of two components:
\begin{itemize}
\item A warm and compact component, located at typical distances of a few 100 to a few 1\,000\,AU from the star, with a typical temperature of a few 100\,K (maximum flux around 10\,$\mu$m),
\item A cold component distributed within a relatively large hollow ellipsoid with a typical radius of a few 100\,000\,AU and temperature around 40\,K (maximum flux around 70\,$\mu$m). This cold component is also the scattering medium for the light echoes observed by Havlen~(\cite{havlen72}) and Kervella et al.~(\cite{kervella08}). It could also host the thermal infrared ``heat echoes" predicted by Mayes et al.~(\cite{mayes85}), although their observation appears extremely challenging.
\end{itemize}
In contrast to RS\,Pup, the CSE of $\ell$\,Car appears to be devoid of a cold dust component, but its warm component (resolved with VISIR and MIDI) is comparable to RS\,Pup's, with a typical size of 500\,AU and a temperature of a few 100\,K. 

\section{Discussion \label{discussion}}

We present in this Section our interpretation of the properties of the CSEs of $\ell$\,Car and RS\,Pup. We then examine the mass loss scenarios predicted by existing Cepheid models, in view of these properties.

\subsection{Origins and evolution of the CSEs}

In spite of their relatively similar physical properties, $\ell$\,Car and RS\,Pup appear to be surrounded by two rather different kinds of CSEs.
The warm (10\,$\mu$m) components of the CSEs of both stars are probably created by a continuous mass loss from the Cepheids themselves, likely driven by pulsation as discussed by M\'erand et al.~(\cite{merand07}).
The hot CSEs observed in the $K$ band around $\ell$\,Car (Sect.~\ref{vinci-2006}), around \object{Polaris} and \object{$\delta$ Cep} (M\'erand et al.~\cite{merand06}), and around \object{Y Oph} (M\'erand et al.~\cite{merand07}) very likely represent the innermost emission (within a few stellar radii of the Cepheid) of the warm component we observed with VISIR and MIDI around our two targets. 
%
Along the same direction, B\"ohm-Vitense \& Love~(\cite{bohm94}) suggested that the fixed (non pulsating) H$\alpha$ absorption feature that they observe in the spectrum of $\ell$\,Car is caused by a CSE with a size of the order of 1000\,AU. This dimension is in good agreement with the typical size of the compact component we detected. More recently, Nardetto et al.~(\cite{nardetto08}) confirmed the presence of a non-pulsating component in the H$_\alpha$ profiles of both $\ell$\,Car and RS\,Pup.

However, the large ellipsoidal cold dust shell surrounding RS\,Pup is probably made of compressed interstellar material. We speculate that the mechanism for this compression could be linked to the past evolution of RS\,Pup. Its evolutionary mass was determined by Caputo et al.~(\cite{caputo05}) to be $12.8 \pm 0.9$\,M$_\odot$, and it thus spent its lifetime on the main sequence as a B2V hot dwarf, most probably fast rotating, with an effective temperature around 20\,000\,K. A good example of such a Cepheid precursor is the nearby Be star Achernar. This spectacularly flattened fast rotating B3 dwarf was shown recently to experience mass loss through a fast stellar wind ejected from its overheated polar caps (Kervella \& Domiciano de Souza~\cite{kervella06b}). Such a preferential mass loss along the polar axis of the star could explain the elongation observed in the 70\,$\mu$m image of RS\,Pup presented in Fig.~\ref{rspup-images}. The interstellar medium (ISM) would be more efficiently compressed along the polar axis, thus elongating the cavity carved by the star in the dusty ISM. Another additional pressure would come from the enhanced ultraviolet flux as seen from above the stellar poles, compared to the darkened equatorial belt (due to the Von Zeipel effect: see Aufdenberg et al.~\cite{aufdenberg06} for an application to Vega).

\begin{figure*}[ht]
\centering
\includegraphics[width=6cm]{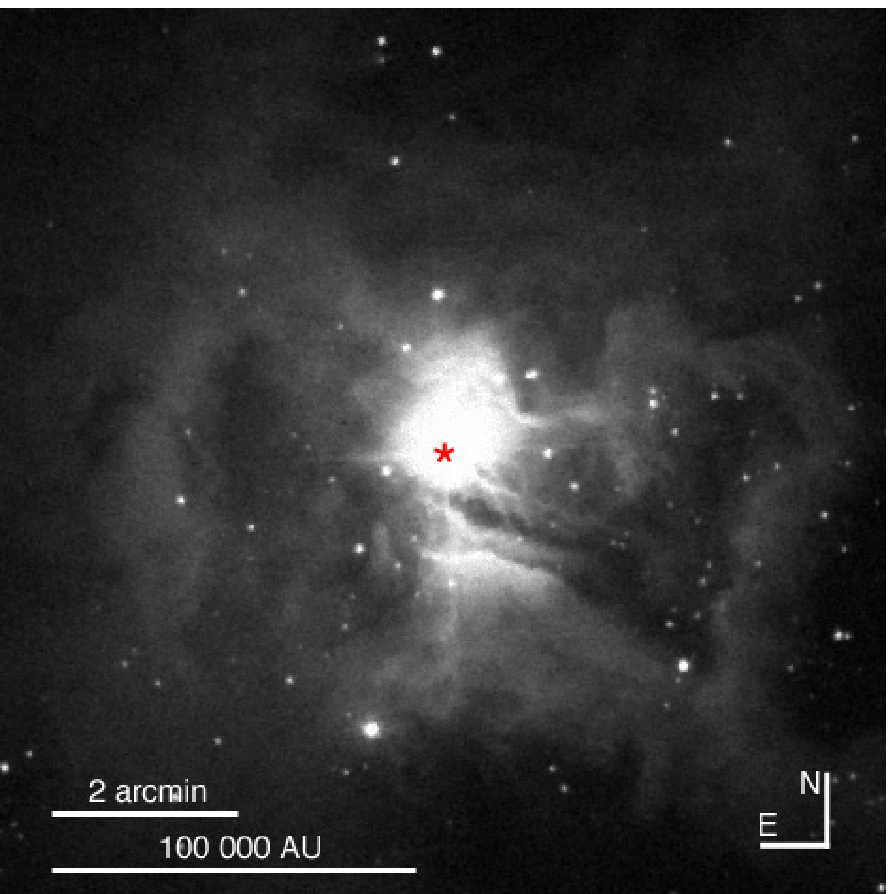}
\includegraphics[width=6cm]{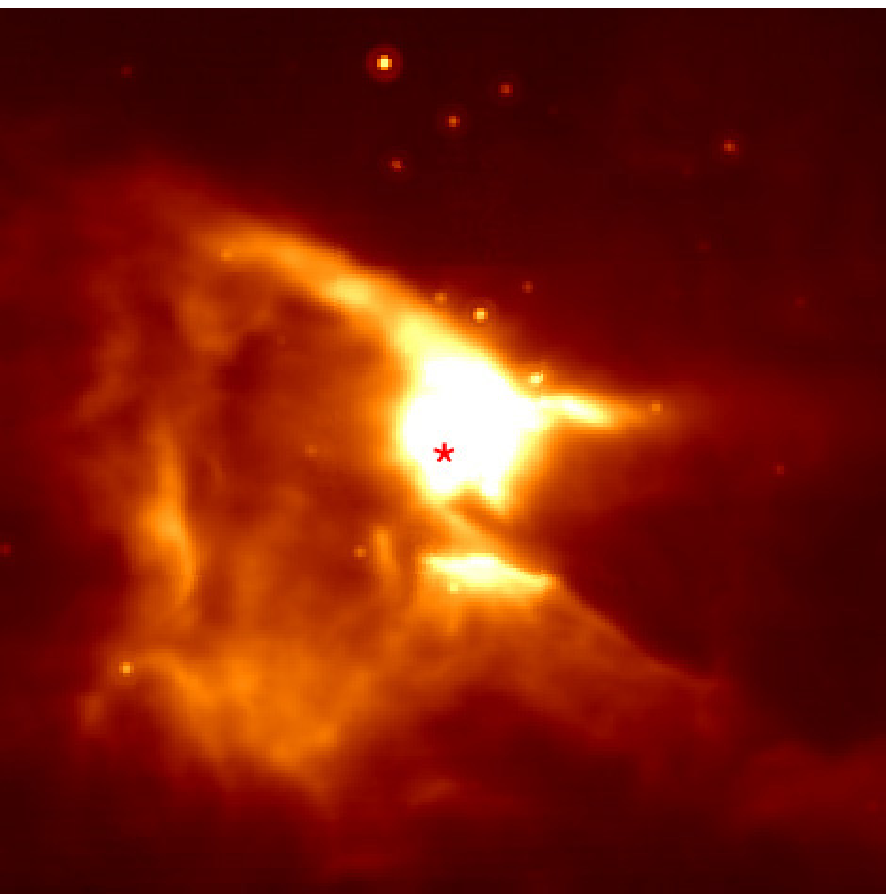}
\includegraphics[width=6cm]{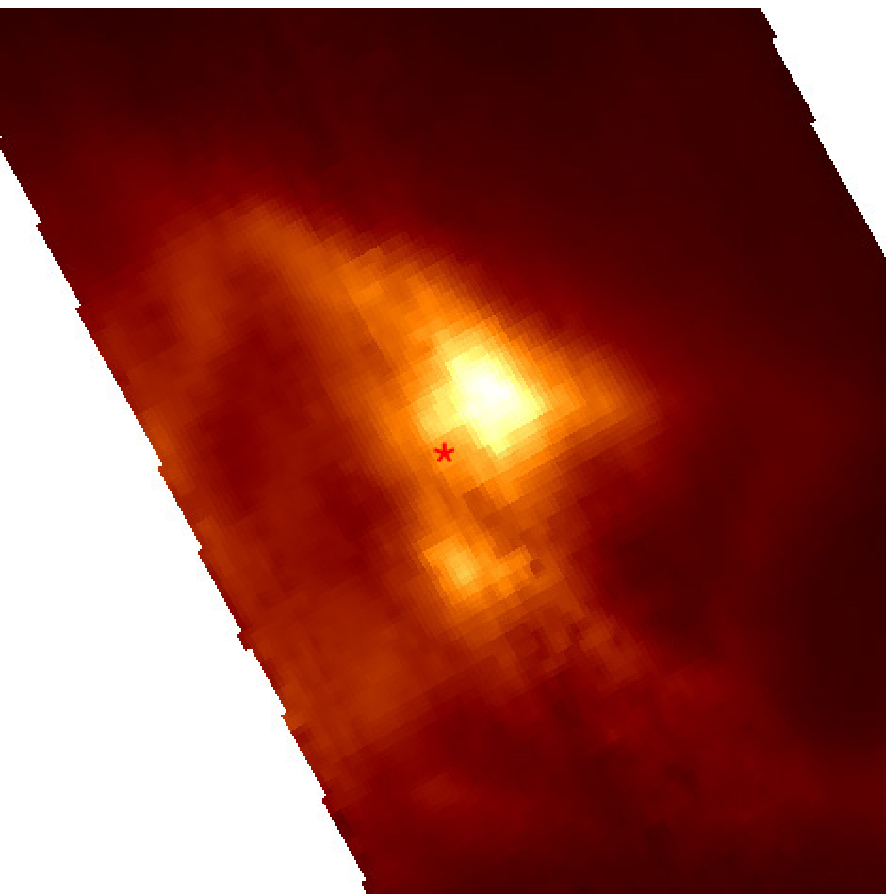}
\caption{Structure of VdB\,139 at visible and infrared wavelengths: POSS-II red image (left), \emph{Spitzer}/MIPS images at 24\,$\mu$m (center) and at 70\,$\mu$m (right). The position of the star is marked with ``$\star$", and the linear scale of the image is given for a distance of 430\,pc. \label{ngc7023}}
\end{figure*}

A plausible analog of a ``young" RS\,Pup is provided by the Herbig Be object \object{HD 200775} and its surrounding dusty reflection nebula \object{VdB 139} (also referred to as \object{NGC 7023}). The central object of this nebula is a binary with a composite spectrum B2Ve. The primary component has a mass of 11\,M$_\odot$ and the secondary component 9\,M$_\odot$, with a system age of only $\approx 100\,000$\,years (Alecian et al.~\cite{alecian08}). Considering their masses, the two stars will cross the instability strip in the HR diagram and become Cepheids within a few tens of million years. The bright pair illuminates a large circumstellar nebula that was extensively studied at visible (Witt et al.~\cite{witt06}; Bern\'e et al.~\cite{berne08}), infrared (Werner et al.~\cite{werner04b}; Sellgren et al.~\cite{sellgren07}), and submillimetric wavelengths (Gerin et al.~\cite{gerin98}).
The distance of HD\,200775 is uncertain. It was estimated at 430\,pc by van den Ancker et al.~(\cite{vda97}) based on the \emph{Hipparcos} parallax, before the star was recognized as a binary. Based on Wolf diagrams, Kun~(\cite{kun98}) find a closer distance of 300\,pc. We adopt here the value from van den Ancker et al.~(\cite{vda97}), although it may be biased by unrecognized orbital motion.
The angular radius of the reflection nebula of about 10 arcminutes correponds to a linear extension of the order of 200\,000\,AU from the central star, which is comparable to the linear radius of the RS\,Pup nebula. In spite of its young age, \object{HD\,200775} already started to carve a cavity in the interstellar dust, and this cavity presents a biconical structure that is well defined at visible, infrared (Werner et al.~\cite{werner04b}) and radio wavelengths (Fuente et al.~\cite{fuente98}).
A visible image from the red POSS-II atlas and \emph{Spitzer}/MIPS images of VdB\,139 at 24 and 70\,$\mu$m (retrieved from the SSC archive) are shown in Fig.~\ref{ngc7023}, to be compared with the images of RS\,Pup presented in Fig.~\ref{rspup-images}. The overall distribution of the thermal infrared emission in VdB\,139 appears similar to RS\,Pup, with irregular filamentary structures and knots. These structures, that are also observable in the visible images of the nebula in scattered light (Witt et al.~\cite{witt06}), are typical of the turbulent ISM in star forming regions (Boulanger~\cite{boulanger07}; Dwarkadas~\cite{dwarkadas08}). 

\subsection{Evolutionary vs. pulsational masses of Cepheids}

From a model of the dust emission based on IRAS photometry, Havlen~(\cite{havlen72}) and McAlary \& Welch~(\cite{mcalary86}) obtain a dust mass of the nebula surrounding RS\,Pup of approximately 0.03\,M$_\odot$, converting to a total gas+dust mass of the order of 3\,M$_\odot$. If the cold component of RS\,Pup's nebula is indeed interstellar in origin, the actual mass originating from the star itself would be much smaller. Such a scenario would strengthen the theoretical results by Caputo et al.~(\cite{caputo05}) and Neilson \& Lester~(\cite{neilson08a}, \cite{neilson08b}) that predict a higher mass loss rate for short period Cepheids.
On the other hand, if the totality of the mass of the RS\,Pup nebula originates from the star, this would be in contradiction with the modeling by Caputo et al.~(\cite{caputo05}), who established that the pulsation and evolutionary masses of RS\,Pup agree well, without evolutionary mass loss.

The conclusion of Caputo et al.~(\cite{caputo05}) is in striking contradiction with the computations by Keller~(\cite{keller08}) using the Padova evolutionary modeling code. This author predicts a uniform mass discrepancy between the evolutionary and pulsation masses, the former being $17 \pm 5$\% larger than the latter, independently of the pulsation period. Applied to RS\,Pup's mass of $\approx 13$\,M$_\odot$, this converts into a total mass loss of 2.2\,M$_\odot$, in excellent agreement with the 3\,M$_\odot$ nebular mass computed by McAlary \& Welch~(\cite{mcalary86}). Keller~(\cite{keller08}) attributes the discrepancy in derived mass with Caputo et al.~(\cite{caputo05}) to a different treatment of convective overshoot between the two models.

However, a problem with Keller's~(\cite{keller08}) model is that the application of the same 17\% mass loss fraction to $\ell$\,Car's (whose mass is also $\approx 13$\,M$_\odot$) leads naturally to the same $\approx 2$\,M$_\odot$ total mass loss, and consequently $\approx 0.02$\,M$_\odot$ of dust. Such a considerable mass should be observable as a CSE similar to RS\,Pup at 24 and 70\,$\mu$m. According to the period-age relationship established by Bono et al.~(\cite{bono05}), $\ell$\,Car and RS\,Pup are between 17 and 19\,million years old. As the main sequence lifetime of a 13\,M$_\odot$ star is $\approx 15$ to 17\,Myr, and assuming that the mass is lost mostly during the He burning phase, the star would have to expel of the order of $10^{-6}$\,M$_\odot$/year continuously over 2\,Myr. Such a rapid and extremely high mass loss rate would necessarily create a massive nebula around $\ell$\,Car, that is not observed at present.  A differential dissipation of their CSEs in the ISM is also difficult to invoke based on the short timescales involved.

\section{Conclusion}

Thanks to a combination of instrumentation in the thermal infrared covering a range of spatial resolutions from arcminutes to milliarcseconds, we presented a characterization of the CSEs of $\ell$\,Car and RS\,Pup both spatially and spectrally. The detected infrared excess relative to the stellar photospheric emission ranges between $10$\% and $50$\% for $\ell$\,Car, from 8 to 70\,$\mu$m. A similar level of emission is also observed around RS\,Pup, and a much more considerable contribution fully outshines the stellar emission longwards of 24\,$\mu$m.
We thus conclude that both CSEs host a warm component with a similar brightness and spatial extension around 10\,$\mu$m, and that RS\,Pup contains in addition a cold dusty ellipsoidal shell located at a few 100\,000\,AU from the star. This cold component is not detected around $\ell$\,Car.

Our interpretation is that the warm CSE components are the consequence of ongoing mass loss from the stars. Based on its spectral and spatial properties, the cold dust shell of the CSE of RS\,Pup is most likely of interstellar origin. We speculate that it was shaped in its present elongated ellipsoidal form during the main sequence lifetime of the then fast rotating B2V progenitor of RS\,Pup. The ISM was compressed by the strong stellar wind and UV radiation preferentially along the polar axis of the star. We also propose that the reflection nebula surrounding the Herbig Be star HD\,200775 is an analogue of RS\,Pup shortly after it formed as a star. In this scenario, the absence of a cold dusty shell around $\ell$\,Car would be explained by a lower ISM density around this star at the time of its formation.
Conversely, the presence of very large dusty envelopes such as that of RS\,Pup would require a high density of dust in the ISM at the time of the formation of the Cepheid progenitor. In this context, their occurrence around Cepheids is probably relatively rare.

The presence of warm CSEs around the two brightest long-period Cepheids shows that many long-period members of this class, if not all, could be surrounded by warm circumstellar envelopes created by ongoing mass loss. This is also the conclusion of M\'erand et al.~(\cite{merand07}) from interferometric observations in the near-IR. In the presently studied cases of $\ell$\,Car and RS\,Pup, the warm CSE flux contribution is negligible in the visible and limited in the near-IR ($\lesssim 5$\% of the photosphere). Their direct photometric impact on the Cepheid distance scale at these wavelengths will thus be minor or negligible. However, as shown by Neilson et al.~(\cite{neilson08c}), the differential behavior of long and short period Cepheids with respect to mass-loss could affect the slope of the P--L relations.
Although weak in the near-infrared, CSEs also have an influence on the determination of Cepheid distances using the Baade-Wesselink method, and particularly its interferometric version (see e.g. M\'erand et al.~\cite{merand06}). If not properly taken into account, the spatially incoherent CSE flux contribution biases the interferometric measurement of the angular diameter, and the derived distance as a consequence.
The detected mid-IR excess is more considerable ($\approx 15\%$ at 10\,$\mu$m) and could noticeably complicate the definition of thermal infrared P--L relations. As suggested by Ngeow \& Kanbur~(\cite{ngeow08}), such relations nevertheless appear highly desirable for future extragalactic Cepheid observations with e.g. the \emph{James Webb Space Telescope}.

\begin{acknowledgements}
We thank Drs. S. Brillant and A. Smette from ESO for their help with the VISIR data processing.
This research made use of the SIMBAD and VIZIER databases at the CDS, Strasbourg (France),
and NASA's Astrophysics Data System Bibliographic Services.
This work is based in part on archival data obtained with the \emph{Spitzer} Space Telescope, which
is operated by the Jet Propulsion Laboratory, California Institute of Technology under a contract with NASA.
This research made use of Tiny Tim/\emph{Spitzer}, developed by John Krist for the \emph{Spitzer} Science Center.
The Second Palomar Observatory Sky Survey (POSS-II) was made by the California Institute of Technology with funds
from the National Science Foundation, the National Geographic Society, the Sloan Foundation, the
Samuel Oschin Foundation, and the Eastman Kodak Corporation. The Oschin Schmidt Telescope is operated by
the California Institute of Technology and Palomar Observatory. 
We also received the support of PHASE, the high angular resolution partnership between 
ONERA, Observatoire de Paris, CNRS, and University Denis Diderot Paris 7.
\end{acknowledgements}

{}

\end{document}